\documentclass[11pt,a4paper]{article}
\pdfoutput=1
\usepackage{jheppub}
\usepackage{amssymb,amsmath,amsthm,amsbsy,amsfonts,mathrsfs,enumerate,float,wrapfig,amsmath, todonotes,MnSymbol,comment, bbm}
\usepackage[utf8]{inputenc}
\usepackage[outline]{contour}
\DeclareUnicodeCharacter{2212}{-}
\usepackage{subfigure}

\newcommand{\nn}{\nonumber}
\newcommand{\be}{\begin{equation}}
	\newcommand{\ee}{\end{equation}}
\newcommand{\bea}{\begin{eqnarray}}
	\newcommand{\eea}{\end{eqnarray}}
\newcommand{\ba}{\begin{align}}
	\newcommand{\ea}{\end{align}}

\newcommand{\SU}{\mathrm{SU}}
\newcommand{\SO}{\mathrm{SO}}
\newcommand{\Spin}{\mathrm{Spin}}
\usepackage{xcolor}

\usepackage{longtable
,multirow}

\usepackage{array}
\newcolumntype{L}[1]{>{\raggedright\let\newline\\\arraybackslash\hspace{0pt}}m{#1}}
\newcolumntype{C}[1]{>{\centering\let\newline\\\arraybackslash\hspace{0pt}}m{#1}}
\newcolumntype{R}[1]{>{\raggedleft\let\newline\\\arraybackslash\hspace{0pt}}m{#1}}

\pdfoutput=1
\usepackage{jheppub}
\usepackage[utf8]{inputenc}
\usepackage{tikz} 
\usepackage{tikz-cd} 

\tikzset{cross/.style={cross out, draw=black,thick, minimum size=2*(#1-\pgflinewidth), inner sep=0pt, outer sep=0pt}, cross/.default={3.5pt}}

\usetikzlibrary{decorations.pathreplacing} 
\usetikzlibrary{decorations.pathmorphing} 
\usetikzlibrary{decorations.markings} 
\usetikzlibrary{arrows} 
\usetikzlibrary{shapes} 
\usetikzlibrary{decorations.pathmorphing}
\usetikzlibrary{positioning} 

\tikzset{7brane/.style={circle, draw=black, fill=black,ultra thick,inner sep=1 pt, minimum size=1 pt,}, c/.default={4pt}}
\tikzset{flavor/.style={regular polygon,regular polygon sides=4,inner sep=2.5pt, draw}}
\tikzset{gauge/.style={circle, draw,inner sep=2.5pt}}
\tikzset{bd/.style={circle, draw=black, inner sep=0pt, fill=black, minimum size=2mm}}
\tikzset{bds/.style={circle, draw=black, inner sep=0pt, fill=black, minimum size=1mm}}
\tikzset{wd/.style={circle, draw=black, inner sep=0pt, fill=white, minimum size=2mm}}
\tikzstyle{ligne}=[draw, very thick] 
\tikzstyle{ligner}=[draw, red, thick]
\tikzstyle{ligneb}=[draw, blue, thick]
\tikzstyle{ligneg}=[draw, green, thick]
\tikzstyle{brane}=[draw] 
\tikzset{sevenb/.style={circle, draw,fill=white,inner sep=2.5pt}}
\tikzstyle{gridline}=[draw, black!20!] 
\tikzset{hasse/.style={circle, fill,inner sep=2pt}}
\tikzset{
tight dash dot/.style={dash pattern=on 2.5pt off 1pt on 1pt off 1pt}
}

\theoremstyle{definition}
\newtheorem{exmp}{Example}[section]

\title{5-brane webs and generalised symmetries}

\author[a]{Mohammad Akhond,} 
\author[b]{Federico Carta,}  
\author[c]{Sung-Soo Kim,}  
\author[d]{and Futoshi Yagi}

\affiliation[a]{Sezione INFN Roma “Tor Vergata” \& Dipartimento di Fisica,\protect\\ \small\it Universit\`a di Roma “Tor Vergata”, Via della Ricerca Scientifica 1, 00133, Roma, Italy}
\affiliation[b]{Istituto Nazionale Di Fisica Nucleare, Sezione di Bologna,\\
Viale Berti Pichat 6/2, 40127 Bologna, Italy}
\affiliation[c]{School of Physics, University of Electronic Science and Technology of China,\\
 No.~2006 Xiyuan Ave, West Hi-Tech Zone, Chengdu, Sichuan 611731, China}
\affiliation[d]{School of Science, Huzhou Normal University, No.~759 East 2nd Road, Huzhou, Zhejiang, 313000, China}

\emailAdd{akhond@roma2.infn.it}
\emailAdd{federico.carta@bo.infn.it}
\emailAdd{sungsoo.kim@uestc.edu.cn}
\emailAdd{60714@huznu.edu.cn}

\abstract{5-brane webs provide a powerful geometric description of five-dimensional quantum field theories, encoding data such as symmetries and dualities. Although the extraction of continuous 0-form symmetries from these systems is well understood, their higher-form symmetries have received less attention. We develop a microscopic prescription for determining 1-form symmetries directly from 5-brane webs. BPS Wilson lines are represented by configurations of D3-branes and $(p,q)$ strings, while their screening by local operators is realised through string junctions, Higgsing, and Hanany-Witten transitions. For webs without orientifold planes, this construction reproduces the Smith normal form prescription based on asymptotic 7-brane charges. We extend the analysis to webs containing O7- and O5- planes. For O7$^-$-planes, the resolved 7-brane charges can be incorporated into the Smith normal form, whereas for O7$^+$-planes we propose an effective charge prescription and test it in both Lagrangian and non-Lagrangian examples. For webs with O5-planes, configurations of half-D3-branes reproduce the centre symmetries of Spin gauge theories and their breaking by matter. Crucially, we find that the 1-form symmetry of an O5-plane system is not determined by asymptotic data alone, but can depend on the local structure of the brane configuration near the orientifold.
}

\begin{document}
	\maketitle
\section{Introduction}\label{sec:introduction}
In recent years, there has been growing interest in generalized global symmetries and in the role they play in quantum field theory~\cite{Gaiotto:2014kfa,Bhardwaj:2020phs} (see~\cite{Cordova:2022ruw,Schafer-Nameki:2023jdn, Bhardwaj:2023kri, Luo:2023ive} for reviews). Unlike ordinary global symmetries, which act on local operators, higher form symmetries act on extended operators. In particular, 1-form symmetries act on line operators and provide useful information about the global structure of gauge theories~\cite{Aharony:2013hda, Kapustin:2014gua}, their phases~\cite{tHooft:1977nqb, Gaiotto:2014kfa}, and possible ’t Hooft anomalies~\cite{Gaiotto:2014kfa}. They are especially valuable in the study of strongly coupled theories, where a conventional Lagrangian description may be absent or incomplete. 

Type IIB 5-brane webs~\cite{Aharony:1997ju, Aharony:1997bh} are an indispensable tool in the study of five-dimensional (5d) supersymmetric quantum field theories (SQFTs)~\cite{Seiberg:1996bd, Morrison:1996xf, Intriligator:1997pq}. Their utility resides in the fact that field theoretic objects have a natural geometric and pictorial representation, making computations more tractable. This correspondence has been thoroughly explored for \emph{local} field theory observables and their associated structures. For instance, the Coulomb branch and the prepotential can be read off from the web~\cite{Aharony:1997bh, Intriligator:1997pq}; Nekrasov partition functions~\cite{Nekrasov:2002qd, Nekrasov:2003rj},  and hence BPS spectra, are computed by applying the (refined) topological vertex to the web~\cite{Aganagic:2003db, Iqbal:2007ii,Awata:2008ed}, also in the presence of O5- and O7$^{+}$-planes~\cite{Kim:2017jqn, Bourgine:2017rik, Hayashi:2020hhb, Kim:2024ufq}; UV completions, symmetry enhancements, and dualities are diagnosed from web deformations, feeding into classification programmes~\cite{Bergman:2013aca, Jefferson:2017ahm, Hayashi:2018lyv, Bhardwaj:2020gyu};  Higgs branches, at finite and infinite coupling, are encoded in magnetic quivers read off from the web, with or without orientifold planes~\cite{Cabrera:2018jxt, Bourget:2019rtl, Bourget:2020gzi, Akhond:2020vhc, vanBeest:2020kou, Akhond:2021ffo, Akhond:2022jts}.
On the other hand, the last decade has seen major field theory developments regarding extended operators and the 1-form symmetry group that acts on them. These issues have received little attention in comparison. See for instance \cite{Bhardwaj:2020phs, Apruzzi:2021nmk, Akhond:2023vlb, Tian:2025yrj} for exceptions to this trend, as well as \cite{Morrison:2020ool, Albertini:2020mdx, Closset:2020scj, DelZotto:2022fnw, Cvetic:2022imb, Apruzzi:2022dlm} for other string theory based techniques. 

It is well known, for instance, that the 0-form symmetry of a 5d theory can be extracted from its 5-brane web~\cite{DeWolfe:1999hj, Bergman:2013aca, Bergman:2015dpa, Apruzzi:2021vcu}. This can be achieved by 7-brane analysis~\cite{DeWolfe:1998eu, DeWolfe:1999hj}, or by extracting the magnetic quiver from the brane web~\cite{Cabrera:2018jxt, Akhond:2020vhc}. For ordinary 5-brane webs, without an orientifold plane, a simple algorithm to compute the 1-form symmetry group based on the Smith normal form of the $[p,q]$ 7-branes was proposed in~\cite{Bhardwaj:2020phs, DelZotto:2022joo}. The goal of this work is to generalise that study to 5-brane webs with orientifold planes~\cite{Sen:1996vd, Bergman:2015dpa, Zafrir:2015ftn, Zafrir:2016jpu, Hayashi:2017btw}. Along the way, a clearer microscopic picture of the screening mechanism is presented, which is valid for brane webs with or without orientifold planes.

Our strategy for detecting the 1-form symmetry, follows the field theory construction reviewed in section \ref{sec:field theory review}. It is similar in strategy to discussions of 1-form symmetries of string setups employed in \cite{Armoni:2022xhy, Amariti:2023hev, Bhardwaj:2020phs}.
We consider the space of BPS Wilson line operators, modulo those that can be screened~\cite{tHooft:1977nqb, Kapustin:2005py, Aharony:2013hda}. Wilson lines can be inserted in 5-brane webs by the addition of D3-branes and $(p,q)$ strings~\cite{Assel:2018rcw} (see also \cite{Gaiotto:2015una, Kim:2021gyj} for computations of their expectation values and \cite{Kim:2021gyj, Huang:2022hdo, Liu:2026mvz, Harding:2026gzj} for related blow-up computations). 
Our proposal is that their screening corresponds to the possibility of breaking the $(p,q)$ strings into segments suspended between D3-branes and $[p,q]$ 7-branes. Since both the D3-branes and $[p,q]$ 7-branes extend along the Higgs branch direction, suspended $(p,q)$ string segments can be \emph{Higgsed} away.\footnote{Here Higgsing refers to the geometric process of moving the string segment along the directions transverse to the web. It should not be confused with Higgsing the 5d theory, in particular the 5d gauge group remains unbroken throughout this process.} The remaining D3-branes can then be moved to infinity in the plane of the web, with Hanany-Witten transitions used to annihilate any remaining $(p,q)$ strings. Our screening mechanism reproduces the known 1-form symmetry of ordinary 5-brane webs and, moreover, naturally generalises to brane systems containing orientifold planes. Furthermore, since the mechanism is phrased in terms of string theoretic objects, it can be applied to non-Lagrangian theories.

To detect the 1-form symmetry group for a gauge theory, one has to consider Wilson lines in the minimal representation of the centre of the gauge group. For $\mathrm{Spin}(N)$ gauge theory, this corresponds to matter in (co)spinor representation~\cite{Zafrir:2015ftn, Hayashi:2018bkd}. The insertion of such Wilson lines has not appeared in the literature, and so a proposal for the corresponding insertion into brane webs is presented here.

A surprising outcome of our study is that the 1-form symmetry group of Spin$(N)$ gauge theories constructed from 5-brane webs with O5-planes~\cite{Zafrir:2015ftn, Hayashi:2017btw} depends not only on the asymptotic charges at infinity, but also on the local structure of the brane configuration near the origin of the web. This observation goes beyond the usual lore in the present literature, 
where the global variants of a quantum field theory~\cite{Aharony:2013hda} are expected to be fixed by the boundary values of supergravity fields at infinity~\cite{Witten:1998wy, Aharony:1998qu, GarciaEtxebarria:2019caf, Bergman:2020ifi, Apruzzi:2021phx, Apruzzi:2021nmk}.

The rest of this paper is organised as follows. In section~\ref{sec:field theory review}, we review line operators and their associated 1-form symmetries. In the subsequent sections, we discuss how 1-form symmetries can be identified and realised in 5-brane webs. In section~\ref{sec:linesinweb}, we discuss 5d $\mathcal{N}=1$ theories whose 5-brane webs do not involve orientifold planes, focusing on the screening of Wilson lines and Smith normal form. In section~\ref{sec:withO7}, we conjecture that the 1-form symmetries of theories engineered by brane configurations containing an $\mathrm{O}7^{+}$-plane can be determined from its effective monodromy charge. 
We then describe how the 1-form symmetries of Spin gauge theories are realised in 5-brane webs involving an O5-plane. Finally, we summarise our results and discuss possible future directions in section~\ref{sec:conclusion}.

\vspace{0.5cm}
\section{Review: Line operators and 1-form symmetries}\label{sec:field theory review}
This section reviews the field theoretic origin of 1-form symmetries, as the symmetry acting on the line operators. 1-form symmetries of 5d gauge theories and SCFTs have been explored in \cite{Apruzzi:2022dlm, Bhardwaj:2020phs, Cvetic:2022imb, DelZotto:2022joo}.

\subsection{1-form symmetry}
\label{sec:oneform}
Consider a field theory, and let $L$ denote the set of all genuine line operators. We define a binary relation on $L$, which we denote by $\sim$ in the following way: we say that $L_1, L_2 \in L$ are \emph{equivalent} if and only if there exists a line-changing pointlike operator that connects them. It is easy to show that this binary relation on $L$ is reflexive, symmetric, and transitive, thus defines an equivalence relation. 

Let $\hat{\Gamma}^{(1)}:=L/\sim$ denote the quotient set of $L$ by the equivalence relation defined above. Elements of $\hat{\Gamma}^{(1)}$ are equivalence classes of lines, such that any representative of a given class can be joined to any other representative of the same class by a pointlike operator. In particular, consider the equivalence class of lines which contains the trivial line. Its elements are all lines that can be connected to the trivial line, therefore, all lines that can end on a given pointlike operator. We call any representative of this latter class as a \emph{screened line}. The 1-form symmetry group $\Gamma^{(1)}$ is by definition the Pontryagin dual group of $\hat{\Gamma}^{(1)}$ (See Appendix~\ref{app:pontryagin} for more details.), namely
\begin{equation}
\Gamma^{(1)}:=Hom(\hat{\Gamma}^{(1)}, U(1))\ .
\label{eq::Pontryagin}
\end{equation}
It holds that whenever $\hat{\Gamma}^{(1)}\simeq \mathbb{Z}_n$, then $$\Gamma^{(1)}=Hom(\mathbb{Z}_n, U(1))\simeq \mathbb{Z}_n\,.$$ Most of the cases we will consider are of this form.

\subsection{Wilson lines}\label{sec:Wilson Lines}

Consider a gauge theory based on the gauge group $G$. There exists a subset $W\subset L$ of lines which are called Wilson lines. They are defined by 
\begin{equation}
W_R[C]=\mathcal{P}\exp\left(i\int_C A\right)\ ,
\label{eq::Wline_def}
\end{equation} 
where $C$ is a curve in spacetime, $\mathcal{P}$ is the path-ordering, and $A=A^aT^a_R$ is the gauge connection in the representation $R$ of the gauge algebra.

Suppose we consider a Wilson line $W_R[C]$, extending from a point $x_0$ to a point $x_1$ (and we take $x_0\neq x_1$ so we don't consider closed Wilson loops). Let us stress that we take $x_0$ and $x_1$ to be properly contained in spacetime, and not located at infinity. Under a local gauge transformation the Wilson line will transform as
\begin{equation}
W_R[C]\to W_R'[C] = g(x_1)W_R[C]g^{-1}(x_0) \ ,
\end{equation}
therefore it is not a gauge invariant operator. In order to make a gauge-invariant operator using a line that extends on a finite path, it must be screened. By the definition of screening given above, this means there must exist two local operators $\psi(x_0)$ and $\psi(x_1)$, transforming in the gauge representation $R$, such that  $\psi^{\dagger}(x_1)W_R[C]\psi(x_0)$ is a gauge invariant operator. Of course there can be many different point operators that transform in the right way under the gauge symmetry, so that they can be used to make gauge invariant lines.

The situation is different for lines that extend to infinity in both directions, as in that case one can put boundary conditions at infinity to make such an operator gauge invariant. We will not enter in details of this construction, but one heuristic way to think about it is first to take spacetime to be a box with periodic boundary conditions and then send the size of the box to infinity.

Let us now consider a screened Wilson line $W_R[C]$ and take $R$ to be a reducible representation of $G$. For simplicity, take $R=R_1\oplus R_2$ with $R_1, R_2$ irreducible and so that the representation space $V_R = V_{R_1}\oplus V_{R_2}$, on which $W_{R_i}[C]$ acts. In a proper basis, Wilson line is block diagonal, $W_R[C] = W_{R_1}[C] \oplus W_{R_2}[C].$ It then follows from the definition \eqref{eq::Wline_def},  it holds that, for $\psi = \psi_1 \oplus \psi_2$ with
$\psi_i \in V_{R_i},$
\begin{align}\label{eq:reducible_screen}
\psi^{\dagger}(x_1)W_R[C]\psi(x_0)=\psi_1^{\dagger}(x_1)W_{R_1}[C]\psi_1(x_0)
+\psi_2^{\dagger}(x_1)W_{R_2}[C]\psi_2(x_0)\ .     
\end{align}
If $\psi_i$ is a dynamical local operator, then the corresponding Wilson line $W_{R_i}[C]$ can end on $\psi_i$ and is therefore  screened.

\subsection{1-form symmetry from congruence numbers}\label{sec:global form}

Wilson lines in gauge theories transform under a group isomorphic to the centre $Z$ of the gauge group. Thus the global 1-form symmetry is determined by examining the subgroup of $Z$ that acts on the set of unscreened lines. This is in turn determined by the charges of the matter fields under the centre $Z$. Mathematically, this is known as the \emph{congruence number} of the gauge representations of the matter fields. This subsection reviews the notion of congruence numbers for simple Lie algebras. This will be used heavily in section~\ref{sec:withO5}, to give a physical representation of the congruence numbers in brane webs, and to determine the 1-form symmetry of theories constructed with brane webs in the presence of an O5-plane.

Let $\mathfrak{g}$ be a simple Lie algebra of rank $r$. Let $G$ be the unique connected and simply connected group that has $\mathfrak{g}$ as a Lie algebra. We denote by $\mathcal{R}$ the set of all irreducible representations of $\mathfrak{g}$, and refer to elements  $R\in\mathcal{R}$ in terms of their Dynkin labels $[a_1,\cdots, a_r]$ (where the quantities $a_i$ are non-negative integer numbers). On the set $\mathcal{R}$, there is a binary relation defined in the following way: two representations $R_i$ and $R_j$ are related (i.e., $R_i\sim R_j$) if and only if the difference of any weight of $R_i$ and $R_j$ is a linear combination of simple roots of $\mathfrak{g}$, with integer coefficients. Such binary relation is an equivalence relation~\cite{Lemire:1979qd}. We can then take the quotient set $\mathcal{C}\simeq \mathcal{R}/\sim$, and we will refer to elements of the quotient set as ``congruence classes" of representations, in the following discussion. Let us see this fact explicitly in one example, before giving the general result.

\begin{exmp}
For this example fix $\mathfrak{g}=\mathfrak{su}(2)$. There is only one positive simple root of the complex Lie algebra, which is given by $\alpha=2$. The negative simple root is $-\alpha=-2$. An irreducible representation $R$ with Dynkin label $[a]$ will have $\dim R=a+1$, and its weights are  given by $\{a, a-2, a-4, \cdots, 2-a, -a  \}$. We can then clearly see that any two representations with even highest weights are equivalent, since the difference between their highest weights is even and can therefore be expressed as an integer multiple of $\alpha$. Similarly, any two irreducible representations with odd highest weights are equivalent, while a representation with an even highest weight is not equivalent to one with an odd highest weight. There are therefore two congruence classes, which may be labelled by $a \mod 2$.
\end{exmp}

A similar procedure can be done for any choice of $\mathfrak{g}$ resulting in the following.  We label congruence classes with a congruence number $c$. The congruence number of a given representation $[a_1,\cdots, a_r]$ can be calculated in terms of the Dynkin labels as \cite{Lemire:1979qd}:
\begin{align}
c&=\sum_{i=1}^{n-1} ia_i \pmod n \quad  \mbox{for}\ \mathrm{SU}(n) \\
c&=a_n \pmod 2 \quad  \mbox{for}\ \mathrm{Spin}(2n+1) \\
c&=\sum_{k=0}^{\lfloor \frac{n-1}{2}\rfloor}  a_{2k+1} \pmod 2 \quad \mbox{for}\ \mathrm{Sp}(n)\\
c&=\bigg(a_{n-1}+a_n, \quad  2\sum_{k=0}^{\lfloor \frac{n-3}{2}\rfloor}  a_{2k+1} +(n-2)a_{n-1}+na_n \bigg) \quad  \mbox{for}\ \mathrm{Spin}(2n) \label{eq:c for Spin(2n)}\\
&\quad \pmod 2 \qquad \qquad\pmod 4\cr
c&=a_1-a_2+a_4-a_5 \pmod 3 \quad \mbox{for}\ \mathrm{E}_6\\
c&=a_4+a_6+a_7 \pmod 2 \quad \mbox{for}\ \mathrm{E}_7\\
c&=0 \quad \mbox{for} \quad \ \mathrm{E}_8, \ \mathrm{F}_4, \ \mathrm{G}_2.
\end{align}

The case of $\mathrm{Spin}(2n)$ requires some further discussion. Both for $n$ even and odd, the congruence classes of irreps are labelled by a vector $(c_1,c_2)$ where $c_1$ is calculated $\pmod 2$ and $c_2$ is calculated $\pmod 4$. The expression of such charges in terms of the Dynkin labels is given in \eqref{eq:c for Spin(2n)}. In particular, let us discuss the two cases separately:
\begin{enumerate}
    \item For $\mathrm{Spin}(4n)$, the only vectors that may occur are $(0,0),(0,2),(1,0),(1,2)$. 
    \item For $\mathrm{Spin}(4n+2)$, the only vectors that may occur are $(0,0),(0,2),(1,1),(1,3)$.
\end{enumerate}

Let us discuss an important point. Let $Z$ be the centre of $G$. On the set of congruence classes $\mathcal{C}$, one can define a group structure, and there exists an isomorphism $\mathcal{C}\simeq Z$. We will see this in more detail in the following example, which treats the case of $\mathrm{Spin}(2n)$.

\begin{exmp}
The centre of $\mathrm{Spin}(2n)$ is given by $Z=\mathbb{Z}_2\times \mathbb{Z}_2$ if $n$ is even, and $Z=\mathbb{Z}_4$ if $n$ is odd. Let us discuss these cases separately.
\begin{enumerate}
    \item When $n$ is even, $\mathcal{C}=\{(0,0),(0,2),(1,0),(1,2)\}$. On the other hand, $Z=\mathbb{Z}_2\times \mathbb{Z}_2$. By denoting $e$ and $\rho$ the identity element and the non-trivial element of $\mathbb{Z}_2$, the elements of the Cartesian product $Z=\mathbb{Z}_2\times \mathbb{Z}_2$ are labelled by ordered couples of elements of the two factors, namely $Z=\{(e_1,e_2),(\rho_1, e_2),(e_1,\rho_2),(\rho_1,\rho_2)\}$. It is easy to prove that the following map $\phi:Z\to \mathcal{C}$ is a group isomorphism:
    \begin{equation}\label{eq:ZtoC-isomorph}
        \phi: 
\begin{array}{c}
(e_1,e_2)\mapsto (0,0) \\
(\rho_1,e_2)\mapsto (1,0) \\
(e_1,\rho_2)\mapsto (0,2) \\
(\rho_1,\rho_2)\mapsto (1,2)
\end{array}
        \end{equation}
    \item When $n$ is odd, the only vectors $c$ that may occur are $(0,0),(0,2),(1,1),(1,3)$. On the other hand, $Z=\mathbb{Z}_4$. By denoting $e$ and $\rho$ the identity element and the non-trivial element of $\mathbb{Z}_4$, the elements of $Z$ can be denoted as $Z=\{e,\rho,\rho^2,\rho^3\}$. It is easy to prove that the following map $\phi:Z\to \mathcal{C}$ is a group isomorphism:
    \begin{equation}
        \phi: 
\begin{array}{c}
e\mapsto (0,0) \\
\rho \mapsto (1,1) \\
\rho^2 \mapsto (0,2) \\
\rho^3 \mapsto (1,3)\ .
\end{array}
\end{equation}
\end{enumerate}
\end{exmp}

The 1-form symmetry is isomorphic to the subgroup of the centre $Z$ obtained after imposing an additional constraint, that sets to identity the element of $Z$ corresponding to the congruence numbers of the matter representations.

\vspace{0.5cm}
\section{Wilson Lines in 5-brane webs {\it without} O-planes}
\label{sec:linesinweb}

In this section, we consider 5d $\mathcal{N}=1$ supersymmetric quantum field theories (SQFTs) that admit a description as the worldvolume theory on a web of $(p,q)$ 5-branes in type IIB superstring theory. See \cite{Aharony:1997ju,Aharony:1997bh,Brunner:1997gk,Benini:2009gi, Bergman:2015dpa,Zafrir:2015ftn,Hayashi:2018lyv} for references. In this picture, the open string (F1) extended along the $x^0,x^5$-directions sources electric flux ending on the D5-branes, which is the same role that the Wilson line plays in the field theory. Namely, it inserts an external, infinitely heavy charge in the fundamental representation. More precisely, BPS Wilson line operators are given by bound states of probe D3-branes along the $x^0,x^7,x^8, x^9$ directions and $(p,q)$ strings. The relative position of strings and branes is given in table \ref{table:brane_positions}. The representation $R$ of the Wilson line will depend explicitly on how many D3-branes are present and the type of $(p,q)$ strings considered.

\begin{table}[ht]
\centering
\begin{tabular}{c|ccccc|cc|ccc}
\hline
Brane & 0 & 1 & 2 & 3 & 4 & 5 & 6 & 7 & 8 & 9 \\
\hline
D5 or (1,0) / O5   & $\times$ & $\times$ & $\times$ & $\times$ & $\times$ &       & $\times$ &       &       &       \\
NS5 or (0,1)  & $\times$ & $\times$ & $\times$ & $\times$ & $\times$ & $\times$ &       &       &       &       \\
F1    & $\times$ &       &       &       &       & $\times$      &       &  &  &  \\
D1    & $\times$ &       &       &       &       &       &  $\times$     & &  &  \\
D3    & $\times$ &       &       &       &       &       &       & $\times$ & $\times$ & $\times$ \\
$[p,q]~7$/ O7    & $\times$ & $\times$      & $\times$      & $\times$      & $\times$      &       &       & $\times$ & $\times$ & $\times$ \\

\hline
\end{tabular}
\caption{Summary of the various constituents of the 5-brane web with the insertion of Wilson lines. A $\times$ means the object is extended in the corresponding direction. Note that the O5-plane (resp. O7) extends along the same dimensions as the $(1,0)$ 5-brane (resp. $[p,q]$ 7-brane). In this section no orientifold planes will be discussed, but they are added to this table for future reference.}
\label{table:brane_positions}
\end{table}

\subsection{\texorpdfstring{$\SU(2)_\theta$}{SU(2)theta}}
\label{sec:E1th}
We consider two $\SU(2)_\theta$ theories with different discrete theta angle $\theta =0,\pi$ and how screening of Wilson lines is realized on their 5-brane web. 

\subsection*{\underline{\texorpdfstring{$\SU(2)_0$}{SU(2)0} theory}}
We first consider the $\SU(2)_0$ theory or the $E_1$ theory. Given a representation of $\mathrm{SU}(2)$, one would like to identify a configuration of $(p,q)$ strings and D3-branes associated with each state of that representation. A Wilson line corresponding to an irreducible representation of $\mathrm{SU}(2)$ with Dynkin label $[n]$ will be denoted by $W_{[n]}$. The character for such a representation takes the form
\begin{equation}
    \chi_{[n]}=\sum_{j=0}^nx^{n-2j}\ .
\end{equation}

Our general proposal is very simple: to a configuration of $n$ D3-branes inserted in the compact face of the $E_1$ web, one associates the weight $x^n$, which is the highest weight appearing in the character, while a D3-brane outside the compact face is a singlet. To lower the states, one introduces a W-boson, namely an F1-string suspended between the two D5-branes. A configuration of $n$ D3-branes and $j$ F1-strings suspended between the two D5-branes is associated with the state $x^{n-2j}$. 

To make this proposal consistent with Hanany-Witten transitions, one has to make a refinement of this statement. Let us first fix an orientation, and without loss of generality assume that F1-strings always point downwards. An F1-string starting (ending) on the upper D5-brane is to be associated with a factor of $x^{-1}$ ($x$). Similarly, an F1-string ending (starting) from the lower D5-brane is associated with a factor of $x^{-1}$ ($x$) respectively. With this convention one recovers the advertised factor of $x^{-2j}$ associated with a W-boson, though this definition is also consistent with Hanany-Witten moves, which take D3-branes out of the compact face in the vertical direction. For example, if one takes a D3-brane from the compact face and moves it beyond the upper D5-brane, it no longer contributes to the character, but instead an F1-string will be created between this D3, and ending on the upper D5, leaving the character invariant.

\begin{figure}[H]
	\centering
	\begin{tikzpicture}
		\node at (0,2) {\raisebox{-.5\height}{ \scalebox{3}{  
					\begin{tikzpicture}[x=.5cm,y=.5cm] 
						\draw (0,0)--(0.6,0.6);
						\draw (0,2)--(0.6,1.4);
						\draw (2,0)--(1.4,0.6);
						\draw (1.4,1.4)--(2,2);
						\draw (0.6,0.6)--(1.4,0.6);
						\draw (0.6,1.4)--(1.4,1.4);
						\draw (0.6,0.6)--(0.6,1.4);
						\draw (1.4,0.6)--(1.4,1.4);
                        \node at (1,1)[cross, red,scale=0.4, line width=0.5pt]{};

						\draw (3,0)--(3.6,0.6);
						\draw (3,2)--(3.6,1.4);
						\draw (5,0)--(4.4,0.6);
						\draw (4.4,1.4)--(5,2);
						\draw (3.6,0.6)--(4.4,0.6);
						\draw (3.6,1.4)--(4.4,1.4);
						\draw (3.6,0.6)--(3.6,1.4);
						\draw (4.4,0.6)--(4.4,1.4);
                            
                            \draw[blue, decorate, decoration={snake, segment length=5pt, amplitude=1pt}] (4,1.4)--(4,0.6);
                            \node at (4,1)[cross, red,scale=0.4, line width=0.5pt]{};
		\end{tikzpicture}}}};
	\end{tikzpicture}
	\caption{Brane-web description of the two states of a Wilson Line in the fundamental representation of $\mathfrak{su}(2)$. Here, D3-branes are marked as red crosses, and fundamental strings are denoted by wiggly lines. In terms of the character $\chi_{[1]}(x)=x+x^{-1}$, these two configurations correspond to $x$ and $x^{-1}$.}
	\label{fig:fundline}
\end{figure}
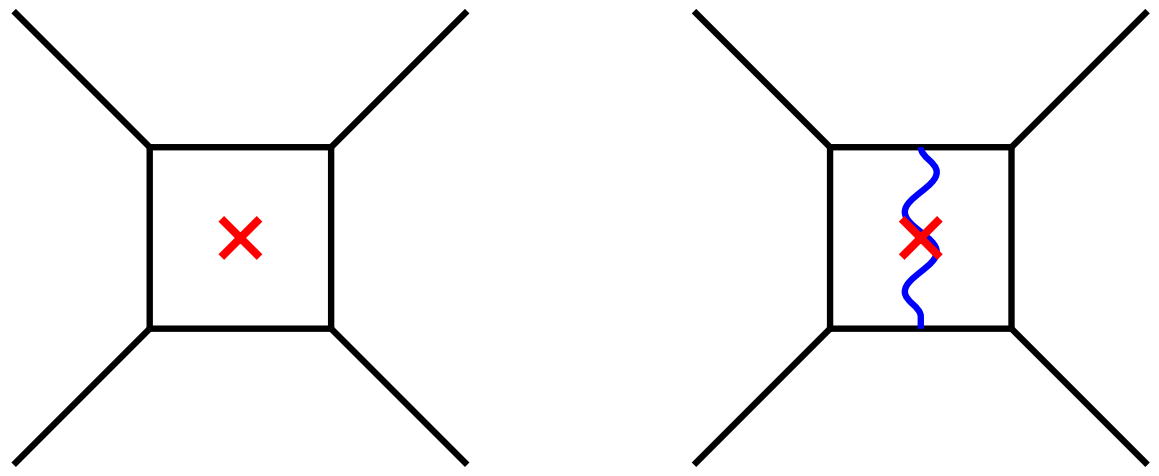

As an illustrative example of the above discussion, consider the fundamental representation $[1]$ of $\mathrm{SU}(2)$. The two states of $W_{[1]}$ are realized as follows: the highest weight is a D3-brane placed between the upper and lower D5-brane. The lowest weight is a F1-string that extends from the lower D5-brane to the upper one, and a D3-brane on top of it. See Figure~\ref{fig:fundline} for clarity and \cite{Assel:2018rcw}.~

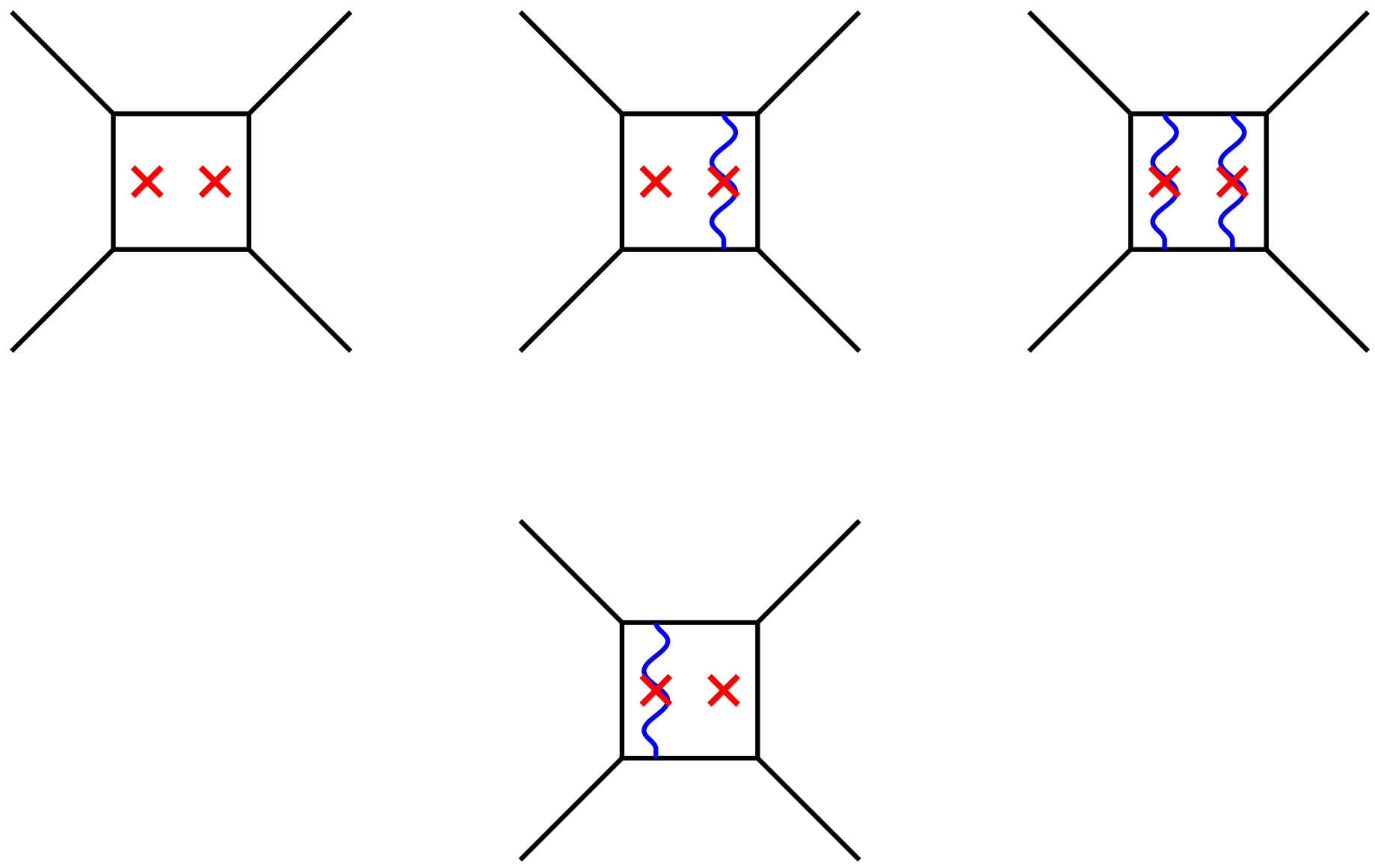
\begin{figure}[H]
	\centering
	\begin{tikzpicture}
		\node at (0,2) {\raisebox{-.5\height}{ \scalebox{3}{  
					\begin{tikzpicture}[x=.5cm,y=.5cm] 
						\draw (0,0)--(0.6,0.6);
						\draw (0,2)--(0.6,1.4);
						\draw (2,0)--(1.4,0.6);
						\draw (1.4,1.4)--(2,2);
						\draw (0.6,0.6)--(1.4,0.6);
						\draw (0.6,1.4)--(1.4,1.4);
						\draw (0.6,0.6)--(0.6,1.4);
						\draw (1.4,0.6)--(1.4,1.4);
                        \node at (0.8,1)[cross, red,scale=0.4, line width=0.5pt]{};
                        \node at (1.2,1)[cross, red,scale=0.4, line width=0.5pt]{};
						\draw (3,0)--(3.6,0.6);
						\draw (3,2)--(3.6,1.4);
						\draw (5,0)--(4.4,0.6);
						\draw (4.4,1.4)--(5,2);
						\draw (3.6,0.6)--(4.4,0.6);
						\draw (3.6,1.4)--(4.4,1.4);
						\draw (3.6,0.6)--(3.6,1.4);
						\draw (4.4,0.6)--(4.4,1.4);

                    \draw[blue, decorate, decoration={snake, segment length=5pt, amplitude=1pt}] (1.2+3,1.4)--(1.2+3,0.6);  
                        \node at (0+3.8,1)[cross, red,scale=0.4, line width=0.5pt]{};
                        \node at (1.2+3,1)[cross, red,scale=0.4, line width=0.5pt]{}; 
						\draw (3+3,0-3+3)--(3.6+3,0.6-3+3);
						\draw (3+3,2-3+3)--(3.6+3,1.4-3+3);
						\draw (5+3,0-3+3)--(4.4+3,0.6-3+3);
						\draw (4.4+3,1.4-3+3)--(5+3,2-3+3);
						\draw (3.6+3,0.6-3+3)--(4.4+3,0.6-3+3);
						\draw (3.6+3,1.4-3+3)--(4.4+3,1.4-3+3);
						\draw (3.6+3,0.6-3+3)--(3.6+3,1.4-3+3);
						\draw (4.4+3,0.6-3+3)--(4.4+3,1.4-3+3);

                        \draw[blue, decorate, decoration={snake, segment length=5pt, amplitude=1pt}] (0.8+3+3,1.4-3+3)--(0.8+3+3,0.6-3+3);
                        \draw[blue, decorate, decoration={snake, segment length=5pt, amplitude=1pt}] (1.2+3+3,1.4-3+3)--(1.2+3+3,0.6-3+3);
                        \node at (0+3.8+3,1-3+3)[cross, red,scale=0.4, line width=0.5pt]{};
                        \node at (1.2+3+3,1-3+3)[cross, red,scale=0.4, line width=0.5pt]{};

                        \draw (0+3,0-3)--(0.6+3,0.6-3);
						\draw (0+3,2-3)--(0.6+3,1.4-3);
						\draw (2+3,0-3)--(1.4+3,0.6-3);
						\draw (1.4+3,1.4-3)--(2+3,2-3);
						\draw (0.6+3,0.6-3)--(1.4+3,0.6-3);
						\draw (0.6+3,1.4-3)--(1.4+3,1.4-3);
						\draw (0.6+3,0.6-3)--(0.6+3,1.4-3);
						\draw (1.4+3,0.6-3)--(1.4+3,1.4-3);

                        \draw[blue, decorate, decoration={snake, segment length=5pt, amplitude=1pt}] (0.8+3,1.4-3)--(0.8+3,0.6-3);  
                        \node at (0.8+3,1-3)[cross, red,scale=0.4, line width=0.5pt]{};
                        \node at (1.2+3,1-3)[cross, red,scale=0.4, line width=0.5pt]{};

		\end{tikzpicture}}}};
	\end{tikzpicture}
	\caption{The four different states of the $W_{[1]\otimes[1]}$ line. Following $[1]\otimes[1] = [2]\oplus[0]$, we organize that the top line represents three states of $[2]$ (or in terms of the character, $\chi_{[2]}(x)=x^2+1+x^{-2}$), and the bottom one represents the singlet state $[0]$ (or in terms of the character, $\chi_{[0]}(x)=1$).}
	\label{fig:tensor_line}
\end{figure}

One can further consider a Wilson line in the tensor product representation $[1]\otimes [1]$. This representation has four states, depicted in figure \ref{fig:tensor_line}.

We want now to discuss the screening mechanism, from the point of view of the brane web. There are many ways to see the screening. In the following we discuss two ways which we call respectively \emph{Wilson line screening} and  \emph{7-brane screening}.\\

\paragraph{Wilson line screening}

\begin{figure}[H]
	\centering
	\begin{tikzpicture}
		\node at (0,2) {\raisebox{-.5\height}{ \scalebox{3}{  
					\begin{tikzpicture}[x=.5cm,y=.5cm] 
						\draw (0,0)--(0.6,0.6);
						\draw (0,2)--(0.6,1.4);
						\draw (2,0)--(1.4,0.6);
						\draw (1.4,1.4)--(2,2);
						\draw (0.6,0.6)--(1.4,0.6);
						\draw (0.6,1.4)--(1.4,1.4);
						\draw (0.6,0.6)--(0.6,1.4);
						\draw (1.4,0.6)--(1.4,1.4);
                        \node at (0.8,1)[cross, red,scale=0.4, line width=0.5pt]{};
                        \node at (1.2,1)[cross, red,scale=0.4, line width=0.5pt]{};
                          \draw[blue, decorate, decoration={snake, segment length=5pt, amplitude=1pt}] (1.2,1.4)--(1.2,0.6);
						\draw (3,0)--(3.6,0.6);
						\draw (3,2)--(3.6,1.4);
						\draw (5,0)--(4.4,0.6);
						\draw (4.4,1.4)--(5,2);
						\draw (3.6,0.6)--(4.4,0.6);
						\draw (3.6,1.4)--(4.4,1.4);
						\draw (3.6,0.6)--(3.6,1.4);
						\draw (4.4,0.6)--(4.4,1.4);

                    \draw[blue, decorate, decoration={snake, segment length=5pt, amplitude=1pt}] (1.2+3,1.4)--(1.2+3,0.6);  
                        \node at (1.2+3,0.8)[cross, red,scale=0.4, line width=0.5pt]{};
                        \node at (1.2+3,1.2)[cross, red,scale=0.4, line width=0.5pt]{};

                \draw[ densely dotted, scale=0.5] (14.2,2.3)--(12,1.8);
                \draw[blue, decorate, decoration={snake, segment length=2.1pt, amplitude=0.8pt}](6.2,0.95)--(6.2,0.55);
                \draw[densely dotted, scale=0.5] (14.2,2.3-0.8)--(12,1.8-0.8);
                \node[label=left:\scalebox{0.4}{$x^{7,8,9}$}] at (6.5,0.8){};

						\draw (3+3,0)--(3.6+3,0.6);
						\draw (3+3,2)--(3.6+3,1.4);
						\draw (5+3,0)--(4.4+3,0.6);
						\draw (4.4+3,1.4)--(5+3,2);
						\draw (3.6+3,0.6)--(4.4+3,0.6);
						\draw (3.6+3,1.4)--(4.4+3,1.4);
						\draw (3.6+3,0.6)--(3.6+3,1.4);
						\draw (4.4+3,0.6)--(4.4+3,1.4);

                    \draw[blue, decorate, decoration={snake, segment length=5pt, amplitude=1pt}] (1.2+3+3,0.8)--(1.2+3+3,0.6);  
                    \draw[blue, decorate, decoration={snake, segment length=5pt, amplitude=1pt}] (1.2+3+3,1.2)--(1.2+3+3,1.4);  
                        \node at (1.2+3+3,0.8)[cross, red,scale=0.4, line width=0.5pt]{};
                        \node at (1.2+3+3,1.2)[cross, red,scale=0.4, line width=0.5pt]{}; 
						\draw (3,-3)--(3.6,0.6-3);
						\draw (3,2-3)--(3.6,1.4-3);
						\draw (5,0-3)--(4.4,0.6-3);
						\draw (4.4,1.4-3)--(5,2-3);
						\draw (3.6,0.6-3)--(4.4,0.6-3);
						\draw (3.6,1.4-3)--(4.4,1.4-3);
						\draw (3.6,0.6-3)--(3.6,1.4-3);
						\draw (4.4,0.6-3)--(4.4,1.4-3);
                        \node at (1.2+3,0.8-3.5)[cross, red,scale=0.4, line width=0.5pt]{};
                        \node at (1.2+3,1.2-2.5)[cross, red,scale=0.4, line width=0.5pt]{};

		\end{tikzpicture}}}};
	\end{tikzpicture}
	\caption{Wilson line screening for the singlet states of $W_{[1]\otimes[1]}$}
	\label{fig:tensor_line_higgs}
\end{figure}
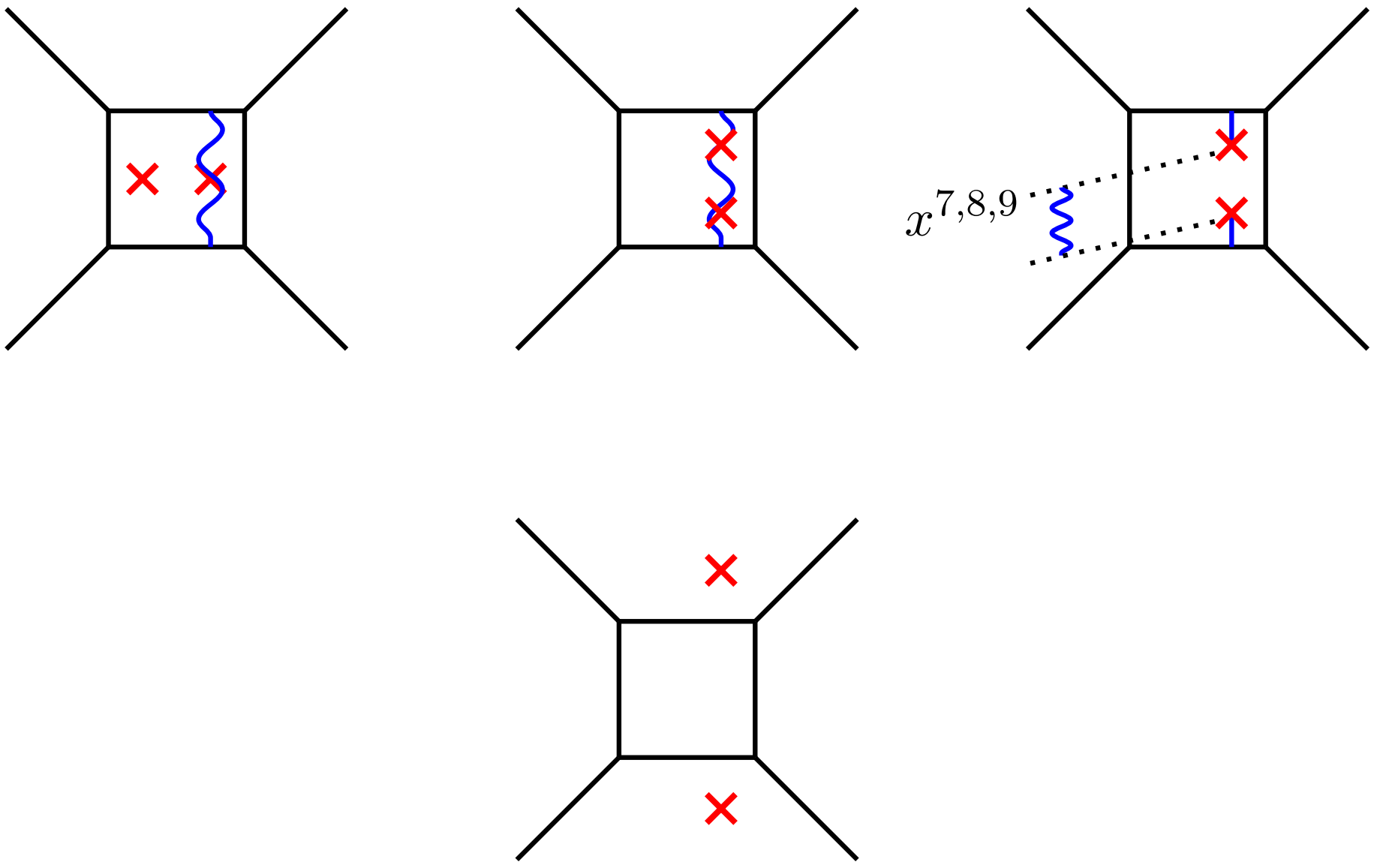

Consider for example the second state depicted in figure \ref{fig:tensor_line}. One can move both D3s on top of the $(0,1)$-string. The middle segment of the string can now be lifted out of the plane of the picture. The remaining D3s and $(0,1)$-string segments can now be moved away to infinity, in the plane of the picture, by Hanany-Witten moves. This mechanism is depicted in figure \ref{fig:tensor_line_higgs}.

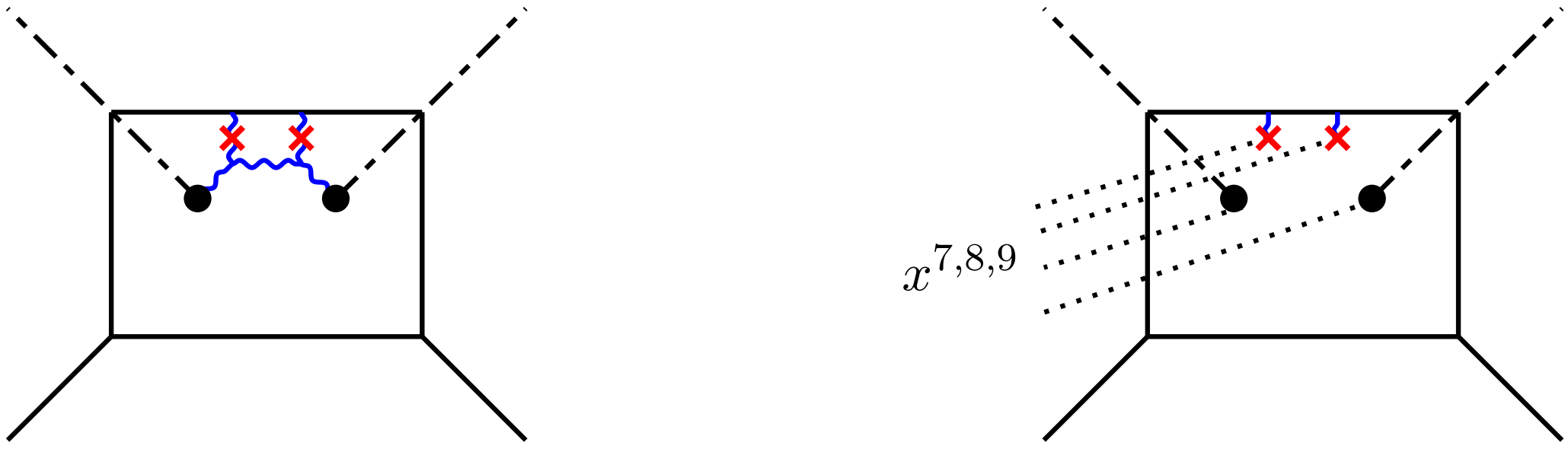
\begin{figure}[H] 
	\centering
	\begin{tikzpicture}
		\node at (0,2) {\raisebox{-.5\height}{ \scalebox{3}{  
					\begin{tikzpicture}[x=.5cm,y=.5cm] 
				        \draw (0,0-0.5)--(0.6,0.6-0.5);
						\draw[tight dash dot] (1+0.1,1-0.1)--(0,2);
						\draw (3,0-0.5)--(2.4,0.6-0.5);
						\draw[tight dash dot] (2-0.1,1-0.1)--(3,2);
						\draw (0.6,0.6-0.5)--(2.4,0.6-0.5);
						\draw (0.6,1.4)--(2.4,1.4);
						\draw (0.6,0.6-0.5)--(0.6,1.4);
						\draw (2.4,0.6-0.5)--(2.4,1.4);

                           \draw[blue, decorate, decoration={snake, segment length=2pt, amplitude=0.3pt}] (1+0.1,1-0.1)--(1.2+0.1,1.2-0.1);
                           
                           \draw[blue, decorate, decoration={snake, segment length=2pt, amplitude=0.3pt}] (1.2+0.1,1.2-0.1)--(1.2+0.1,1.4);
                           
                           \draw[blue, decorate, decoration={snake, segment length=2pt, amplitude=0.3pt}] (1.2+0.1,1.2-0.1)--(1.8-0.1,1.2-0.1);
                           
                           \draw[blue, decorate, decoration={snake, segment length=2pt, amplitude=0.3pt}] (1.8-0.1,1.2-0.1)--(1.8-0.1,1.4);

                           \draw[blue, decorate, decoration={snake, segment length=2pt, amplitude=0.3pt}] (1.8-0.1,1.2-0.1)--(2-0.1,1-0.1);

                        \node at (1.2+0.1,1.25)[cross, red,scale=0.3, line width=0.5pt]{};
                            
                        \node at (1.8-0.1,1.25)[cross, red,scale=0.3, line width=0.5pt]{};

                            \fill (1+0.1,1-0.1) circle (0.08);
                            \fill (2-0.1,1-0.1) circle (0.08); 
				        \draw (0+6,0-0.5)--(0.6+6,0.6-0.5);
						\draw[tight dash dot] (1+0.1+6,1-0.1)--(0+6,2);
						\draw (3+6,0-0.5)--(2.4+6,0.6-0.5);
						\draw[tight dash dot] (2-0.1+6,1-0.1)--(3+6,2);
						\draw (0.6+6,0.6-0.5)--(2.4+6,0.6-0.5);
						\draw (0.6+6,1.4)--(2.4+6,1.4);
						\draw (0.6+6,0.6-0.5)--(0.6+6,1.4);
						\draw (2.4+6,0.6-0.5)--(2.4+6,1.4);

                           \draw[blue, decorate, decoration={snake, segment length=2pt, amplitude=0.3pt}] (1.2+0.1+6,1.25)--(1.2+0.1+6,1.4);

                           \draw[blue, decorate, decoration={snake, segment length=2pt, amplitude=0.3pt}] (1.8-0.1+6,1.25)--(1.8-0.1+6,1.4);

                            \fill (1+0.1+6,1-0.1) circle (0.08);
                            \fill (2-0.1+6,1-0.1) circle (0.08); 

                    \draw[densely dotted, scale=0.5] (14.2+0.4,2.3+0.2)--(11.9,1.7);
                    
                    \draw[densely dotted, scale=0.5] (14.2+0.4+0.8,2.3+0.2)--(11.9,1.4);

                    \draw[densely dotted, scale=0.5] (14.2+0.1,2.3-0.8+0.2)--(12,1); 

                    \draw[densely dotted, scale=0.5] (15.6,1.7)--(12,0.48); 

                    \node at (1.2+0.1+6,1.25)[cross, red,scale=0.3, line width=0.5pt]{};
                            
                    \node at (1.8-0.1+6,1.25)[cross, red,scale=0.3, line width=0.5pt]{};

                \node[label=left:\scalebox{0.4}{$x^{7,8,9}$}] at (6.4,0.5){};

		\end{tikzpicture}}}};
	\end{tikzpicture}
	\caption{7-brane screening. Here, 7-branes are denoted by thick black dots and the corresponding monodromy cuts are denoted dash-dot lines. Two red crosses are D3-branes. The string junction is denoted in blue. }
	\label{fig:7-brane screening}
\end{figure}

\paragraph{7-brane screening}
We can also screen the Wilson line with the help of 7-branes. As depicted in Figure \ref{fig:7-brane screening}, relevant 7-branes are brought inside the Coulomb branch chamber. A string junction is formed, whose end points are located at the 7-branes and also 5-branes position. It follows that a Higgsing arises when D3-branes are on top of the string junction. The junction is then fragmented and reconnected to D3-branes, so that the fragmented part of the junction can be taken to the $x^{7,8,9}$ directions. The resulting configuration is equivalent to the Wilson line screening.

We have seen (in two different but equivalent ways) that one specific state of the Wilson line $W_{[1]\otimes [1]}$ can be screened. The analysis can be repeated for any other state of $W_{[1]\otimes [1]}$. This means that there exists a point-like operator on which $W_{[1]\otimes [1]}$ can end, or equivalently that $W_{[1]\otimes[1]}\sim W_{[0]}$, where $W_{[0]}$ is the trivial line. 
It follows from \eqref{eq:reducible_screen} that all the irreducible components of $W_{[1]\otimes [1]}$ must be screened. The tensor product decomposition relation
\begin{equation}
W_{[1]\otimes [1]}=W_{[2]}\oplus W_{[0]}
\end{equation} implies then that $W_{[2]}$ can also be screened, namely $W_{[2]}\sim W_{[0]}$. With a similar logic one concludes that the fundamental Wilson line $W_{[1]}$ cannot be screened.

We can now proceed by considering a Wilson line in the representation $W_{[1]\otimes [1]\otimes [1]}$. Since $W_{[1]\otimes [1]\otimes [1]}=W_{[3]}\oplus 2 W_{[1]}$ one has that $W_{[3]}\sim W_{[1]}$, and since we established that $W_{[1]}$ cannot be screened, nor can $W_{[3]}$.

We can now proceed by considering a Wilson line in the representation $W_{[1]\otimes [1]\otimes [1]\otimes [1]}$. One can see from the web that $W_{[1]\otimes [1]\otimes [1]\otimes [1]}$ can be screened. Since $W_{[1]\otimes [1]\otimes [1]\otimes [1]}=W_{[4]}\oplus 3W_{[2]}\oplus 2W_{[0]}$ one has that $W_{[4]}\sim W_{[2]}\sim W_{[0]}$ belong to the same class: both can be screened. 

The logic above can be iterated, and one arrives at the conclusion that there are just two equivalence classes. All lines $W_{[2n]}$, ($n\in \mathbb{N}$) can be screened, while lines $W_{[2n+1]}$ can not. Therefore, we can conclude from this analysis that $\hat{\Gamma}^{(1)}\simeq \mathbb{Z}_2$ and by using \eqref{eq::Pontryagin} the 1-form symmetry is $\Gamma^{(1)}\simeq \mathbb Z_2.$

\subsubsection*{\underline{\texorpdfstring{$\SU(2)_\pi$}{SU2pi} theory}}

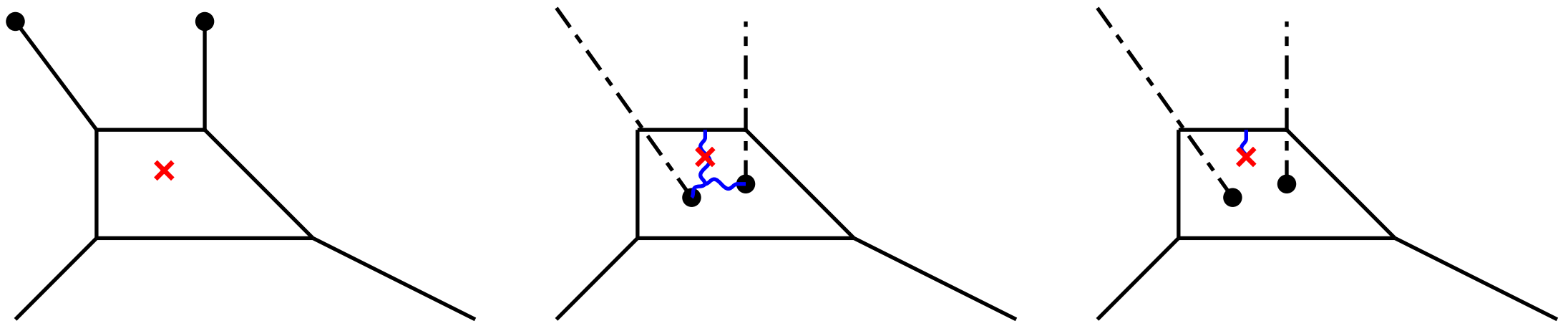
\begin{figure}[H]
	\centering
	\begin{tikzpicture}
		\node at (0,2) {\raisebox{-.5\height}{ \scalebox{2.5}{  
					\begin{tikzpicture}[x=.5cm,y=.5cm] 
						\draw (0,0)--(0.6,0.6);
						\draw (0,2.2)--(0.6,1.4);
						\draw (1.4,1.4)--(2.2,0.6);
						\draw (0.6,0.6)--(2.2,0.6);
						\draw (0.6,1.4)--(1.4,1.4);
						\draw (0.6,0.6)--(0.6,1.4);
						\draw (1.4,1.4)--(1.4,2.2);
						\draw (2.2,0.6)--(3.4,0);
                        \fill (0,2.2) circle (0.07);
                        \fill (1.4,2.2) circle (0.07);
                        
                        \node at (1.1,1.1)[cross, red, scale=0.3, line width=0.5pt]{};

						
						\draw (4,0)--(4.6,0.6);
						\draw[tight dash dot] (4,2.3)--(5,0.9);
						\draw (5.4,1.4)--(6.2,0.6);
						\draw (4.6,0.6)--(6.2,0.6);
						\draw (4.6,1.4)--(5.4,1.4);
						\draw (4.6,0.6)--(4.6,1.4);
						\draw[tight dash dot] (5.4,1)--(5.4,2.2);
						\draw (6.2,0.6)--(7.4,0);

                        \fill (5,0.9) circle (0.07);
                        \fill (5.4,1) circle (0.07);
                       
						\draw[blue, decorate, decoration={snake, segment length=3pt, amplitude=0.5pt}] (5,0.9)--(5.1,1);

						\draw[blue, decorate, decoration={snake, segment length=3pt, amplitude=0.5pt}] (5.1,1)--(5.1,1.4);
                        \draw[blue, decorate, decoration={snake, segment length=3pt, amplitude=0.5pt}] (5.1,1)--(5.4,1);
                        \node at (5.1,1.2)[cross, red, scale=0.3, line width=0.5pt]{};


						\draw (4+4,0)--(4.6+4,0.6);
						
                        \draw[tight dash dot] (4+4,2.3)--(5+4,0.9);
						
                        \draw (5.4+4,1.4)--(6.2+4,0.6);
						\draw (4.6+4,0.6)--(6.2+4,0.6);
						\draw (4.6+4,1.4)--(5.4+4,1.4);
						\draw (4.6+4,0.6)--(4.6+4,1.4);
						
                        \draw[tight dash dot] (5.4+4,1)--(5.4+4,2.2);
						\draw (6.2+4,0.6)--(7.4+4,0);

                        \fill (5+4,0.9) circle (0.07);
                        \fill (5.4+4,1) circle (0.07);

						\draw[blue, decorate, decoration={snake, segment length=3pt, amplitude=0.5pt}] (5.1+4,1.2)--(5.1+4,1.4);

                        \node at (5.1+4,1.2)[cross, red, scale=0.3, line width=0.5pt]{};
						
		\end{tikzpicture}}}};
	\end{tikzpicture}
	\caption{On the left, the highest weight state of a Wilson line $W_{[1]}$ is pictured. On the centre, the 7-brane screening effect for this state of the $W_{[1]}$ line. On the right, the brane junction fragment is removed into the $x^{7,8,9}$ direction.}
	\label{fig:E1t_screen}
\end{figure}

Let us consider the $\SU(2)_{\pi}$ theory or $\tilde{E}_1$ theory whose 5-brane configuration is given in Figure~\ref{fig:E1t_screen}.
Performing a similar analysis as the one in section \ref{sec:E1th}, one sees that the fundamental Wilson line $W_{[1]}$ can be screened. We depict this in figure \ref{fig:E1t_screen}. Now, consider a generic irrep $R$ of $\mathrm{SU}(2)$ of Dynkin label $R=[n]$. Since the highest weight is additive, $R$ will appear in the irreducible representation decomposition of the tensor product $\bigotimes^n [1]=[n]\oplus\cdots$. This implies that all the BPS Wilson lines $W_{[n]}$ can be screened and therefore are in the same class. Therefore, for this theory, the 1-form symmetry is trivial.

\subsection{\texorpdfstring{$\mathrm{SU}(N)$}{SU(N)} Chern-Simons level \texorpdfstring{$k$}{k=N}}
In this section we consider the $\mathrm{SU}(N)_k$ theory. In the absence of hypermultiplets, it has a 1-form symmetry $\Gamma^{(1)}=\mathbb{Z}_{\mathrm{gcd}(N,k)}$ \cite{BenettiGenolini:2020doj, Gukov:2020btk}. If  hypermultiplets are present in a representation with $N$-ality $n$, the 1-form symmetry is further broken to $\Gamma^{(1)}=\mathbb{Z}_{\mathrm{gcd}(N,k, n)}$.

We will now restrict ourselves to the case $k=N$. The 5-brane web for $\mathrm{SU}(N)_N$ is given by intersecting an $(N,1)$, an $(N,-1)$, a $(1,1)$ and a $(1,-1)$ 5-brane. The 1-form symmetry for this theory is $\mathbb{Z}_N$.  

We will discuss how to see this 1-form symmetry from the Wilson line screening mechanisms, as explained in the previous sections. For simplicity, we restrict ourselves to the case $N=3$. Generalisation to arbitrary $N$ is straightforward. 

Insertion of 3 Wilson lines in the fundamental representation can be achieved by inserting 3 D3-branes, denoted by a cross, in the 5-brane web \cite{Assel:2018rcw}
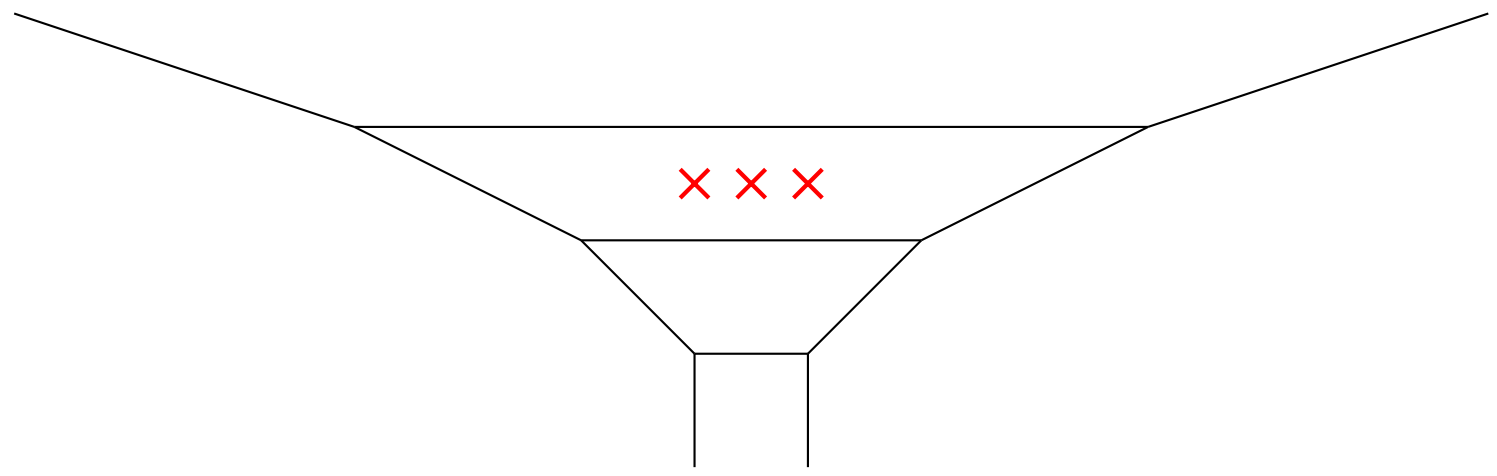
\begin{equation}
    \begin{array}{c}
    \begin{tikzpicture}[scale=.75]
    \draw(0,0)--(0,1);
    \node[cross, red]at(0.5,2.5){};
    \node[cross, red]at(0,2.5){};
    \node[cross, red]at(1,2.5){};
    \draw(1,0)--(1,1);
    \draw(0,1)--(1,1);
    \draw(1,1)--(2,2);
    \draw(0,1)--(-1,2);
    \draw(-1,2)--(2,2);
    \draw(4,3)--(2,2);
    \draw(-1,2)--(-3,3);
    \draw(4,3)--(-3,3);
    \draw(-6,4)--(-3,3);
    \draw(4,3)--(7,4);
    \end{tikzpicture}
    \end{array}\; .
    \label{Pure SU(3)_3 web-1}\nonumber
\end{equation}
The above configuration is Hanany-Witten equivalent to
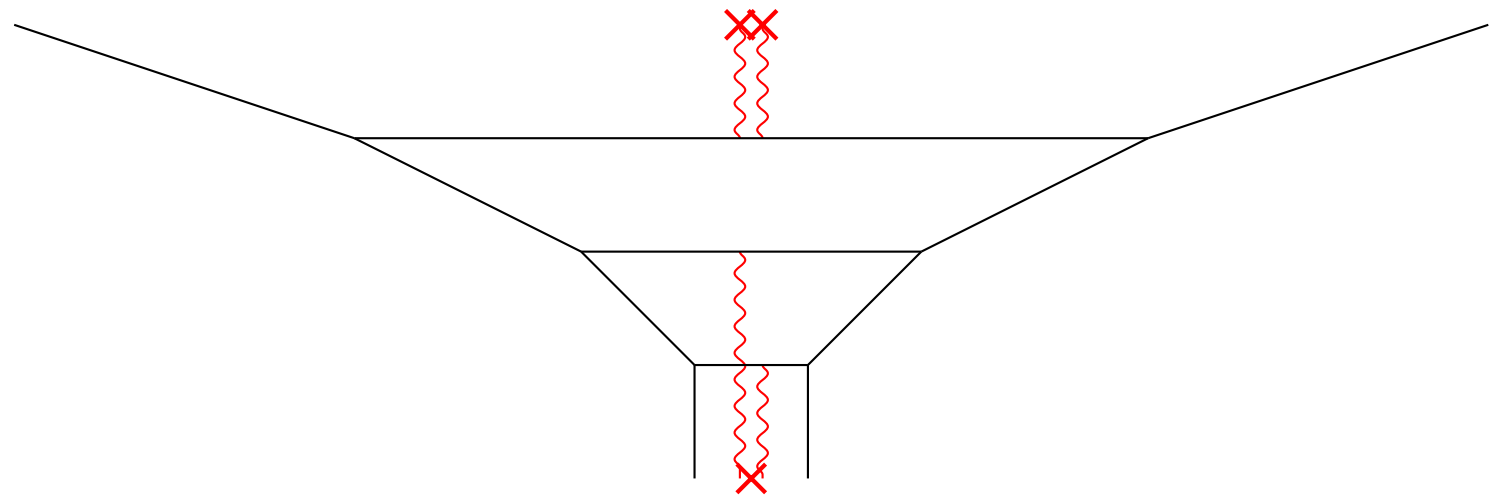
\begin{equation}
    \begin{array}{c}
    \begin{tikzpicture}[scale=.75]
    \draw(0,0)--(0,1);
    \draw[red, decorate, decoration={snake, segment length=5pt, amplitude=1pt}](0.6,3)--(0.6,4);
    \draw[red, decorate, decoration={snake, segment length=5pt, amplitude=1pt}](0.4,3)--(0.4,4);
    \draw[red, decorate, decoration={snake, segment length=5pt, amplitude=1pt}](0.4,2)--(0.4,0);
    \draw[red, decorate, decoration={snake, segment length=5pt, amplitude=1pt}](0.6,1)--(0.6,0);
    \node[cross, red]at(0.5,0){};
    \node[cross, red]at(0.4,4){};
    \node[cross, red]at(0.6,4){};
    \draw(1,0)--(1,1);
    \draw(0,1)--(1,1);
    \draw(1,1)--(2,2);
    \draw(0,1)--(-1,2);
    \draw(-1,2)--(2,2);
    \draw(4,3)--(2,2);
    \draw(-1,2)--(-3,3);
    \draw(4,3)--(-3,3);
    \draw(-6,4)--(-3,3);
    \draw(4,3)--(7,4);
    \end{tikzpicture}
    \end{array}\; ,
    \label{Pure SU(3)_3 web-2}\nonumber
\end{equation}
where red lines denote fundamental strings.
We can screen the 3 Wilson lines in the fundamental representation of $\mathrm{SU}(3)$ by the W-bosons indicated in blue in the following web diagram
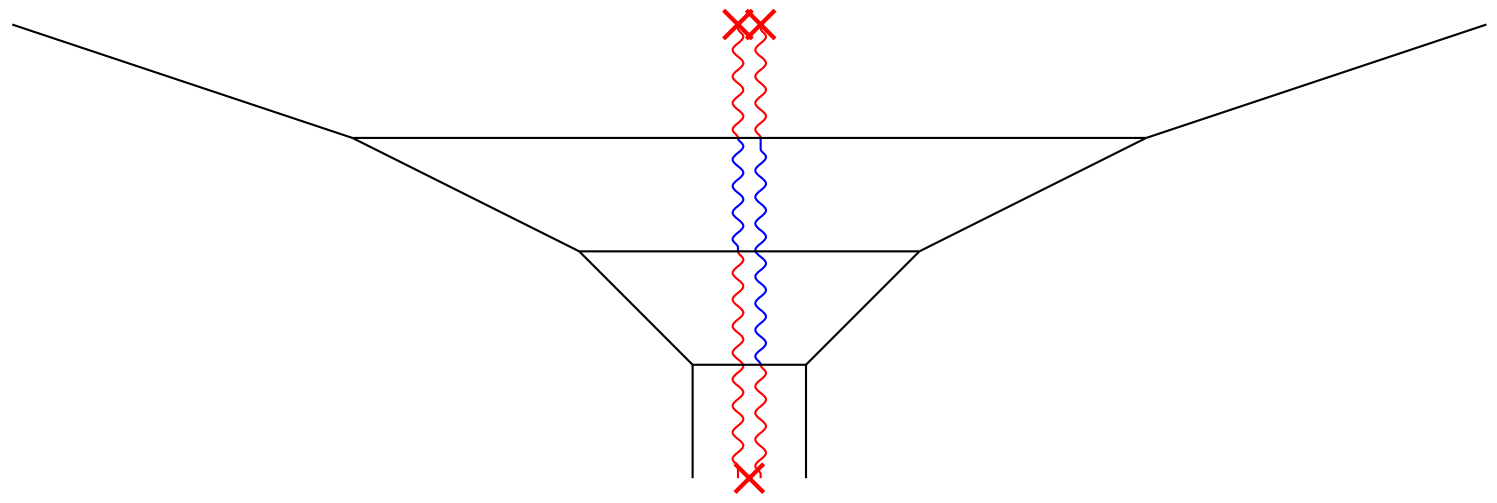
\begin{equation}
    \begin{array}{c}
    \begin{tikzpicture}[scale=.75]
    \draw(0,0)--(0,1);
    \draw[blue, decorate, decoration={snake, segment length=5pt, amplitude=1pt}](.4,3)--(.4,2);
    \draw[blue, decorate, decoration={snake, segment length=5pt, amplitude=1pt}](.6,1)--(.6,3);
    \draw[red, decorate, decoration={snake, segment length=5pt, amplitude=1pt}](0.6,3)--(0.6,4);
    \draw[red, decorate, decoration={snake, segment length=5pt, amplitude=1pt}](0.4,3)--(0.4,4);
    \draw[red, decorate, decoration={snake, segment length=5pt, amplitude=1pt}](0.4,2)--(0.4,0);
    \draw[red, decorate, decoration={snake, segment length=5pt, amplitude=1pt}](0.6,1)--(0.6,0);
    \node[cross, red]at(0.5,0){};
    \node[cross, red]at(0.4,4){};
    \node[cross, red]at(0.6,4){};
    \draw(1,0)--(1,1);
    \draw(0,1)--(1,1);
    \draw(1,1)--(2,2);
    \draw(0,1)--(-1,2);
    \draw(-1,2)--(2,2);
    \draw(4,3)--(2,2);
    \draw(-1,2)--(-3,3);
    \draw(4,3)--(-3,3);
    \draw(-6,4)--(-3,3);
    \draw(4,3)--(7,4);
    \end{tikzpicture}
    \end{array}\;
    \label{Pure SU(3)_3 web-3} \nonumber
\end{equation}
leading us to conclude that the 1-form symmetry for this theory is $\mathbb{Z}_3$. 
\newpage

\subsection{Smith normal form and 1-form symmetry}

Let $A$ be an $m\times n$ matrix. Let $d_i(A)$ be the greatest common divisor of the determinants of all minors of $A$ of dimension $i\times i$. Define $d_0(A):=1$, and  
\begin{equation}
\alpha_i:=\dfrac{d_i(A)}{d_{i-1}(A)} \ .
\end{equation}
Then, there exist a couple of matrices $S, T$ of size $m\times m$ and $n\times n$, respectively, such that the product $SAT$ is a $m\times n$ matrix of the form, 
\begin{equation}
SAT=\begin{pmatrix}
\alpha_1 & 0 & 0 & \cdots & 0 & \cdots & 0 \\
0 & \alpha_2 & 0 &  &  &  &  \\
0 & 0 & \ddots &  & \vdots &  & \vdots \\
\vdots &  &  & \alpha_r &  &  &  \\
0 &  & \cdots & & 0 & \cdots & 0\\
\vdots &  &  &  & \vdots &  & \vdots \\
0 &  & \cdots & & 0 & \cdots & 0\\
\end{pmatrix} \ .
\end{equation} 

We call the matrix $SAT$, the \emph{Smith Normal Form} (SNF) of $A$.

\paragraph{Example of SNF computation.}
Consider the matrix 
\begin{equation}
A=\begin{pmatrix}
2 & 4 \\
3 & 1 \\
1 & 2
\end{pmatrix}\ .
\end{equation}
All the minors of size $1 \times 1$ are the elements of the set $\mbox{Min}_1=\{2,4,3,1,1,2\}$. Therefore, $d_1(A)=\gcd(2,4,3,1,1,2)=1$. All the minors of size $2 \times 2$ are the elements of the set
$\mbox{Min}_2=\left\{ 
\begin{pmatrix} 2 & 4 \\ 3 & 1 \end{pmatrix},
\begin{pmatrix} 3 & 1 \\ 1 & 2 \end{pmatrix},
\begin{pmatrix} 2 & 4 \\ 1 & 2 \end{pmatrix}
\right\}
.$
Therefore, $d_2(A)=\gcd(-10,5)=5$. In this example, there are no minors of size $i\times i$ for $i>3$. Therefore, we have $\alpha_1=1, \alpha_2=5$.
\begin{equation}
\mathrm{SNF}(A) := SAT=\begin{pmatrix}
1 & 0  \\
0 & 5  \\
0 & 0 
\end{pmatrix} \ .
\end{equation} 

\subsubsection*{Example: the \texorpdfstring{$\mathrm{SU}(N)_{k}$}{SU(N)k} theory}
Consider a 5-brane web for $\mathrm{SU}(N)_{k}$ whose external 5-branes are of charges $(N,-1)$, $(k,1)$, $(N-k, -1)$, and $(0,1)$.
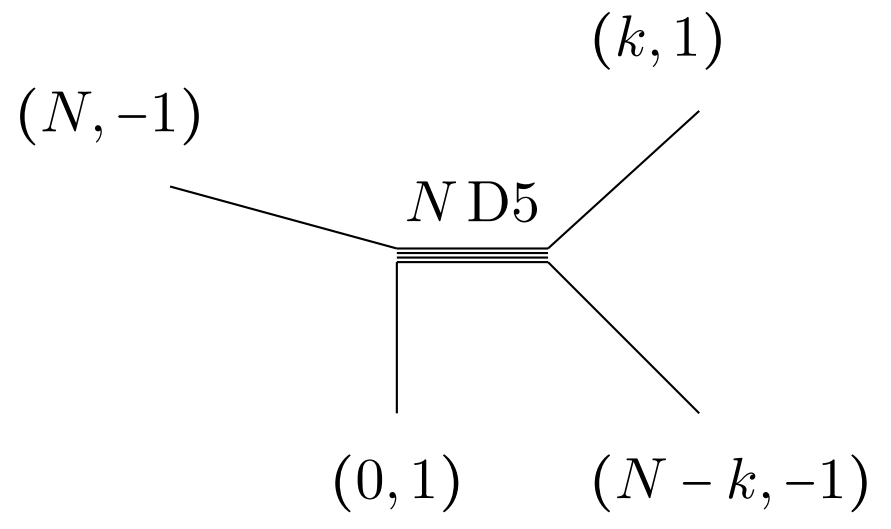
\begin{equation}
    \begin{tikzpicture}
             \draw(0,0)--(1,0);
             \draw(0,0.03)--(1,0.03);
             \draw(0,0.06)--(1,0.06);
             \draw(0,0.09)--(1,0.09);
             \draw(0,0)--(0,-1);
             \draw(0,0.09)--(-1.5,.5);
             \draw(2,1)--(1,0.09);
             \draw(2,-1)--(1,0);
         \node[label=below:{$(0,1)$}]at(0,-1){};
         \node[label=below right:{$(N-k,-1)$}]at(1,-1){};
         \node[label=above right:{$(k,1)$}]at(1,1){};
         \node[label=above left:{$(N,-1)$}]at(-1,.5){};
         \node[label=above:{$N\,\mathrm{D5}$}]at(0.5,0){};
         \end{tikzpicture}\nn
\end{equation}
 We consider that the 7-branes of the same charges end on these 5-branes. Following \citep{Bhardwaj:2020phs}, we find that the corresponding 1-form symmetry is given by the SNF of the 7-brane charges. More explicitly, form a matrix $A$ with these 7-brane charges: 
\begin{equation}
A=  \begin{pmatrix}
        N & ~-1 \\
        k & ~1 \\
        N-k &~ -1\\
        0 &~ 1
    \end{pmatrix}\ .
\end{equation}
From this, one can compute SNF as follows:
all the minors of size $1 \times 1$ are the elements of the set $\mbox{Min}_1=\{N,-1,k,N-k,-1,0,1\}$. Therefore, $d_1(A)=1$. All the minors of size $2 \times 2$ are the elements of the set
$\mbox{Min}_2=\left\{ 
\begin{pmatrix} N & ~-1 \\ k & ~1 \end{pmatrix},
\begin{pmatrix} N & ~-1 \\ N-k & ~-1 \end{pmatrix},
\begin{pmatrix} N & ~-1 \\ 0 & ~1 \end{pmatrix},
\begin{pmatrix} k & ~1 \\ N-k & ~-1 \end{pmatrix},
\begin{pmatrix} k & ~1 \\ 0 & ~1 \end{pmatrix},
\begin{pmatrix} N-k &~ -1 \\ 0 &~ 1 \end{pmatrix}
\right\}
.$
Therefore, $d_2(A)=\gcd(N+k,-k,N,-N,k,N-k)=\gcd(N,k)$. So one has
\begin{equation}\mathrm{SNF}(A) =\begin{pmatrix}
~1 & ~0  \\
~0 & ~\gcd(N, k)  \\
~0 & ~0 
\end{pmatrix} \ .
\end{equation} 

The 1-form symmetry is accordingly $\Gamma^{(1)}(SU(N)_{k})=\mathbb{Z}_{\gcd(N,k)}$.

\vspace{0.5cm}

\section{Wilson Lines in 5-brane webs {\it with} O7-planes}\label{sec:withO7}

In this section we discuss brane webs, in which orientifold 7-planes are present. We split the discussion in three steps. First, we review how to recover the 1-form symmetry from a web without orientifold 7-planes, by the method of taking the Smith normal form of the matrix of brane charges, following \cite{Gaiotto:2014kfa, Morrison:2020ool, Bhardwaj:2020phs, DelZotto:2022joo}. Second, we discuss how to use this method for webs in which an O7$^-$-plane is present. Third, we propose how to generalize this method to webs in which an O7$^+$-plane is present. We check our conjectured rule against various examples.

\subsection{Rule for \texorpdfstring{O7$^-$}{O7-}}

Some 5d $\mathcal{N}=1$ theories are realized by brane webs in which an orientifold O7$^-$-plane is present. Such an O7$^-$-plane can be resolved into two distinct 7-branes of  $(p_1,q_1)$ and $(p_2,q_2)$ type. The allowed values for $p_i, q_i$ are such  that the monodromy matrix of the O7$^-$-plane is equal to the product of the monodromy matrices of the two 7-branes of type $(p_i,q_i)$. For example, some allowed choices are $(1,-1)\ , (1,1)$ and $(2,-1)\ , (0,1)$ \citep{Sen:1996vd}.

The procedure to determine the 1-form symmetry of the theory is then to compute the SNF of the matrix of 7-brane charges, in which two extra lines are included, corresponding to the two 7-branes that resolve the orientifold plane. 

We elucidate this with some examples.

\subsubsection{\texorpdfstring{$\mathrm{SU}(N)_{k}+1\mathbf{AS}$}{}}
As the first example, consider the $\mathrm{SU}(N)_{k}+1\mathbf{AS}$ theory based on the SNF. Because the result depends on whether $N$ is even or not, we discuss $N$ even/odd cases separately.  
\vspace{0.5cm}
\paragraph{$\bullet$~$\mathrm{SU}(2n)_{k}+1\mathbf{AS}$}

An example of 5-brane for $\mathrm{SU}(2n)_{k}+1\mathbf{AS}$ is depicted as:
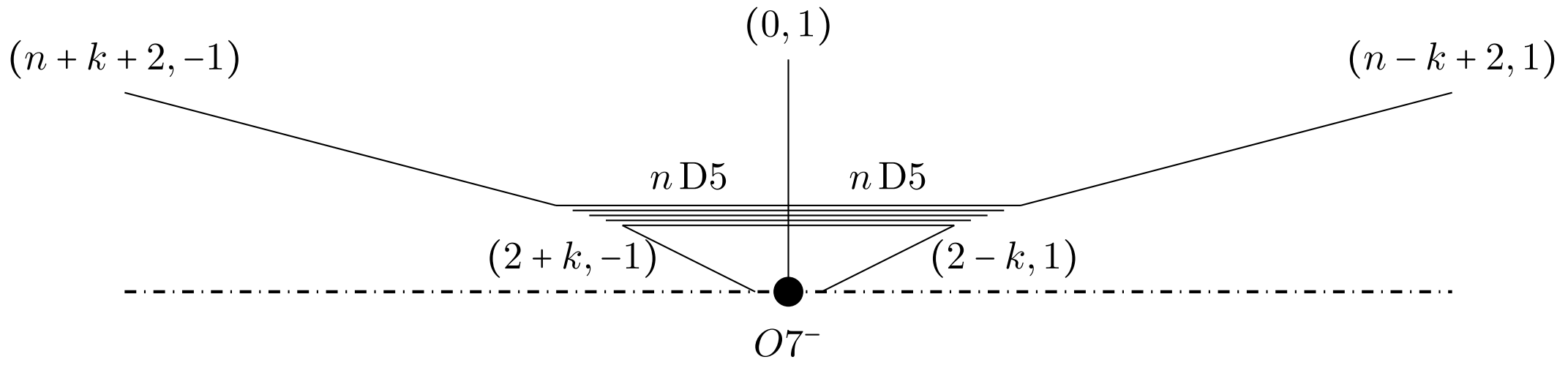
\begin{center}
    \begin{tikzpicture}[scale=1.5]
             \draw(0,0)--(0.8,-0.4);
             \draw(2,0)--(1.2,-0.4);
             \draw(1,-0.4)--(1,1);
             \draw(0,0)--(2,0); 
             \draw(-0.1,0.03)--(2.1,0.03);
             \draw(-0.2,0.06)--(2.2,0.06);
             \draw(-0.3,0.09)--(2.3,0.09);
             \draw(-0.4,0.12)--(2.4,0.12);
             \draw(-0.4,0.12)--(-3,0.8);
             \draw(2.4,0.12)--(5,0.8);
             \draw[dash dot, thick](-3,-0.4)--(5,-0.4);
             \fill (1,-0.4) circle (0.09);
          
          \node[font=\small] at (1,-0.7){$O7^{-}$};
          \node[font=\small] at (2.3,-0.2){$(2-k,1)$};
          \node[font=\small] at(-0.3,-0.2){$(2+k,-1)$};
          \node[font=\small] at (5,1){$(n-k+2,1)$};
          
          \node[font=\small] at (-3,1){$(n+k+2,-1)$};
          \node[font=\small] at (1,1.2){$(0,1)$};
          \node[font=\small]at(0.4,0.3){$n\,\mathrm{D5}$};
          \node[font=\small]at(1.6,0.3){$n\,\mathrm{D5}$};
          
         \end{tikzpicture}
\end{center}
where the external 7-branes are
$(n+k+2,  ~-1), ( n-k+2,  ~1), (0 ,~ 1)$. 
We stress that the global form of the gauge group is $\mathrm{SU}(2n)$ and not $\mathrm{SU}(2n)/\mathbb{Z}_{2}$, as fundamental matter could be added to the theory, as seen from the web.

One then reads the matrix of 7-brane charges:

\begin{equation}
A=  \begin{pmatrix}
        n+k+2 & ~-1 \\
        n-k+2 & ~1 \\
        0 &~ 1 \\
        0 &~ 1\\
        2 &~ -1 
    \end{pmatrix}\ , 
\end{equation}
where in particular the last two lines are due to the fact that the $O7^-$ resolves nonperturbatively into a $[2,-1]$ and a $[0,1]$ brane~\cite{Sen:1996vd}. From the SNF computation, one finds $\Gamma^{(1)}=\mathbb{Z}_{\gcd(2 n, k + n - 2, 2)}$, where the last entity $2$ is the charge of the antisymmetric matter fields under the centre of the gauge group. 
\vspace{0.5cm}

\paragraph{$\bullet$~$\mathrm{SU}(2n+1)_{k+\frac12}+1\mathbf{AS}$}
\vspace{0.5cm}
An example of 5-brane for $\mathrm{SU}(2n+1)_{k+\frac{1}{2}}+1\mathbf{AS}$ is depicted as:

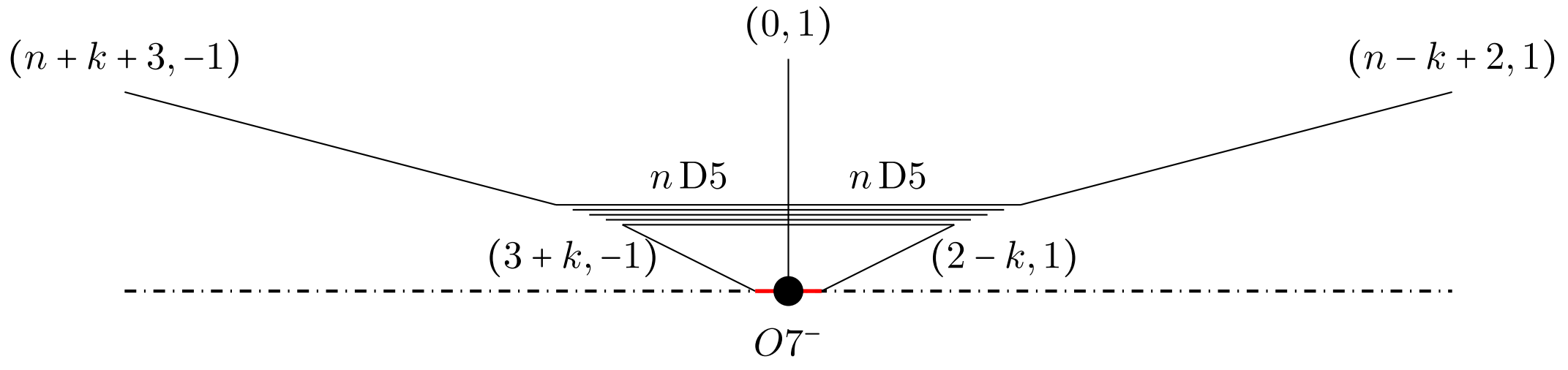
\begin{center}
    \begin{tikzpicture}[scale=1.5]
             \draw(0,0)--(0.8,-0.4);
             \draw(2,0)--(1.2,-0.4);
             \draw(1,-0.4)--(1,1);
             \draw(0,0)--(2,0); 
             \draw(-0.1,0.03)--(2.1,0.03);
             \draw(-0.2,0.06)--(2.2,0.06);
             \draw(-0.3,0.09)--(2.3,0.09);
             \draw(-0.4,0.12)--(2.4,0.12);
             \draw(-0.4,0.12)--(-3,0.8);
             \draw(2.4,0.12)--(5,0.8);
             \draw[dash dot, thick](-3,-0.4)--(5,-0.4);
             \draw[red, line width=1pt](0.8,-0.4)--(1.2,-0.4);  
             \fill (1,-0.4) circle (0.09);
          
          \node[font=\small] at (1,-0.7){$O7^{-}$};
          \node[font=\small] at (2.3,-0.2){$(2-k,1)$};
          \node[font=\small] at(-0.3,-0.2){$(3+k,-1)$};
          \node[font=\small] at (5,1){$(n-k+2,1)$};
          
          \node[font=\small] at (-3,1){$(n+k+3,-1)$};
          \node[font=\small] at (1,1.2){$(0,1)$};
          \node[font=\small]at(0.4,0.3){$n\,\mathrm{D5}$};
          \node[font=\small]at(1.6,0.3){$n\,\mathrm{D5}$};
          
         \end{tikzpicture}
\end{center}
where the thick red line denotes half D5-brane at the monodromy cut location.

The matrix of 7-brane charges reads:

\begin{equation}
A=  \begin{pmatrix}
        n+k+3 & ~-1 \\
        n-k+2 & ~1 \\
        0 &~ 1\\
        1 &~ 1 \\
        1& -1
    \end{pmatrix}\ .
\end{equation}
where in particular the last two lines are due to the fact that the $O7^-$ resolves nonperturbatively into a $[1,-1]$ and a $[1,1]$ brane. From the SNF computation, one finds $\Gamma^{(1)}=\mathbb{Z}_{\gcd(2 n+1, k + n - 1, 2)}$, which is trivial.

\subsubsection{\texorpdfstring{$\mathrm{Sp}(N)_\theta$}{Sp(N)theta} of the discrete theta \texorpdfstring{$\theta=0,\pi$}{theta=0}}
The 5-brane configuration for $\mathrm{Sp}(N)_\theta$ for both $\theta=0,\pi$ is given by
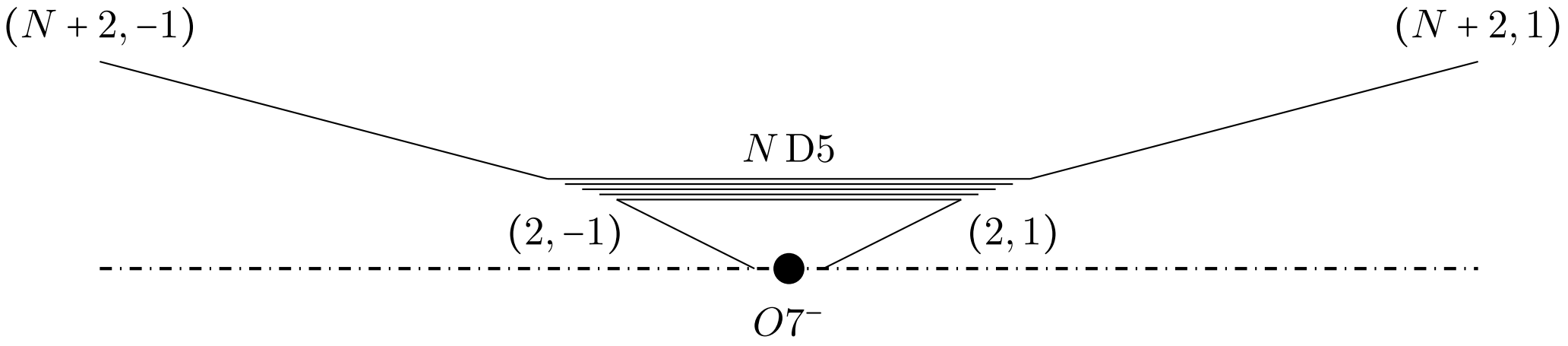
\begin{center}
    \begin{tikzpicture}[scale=1.5]
             \draw(0,0)--(0.8,-0.4);
             \draw(2,0)--(1.2,-0.4);
             \draw(0,0)--(2,0); 
             \draw(-0.1,0.03)--(2.1,0.03);
             \draw(-0.2,0.06)--(2.2,0.06);
             \draw(-0.3,0.09)--(2.3,0.09);
             \draw(-0.4,0.12)--(2.4,0.12);
             \draw(-0.4,0.12)--(-3,0.8);
             \draw(2.4,0.12)--(5,0.8);
             \draw[dash dot, thick](-3,-0.4)--(5,-0.4);
             \fill (1,-0.4) circle (0.09);
          
          \node[font=\small] at (1,-0.7){$O7^{-}$};
          \node[font=\small] at (2.3,-0.2){$(2,1)$};
          \node[font=\small] at(-0.3,-0.2){$(2,-1)$};
          \node[font=\small] at (5,1){$(N+2,1)$};
          
          \node[font=\small] at (-3,1){$(N+2,-1)$};
          %
          \node[font=\small]at(1,0.3){$N\,\mathrm{D5}$};
          
         \end{tikzpicture}
\end{center}
We note that by resolving the $O7^-$-plane into two distinct 7-branes~\cite{Sen:1996vd}, we obtain the corresponding asymptotic 7-brane charges, which can be summarized as
\begin{align}
    \mathrm{Sp}(2n-1)_{\theta=0}:& \quad [2n+1, -1],\quad [2n+1, 1],\quad [1, -1],\quad [1, 1],\\
   \mathrm{Sp}(2n)_{\theta=0}:& \quad [2n+2, -1],\quad [2n+2, 1],\quad [2, -1],\quad [0, 1],\\
   \mathrm{Sp}(2n-1)_{\theta=\pi}:& \quad [2n+1, -1],\quad [2n+1, 1],\quad [2, -1],\quad [0, 1],\\
   \mathrm{Sp(2n)}_{\theta=\pi}:& \quad [2n+2, -1],\quad [2n+2, 1],\quad [1, -1],\quad [1, 1]. 
\end{align}

The matrix of 7-brane charges is reported in the following. \\
\noindent For $\theta=0$, 
\begin{equation}
A_{\mathrm{Sp}(2n-1)_0}=  \begin{pmatrix}
        2n+1 & ~-1 \\
        2n+1& ~1 \\
        1 &~ -1\\
        1 &~ 1
    \end{pmatrix}\ , \qquad
A_{\mathrm{Sp}(2n)_0}=  \begin{pmatrix}
        2n+2 & ~-1 \\
        2n+2 & ~1 \\
        2 &~ -1\\
        0 &~ 1
    \end{pmatrix}\ .
\end{equation}
For $\theta=\pi$, 
\begin{equation}
A_{\mathrm{Sp}(2n-1)_\pi}
=  \begin{pmatrix}
        2n+1 & ~-1 \\
        2n+1& ~1 \\
        2 &~ -1\\
        0 &~ 1
    \end{pmatrix}\ , \qquad
A_{\mathrm{Sp}(2n)_\pi}=  \begin{pmatrix}
        2n+2 & ~-1 \\
        2n+2 & ~1 \\
        1 &~ -1\\
        1 &~ 1
    \end{pmatrix}\ .
\end{equation}

Applying the SNF procedure we get to the following result. 
Pure $\mathrm{Sp}(N)_0$ has $\Gamma^{(1)}=\mathbb{Z}_{2}$ 1-form symmetry, while pure $\mathrm{Sp}(N)_\pi$ has trivial $\Gamma^{(1)}$ 1-form symmetry. This is consistent with field theory results in  
\cite{Gukov:2020btk}.

\subsection{Rule for \texorpdfstring{O7$^+$}{O7+}}
Reading off 1-form symmetry from SNF from the asymptotic charges including the resolved charges of an O7$^-$ is quite powerful. However, it is well-known that  an O7$^+$-plane cannot be resolved into 7-branes. In this subsection, we propose an effective way of implementing an O7$^+$ into SNF. To this end, let us recall 
the monodromy matrix for a $[p,q]$ 7-brane,
\begin{equation}
M{[p,q]}=\left(\begin{array}{cc}
1-pq & p^2 \\
-q^2& 1+pq \\
\end{array}\right)\ . 
\end{equation}
Here, we note that we assume that an $[r,s]$ 7-brane crosses clockwise the cut of the $[p,q]$ 7-brane. In this convention, it is also well-known that the monodromy matrix for an O7$^{+}$ plane is given by
\begin{equation}
\left(\begin{array}{cc}
1& 4 \\
0& 1 \\
\end{array}\right)\ .
\end{equation}
One may notice that the same monodromy matrix can be obtained by introducing a 7-brane charge $[\pm2,0]$, 
\begin{equation}
M{[p,q]}{}\Big|_{p=\pm2, q=0}=\left(\begin{array}{cc}
1& 4 \\
0& 1 \\
\end{array}\right)\ .
\end{equation}

 By comparing the two, we define the ``{\it effective} monodromy charges of the orientifold plane" as being $``[2,0]"$. We then conjecture that we can read the 1-form symmetry of brane web with O7$^+$ by modifying the Smith normal form computation to include the orientifold effective charges. 
 
 We check our procedure with some examples.

\subsubsection{\texorpdfstring{$\mathrm{SU}(N)$}{SU(N)} theories}

As discussed in the preceding subsection, from the perspective of $\mathrm{SU}(N)+1\mathbf{Sym}$ gauge theory, the 1-form symmetry is expected to be of the form $\mathbb{Z}_{\mathrm{gcd}(N, 2, k_{eff})}$, where the first entry $N$ is associated with the rank of gauge group, the second entry, $2$, comes from the ``$N$-ality'' of the hypermultiplet in the symmetric 2nd rank tensor representation, and the last entry represents the effective CS level. The effective CS level $k_{eff}$ is obtained by decoupling the symmetric matter, giving rise to the shifted CS level,
\begin{align}
    k~\to~k+\frac{1}{2} I^{(3)} = k \pm \frac{N+4}{2}=: k_{eff}\ ,
\end{align}
where $I^{(3)}$ is the cubic Dynkin index for the representation associated with hypermultiplets~\cite{Hayashi:2023boy}, and $I^{(3)}=N+4$ for the symmetric hypermultiplet. Hence, the expected 1-form symmetry reads
\begin{align}
   \Gamma^{(1)} = \mathbb{Z}_{\mathrm{gcd}(\,N,\, 2,\, k \pm \frac{N+4}{2}\,)}\ .
\end{align}

\paragraph{$\mathrm{SU}(2n)_k+1\mathbf{Sym}$.}

An example of 5-brane for $\mathrm{SU}(2n)_k+1\mathbf{Sym}$ is depicted as:

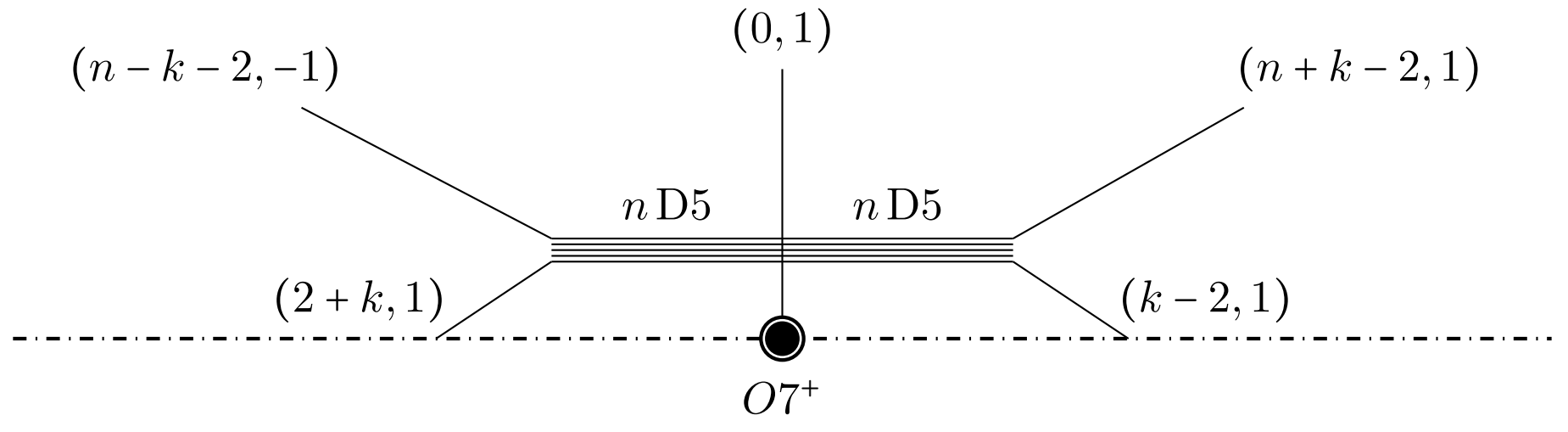
\begin{center}
    \begin{tikzpicture}[scale=1.5]
             \draw(2-2.2,0)--(1.2-2,-0.4); 
             \draw(2.8,-0.4)--(2.2,0);
             \draw(1,-0.4)--(1,1);
             \draw(-0.2,0)--(2.2,0);  
             \draw(-0.2,0.03)--(2.2,0.03); 
             \draw(-0.2,0.06)--(2.2,0.06); 
             \draw(-0.2,0.09)--(2.2,0.09); 
             \draw(-0.2,0.12)--(2.2,0.12); 
             \draw(-0.2,0.12)--(-1.5,0.8);
             \draw(2.2,0.12)--(3.4,0.8);
             \draw[dash dot, thick](-3,-0.4)--(5,-0.4);
              \draw[double, double distance=1pt, thick, fill=white] (1,-0.4) circle (0.09);
             \fill (1,-0.4) circle (0.09);
          
          \node[font=\small] at (1,-0.7){$O7^{+}$};
          \node[font=\small] at (3.2,-0.2){$(k-2,1)$};
          \node[font=\small] at(-1.2,-0.2){$(2+k,1)$};
          \node[font=\small] at (4,1){$(n+k-2,1)$};
          
          \node[font=\small] at (-2,1){$(n-k-2,-1)$};
          \node[font=\small] at (1,1.2){$(0,1)$};
          \node[font=\small]at(0.4,0.3){$n\,\mathrm{D5}$};
          \node[font=\small]at(1.6,0.3){$n\,\mathrm{D5}$};
          
         \end{tikzpicture}
\end{center}
We stress that the global form of the gauge group is $\mathrm{SU}(2n)$ and not $\mathrm{SU}(2n)/\mathbb{Z}_{2}$, as fundamental matter could be added to the theory, as seen from the web.

The matrix of external $(p,q)$-charges together with the O7$^{+}$ contribution is
\begin{equation}
A=  \begin{pmatrix}
        n-k-2 & ~-1 \\
        n+k-2 & ~1 \\
        0 &~ 1\\
        2 &~ 0 \\
    \end{pmatrix}\ .
\end{equation}

From this, one can compute SNF as 
\begin{equation}\mathrm{SNF}(A) =\begin{pmatrix}
~1 & ~0  \\
~0 & ~\gcd(2n-4,n+k-2, 2)  \\
~0 & ~0 
\end{pmatrix} \ .
\end{equation} 

The 1-form symmetry is accordingly $\Gamma^{(1)}\big(\mathrm{SU}(2n)_k+1\mathbf{Sym} \big) =\mathbb{Z}_{gcd(2n-4,n+k-2, 2)}$. 
It is possible to show that this coincides with the gauge theory expectation, which is $\mathbb{Z}_{\gcd(\,2n,\, 2,\, k \pm (n+2)\,)}$ 1-form symmetry. Also, the expression for the 1-form symmetry can be simplified as 
\begin{align}
   \Gamma^{(1)}\big(\mathrm{SU}(2n)_k+1\mathbf{Sym} \big) =\mathbb{Z}_{\gcd(n+k, 2)}\ ,
\end{align}
namely the 1-form symmetry is $\mathbb{Z}_2$ for $n+k$ even, and trivial otherwise.

\paragraph{$\mathrm{SU}(2n+1)_{k+\frac12}+1\mathbf{Sym}$.} 

An example of 5-brane for $\mathrm{SU}(2n+1)_{k+\frac12}+1\mathbf{Sym}$ is depicted as:
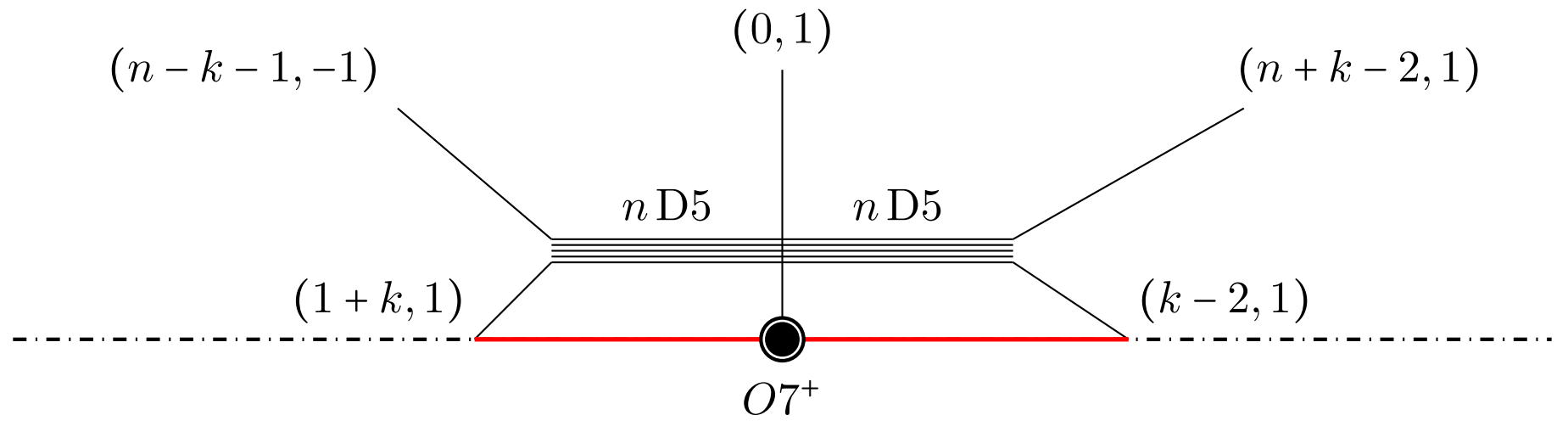
\begin{center}
    \begin{tikzpicture}[scale=1.5]
             \draw(2-2.2,0)--(-0.6,-0.4); 
             \draw(2.8,-0.4)--(2.2,0);
             \draw(1,-0.4)--(1,1);
             \draw(-0.2,0)--(2.2,0);  
             \draw(-0.2,0.03)--(2.2,0.03); 
             \draw(-0.2,0.06)--(2.2,0.06); 
             \draw(-0.2,0.09)--(2.2,0.09); 
             \draw(-0.2,0.12)--(2.2,0.12); 
             \draw(-0.2,0.12)--(-1.,0.8);
             \draw(2.2,0.12)--(3.4,0.8);
             \draw[dash dot, thick](-3,-0.4)--(5,-0.4);
            \draw[red, line width=1pt](-0.6,-0.4)--(2.8,-0.4);  

            \draw[double, double distance=1pt, thick, fill=white]  (1,-0.4) circle (0.09);
             \fill (1,-0.4) circle (0.09);
          
          \node[font=\small] at (1,-0.7){$O7^{+}$};
          \node[font=\small] at (3.3,-0.2){$(k-2,1)$};
          \node[font=\small] at(-1.1,-0.2){$(1+k,1)$};
          \node[font=\small] at (4,1){$(n+k-2,1)$};
          
          \node[font=\small] at (-1.8,1){$(n-k-1,-1)$};
          \node[font=\small] at (1,1.2){$(0,1)$};
          \node[font=\small]at(0.4,0.3){$n\,\mathrm{D5}$};
          \node[font=\small]at(1.6,0.3){$n\,\mathrm{D5}$};
          
         \end{tikzpicture}
\end{center}
Here, the 5-brane configuration has a half D5-brane, shown in red, stuck at the position of the $O7^+$-plane. 

The matrix of the external $(p,q)$-charges together with the $O7^{+}$ contribution is
\begin{equation}
A=  \begin{pmatrix}
        n-k-1 & ~-1 \\
        n+k-2 & ~1 \\
        0 &~ 1\\
        2 &~ 0 \\
    \end{pmatrix}\ .
\end{equation}

From this, one can compute SNF as follows

\begin{equation}\mathrm{SNF}(A) =\begin{pmatrix}
~1 & ~0  \\
~0 & ~1  \\
~0 & ~0 
\end{pmatrix} \ .
\end{equation} 
The 1-form symmetry is accordingly
\begin{align}
 \Gamma^{(1)}\big(\mathrm{SU}(2n+1)_{k+\frac12}+1\mathbf{Sym}\big)=1\ .   
\end{align}
It is possible to show that this coincides with the gauge theory expectation,
which is $\mathbb{Z}_{\mathrm{gcd}(\,2n+1,\, 2,\, k_{eff} \,)}$ that is trivial regardless of $k_{eff}$.

\subsubsection{\texorpdfstring{$\SO(N)$}{SO(N)} theories}
The centre of $\SO(N)$ is trivial for odd $N$, and $\mathbb{Z}_2$ for even $N$. Therefore, from the perspective of the $\SO(N)$ gauge theory, the 1-form symmetry is 
\begin{align}
   \Gamma^{(1)} = \mathbb{Z}_{\mathrm{gcd}(N,\, 2)}\ .
\end{align}

\paragraph{\texorpdfstring{SO($2n$) theory.}{SO(2n)}}

An example of 5-brane for SO($2n$) theory is depicted as:

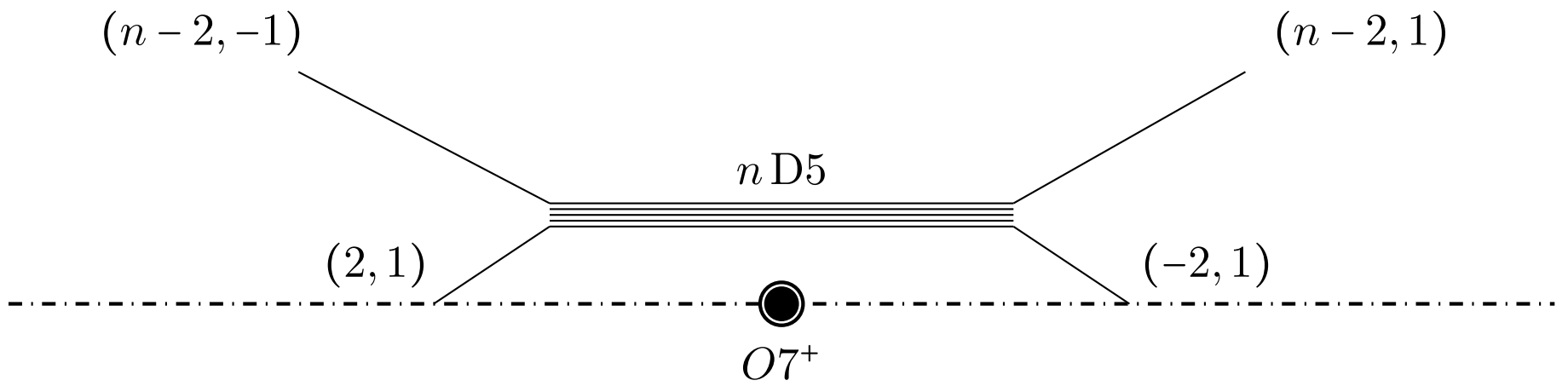
\begin{equation}
    \begin{tikzpicture}[scale=1.5]
             \draw(2-2.2,0)--(1.2-2,-0.4); 
             \draw(2.8,-0.4)--(2.2,0);
             \draw(-0.2,0)--(2.2,0);  
             \draw(-0.2,0.03)--(2.2,0.03); 
             \draw(-0.2,0.06)--(2.2,0.06); 
             \draw(-0.2,0.09)--(2.2,0.09); 
             \draw(-0.2,0.12)--(2.2,0.12); 
             \draw(-0.2,0.12)--(-1.5,0.8);
             \draw(2.2,0.12)--(3.4,0.8);
             \draw[dash dot, thick](-3,-0.4)--(5,-0.4);
             \draw[double, double distance=1pt, thick, fill=white] (1,-0.4) circle (0.09);
             \fill (1,-0.4) circle (0.09);
          
          \node[font=\small] at (1,-0.7){$O7^{+}$};
          \node[font=\small] at (3.2,-0.2){$(-2,1)$};
          \node[font=\small] at(-1.1,-0.2){$(2,1)$};
          \node[font=\small] at (4,1){$(n-2,1)$};
          
          \node[font=\small] at (-2,1){$(n-2,-1)$};
          %
          \node[font=\small]at(1,0.3){$n\,\mathrm{D5}$};
          
         \end{tikzpicture}
\end{equation}

The matrix of $(p,q)$ charges together with the O7$^{+}$ contributions is

\begin{equation}
A=  \begin{pmatrix}
        n-2 & ~-1 \\
        n-2 & ~1 \\
        2 &~ 0 \\
    \end{pmatrix}\ .
\end{equation}
The corresponding SNF leads to $\mathbb{Z}_2$ 1-form symmetry, which agrees with the expectation from the field theory consideration.

\paragraph{\texorpdfstring{SO($2n+1$) theory.}{SO(2n+1)}}
An example of 5-brane for SO($2n+1$) theory is depicted as:

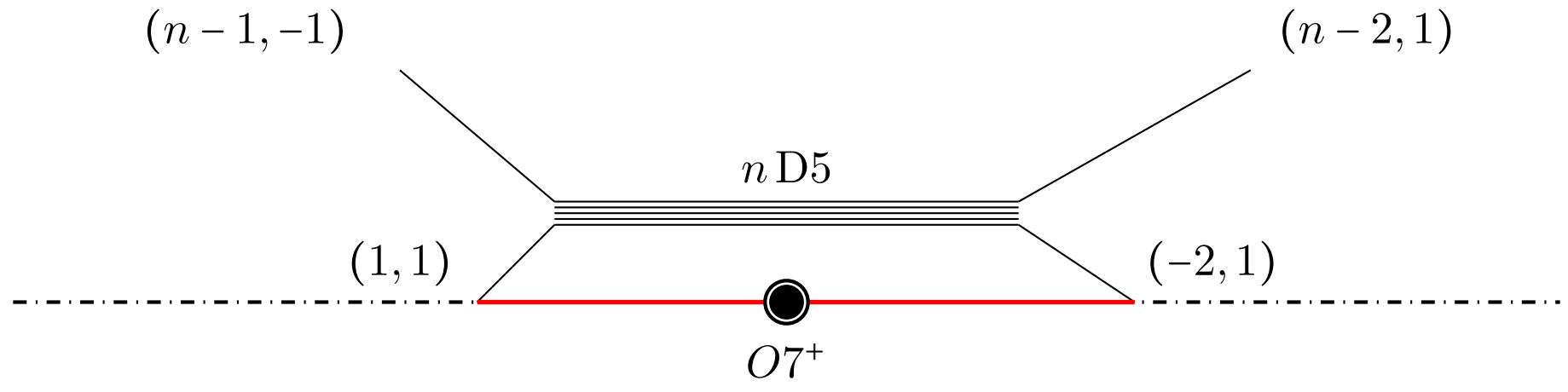
\begin{center}
    \begin{tikzpicture}[scale=1.5]
             \draw(2-2.2,0)--(-0.6,-0.4); 
             \draw(2.8,-0.4)--(2.2,0);
             \draw(-0.2,0)--(2.2,0);  
             \draw(-0.2,0.03)--(2.2,0.03); 
             \draw(-0.2,0.06)--(2.2,0.06); 
             \draw(-0.2,0.09)--(2.2,0.09); 
             \draw(-0.2,0.12)--(2.2,0.12); 
             \draw(-0.2,0.12)--(-1.,0.8);
             \draw(2.2,0.12)--(3.4,0.8);
             \draw[dash dot, thick](-3,-0.4)--(5,-0.4);
            \draw[red, line width=1pt](-0.6,-0.4)--(2.8,-0.4);  
            
            \draw[double, double distance=1pt, thick, fill=white]  (1,-0.4) circle (0.09);
            \fill (1,-0.4) circle (0.09);
             
          \node[font=\small] at (1,-0.7){$O7^{+}$};
          \node[font=\small] at (3.2,-0.2){$(-2,1)$};
          \node[font=\small] at(-1.,-0.2){$(1,1)$};
          \node[font=\small] at (4,1){$(n-2,1)$};
          
          \node[font=\small] at (-1.8,1){$(n-1,-1)$};
          %
          \node[font=\small]at(1,0.3){$n\,\mathrm{D5}$};
          
         \end{tikzpicture}
\end{center}

The matrix of $(p,q)$ charges together with the O7$^{+}$ contributions is

\begin{equation}
A=  \begin{pmatrix}
        n-1 & ~-1 \\
        n-2 & ~1 \\
        2 &~ 0 \\
    \end{pmatrix}\ .
\end{equation}
The corresponding SNF yields trivial 1-form symmetry, which also agrees with the expectation from the field theory consideration.

\subsubsection{Local \texorpdfstring{$\mathbb{P}^2+1\mathbf{Adj}$}{P2+1Adj} or the Bhardwaj theory}
Having gained some confidence about the validity of our conjectured rule for computing the 1-form symmetry for SCFTs constructed from 5-brane webs with an O7$^+$-plane, we now proceed to compute the 1-form symmetry for a family of non-Lagrangian theories. To the best of our knowledge, these results have not previously appeared in the literature.

Consider the 5-brane web for the rank-1 Bhardwaj theory~\cite{Bhardwaj:2019jtr}, also known as the local $\mathbb{P}^2+1\textbf{Adj}$\footnote{$1\textbf{Adj}$ records that the additional hypermultiplet descends from the adjoint of the parent $\mathrm{SU}(2)_\pi+1\mathbf{Adj}$ theory~\cite{Kim:2020hhh}, as $\mathbb{P}^2+1\textbf{Adj}$ can be obtained by decoupling the instantonic hypermultiplet from $\mathrm{SU}(2)_\pi+1\mathbf{Adj}$.}, whose 5-brane web~\cite{Kim:2020hhh,Hayashi:2023boy} is equipped with an O7$^+$-plane,
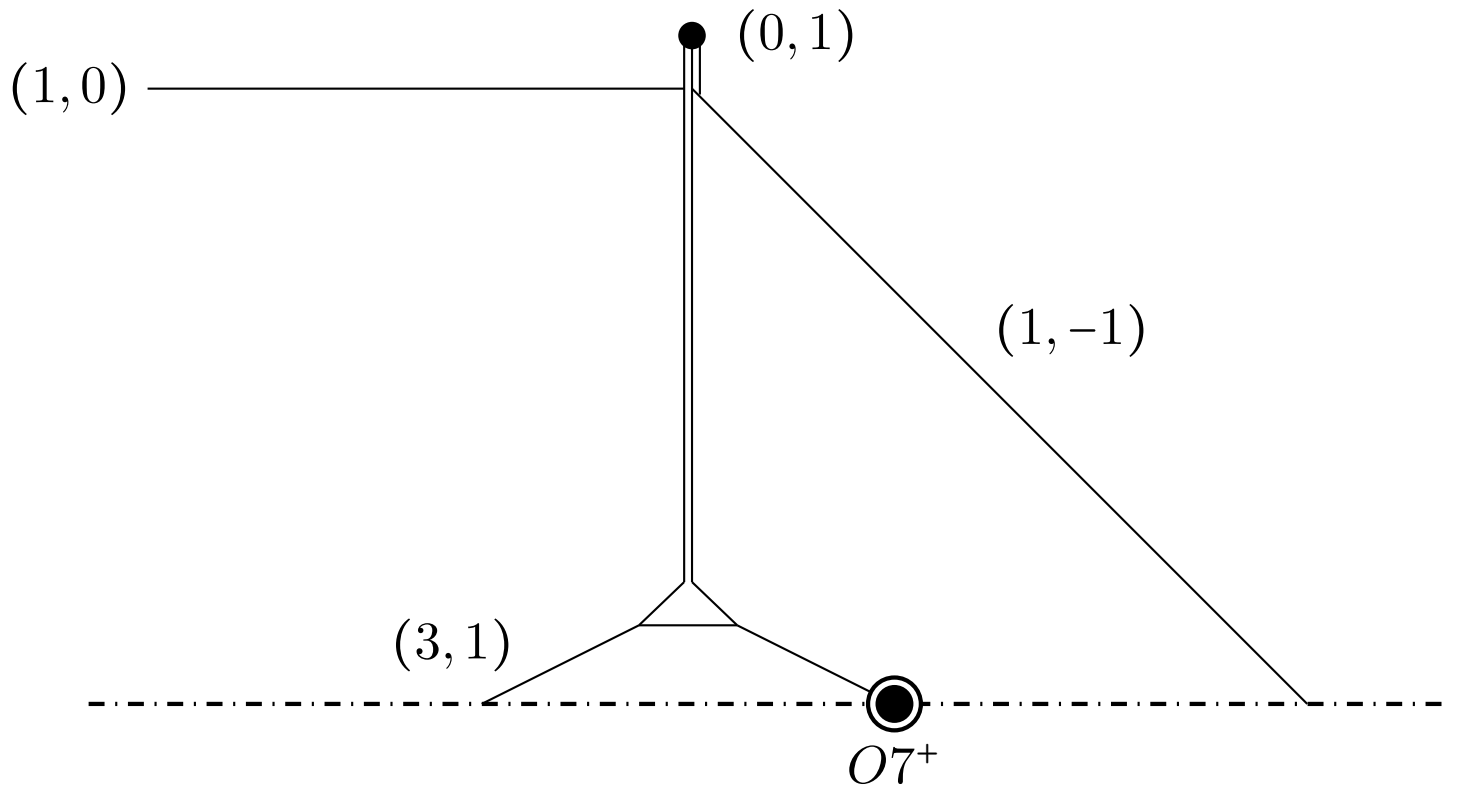
\begin{center}
    \begin{tikzpicture}[scale=1.3]
             \draw(0.1,-0.4)--(0.9,0); 
             \draw(2.2,-0.4)--(1.4,0);
             \draw(0.9,0)--(1.13,0.22);
             \draw(1.4,0)--(1.17,0.22);
             \draw(0.9,0)--(1.4,0); 
             
             \draw(1.17,0.22)--(1.17,3);
             \draw(1.13,0.22)--(1.13,3); 
             \draw(1.21,2.70)--(1.21,3);
             
             \draw[dash dot, thick](1.1-3,-0.4)--(5,-0.4);

              \draw[double, double distance=1pt, thick, fill=white]
    (2.2,-0.4) circle (0.11);

             \fill (2.2,-0.4) circle (0.09);
             \fill (1.17,3) circle (0.07);
          \node[font=\small] at (1.7,3){$(0,1)$};
          
          \node[font=\small] at (2.2,-0.7){$O7^{+}$};
          \node[font=\small] at (-0.05,-0.1){$(3,1)$};
          \node[font=\small] at (3.1,1.5){$(1,-1)$};
          \draw (1.17,2.73)--(4.3,-0.4);

        \draw (-1.6,2.73)--(1.13,2.73);
        \node[font=\small] at (-2,2.73){$(1,0)$};
  
         \end{tikzpicture}
\end{center}
It follows from this diagram that the charge matrix is given by
\begin{equation}
A=  \begin{pmatrix}
        ~0 & ~1~ \\
        ~1 & ~0~ \\
        ~2 &~ 0~ \\
    \end{pmatrix}\ ,
\end{equation}
which yields a trivial 1-form symmetry. One can further consider the higher rank generalisations of Bhardwaj's theory, which can be thought of as $\mathrm{SU}(N)_{\frac{N-1}{2}}``-"1\textbf{F}+\textbf{Sym}$. The resolved brane system can be found in \cite{Akhond:2024nyr}. In particular, the charge matrix is
\begin{equation}
    \begin{pmatrix}
        1&~1~\\
        N-2&~1~\\
        2&~0~
    \end{pmatrix}
\end{equation}
A simple computation predicts that the 1-form symmetry is
\begin{equation}
    \Gamma\left(\SU(N)_{\frac{N-1}{2}}-1\textbf{F}+\textbf{Sym}\right)=\begin{cases}
        1& N\in 2\mathbb{Z}\\
        \mathbb{Z}_2&N\in 2\mathbb{Z}+1
    \end{cases}\;.
\end{equation}

\vspace{0.5cm}

\section{Wilson Lines in 5-brane webs {\it with} O5-planes: Spin gauge theories}\label{sec:withO5}

In this section, we consider Wilson lines associated with Spin gauge groups based on 5-brane constructions. 5d $\mathcal{N}=1$ gauge theories with algebra $\mathfrak{so}(N)$ can be described either by an O7$^+$- or by an O5-plane, but their orientifolding natures are different. Both O7$^+$-planes and O5-planes are fixed loci of $\mathbb{Z}_2$ projections: however, as can seen from table \ref{table:brane_positions}, in the 5-6 plane the O7$^+$ is a pointwise object associated with a $\pi$-rotation, while an O5-plane is a 1d object associated to reflection. We will argue that these two orientifolds describe distinct global structure of the gauge group. A noticeable difference is that an O5-plane can describe matter in the spinor representation~\cite{Zafrir:2015ftn}, which an O7$^+$-plane cannot. Therefore, an O5-plane gives rise to brane descriptions with a Spin($N$) group, while an O7$^+$-plane only describes theories with an SO($N$) group, which we discussed in the preceding section. 

From now on, we consider theories of the Spin($N$) gauge group with $N\in 2\mathbb{Z}$ that are described by an O5-plane. It is known that for spin groups, there exist two distinct possibilities for the centre: for $n\ge 2$, 
\begin{align}
\begin{split}
\mathbb{Z}_2\times \mathbb{Z}_2&\quad \mathrm{for}\quad \mathrm{Spin}(4n)\ ,\cr
\mathbb{Z}_4&\quad \mathrm{for}\quad \mathrm{Spin}(4n+2)\ .
\end{split}
\end{align}
Hence, one finds that the 1-form symmetry for pure Spin($4n$) gauge theories is $\mathbb{Z}_2\times \mathbb{Z}_2$, whereas that for pure Spin($4n+2$) gauge theories is $\mathbb{Z}_4$.

The introduction of matter in the vector ($\mathbf{V}$), spinor ($\mathbf{S}$), or conjugate spinor ($\mathbf{C}$) representation breaks the 1-form symmetry. The resulting symmetry breaking is as follows. For Spin($4n$) gauge theories, 
\begin{align}
\begin{split}
\mathrm{Spin}(4n)+1 \mathbf{S}\, &\quad\rightarrow \quad\mathbb{Z}_2^{\rm C},\\
\mathrm{Spin}(4n)+1 \mathbf{C} &\quad\rightarrow \quad \mathbb{Z}_2^{\rm S},\\
\mathrm{Spin}(4n)+ 1\mathbf{V}&\quad \rightarrow \quad\mathbb{Z}_2^{\rm diag.} .
\end{split}
\end{align}
Here, $\mathbb{Z}_2^{\rm C/S}$ is the remaining $\mathbb{Z}_2$ symmetry after introducing a spinor/conjugate spinor hypermultiplet. On the other hand, in the presence of a hypermultiplet in the vector representation, the 1-form symmetry is broken to the diagonal subgroup $\mathbb{Z}_2^{\rm diag.}\subset\mathbb{Z}_2\times\mathbb{Z}_2$. If the theory has two distinct hypermultiplets, for instance,  $\mathrm{Spin}(4n)+ 1\mathbf{S}+1\mathbf{C}$ or $\mathrm{Spin}(4n)+ 1\mathbf{V}+1\mathbf{S/C}$, then the corresponding 1-form symmetry becomes trivial.  

For Spin($4n+2$) gauge theories, the 1-form symmetry with matter is broken as follows:
\begin{equation}
\begin{split}
    \mathrm{Spin}(4n+2)+1\textbf{S/C}&\quad\to\quad \varnothing,\\
    \mathrm{Spin}(4n+2)+1\textbf{V}&\quad\to \quad\mathbb{Z}_2.
\end{split}
\end{equation}

\begin{figure}[H]
    \centering
    \begin{tikzpicture}
        \draw(0,0)--(2,1);
        \draw(2,1)--(3,2);
        \draw(3,2)--(3,2.5);
        \draw(5,1)--(2,1);
        \draw(4,2)--(3,2);
        \draw(4,2)--(4,2.5);
        \draw(2.5,3)--(3,2.5)--(4,2.5)--(4.5,3);
        \draw(1.5, 3.5)--(2.5,3)--(4.5,3)--(5.5,3.5) ;        
        \draw(5,1)--(4,2);
        \draw(5,1)--(7,0);
        \draw[dashed](-1,0)--(8,0);
        \draw[dotted, thick] (3.5,3.2)--(3.5,3.56);
         \node[label=left:{O5$^+$}] at (0,-0.3){};
         \node[label=left:{O5$^-$}] at (4.2,-0.3){};
         \node[label=left:{O5$^+$}] at (8.2,-0.3){};
        \draw[<->, double] (8.5,2) -- (9.5,2);  
        \draw[dashed](10,0)--(14,0);
        \draw(11,0)--(11,3.5);
        \draw(13,0)--(13,3.5);
        \draw(11,1)--(13,1);
        \draw(11,2)--(13,2);
        \draw(11,2.5)--(13,2.5);
        \draw(11,3)--(13,3);
        \draw[dotted, thick] (12,3.2)--(12,3.56);
    \end{tikzpicture}
    \caption{A simplified diagram with an O5-plane. A 5-brane web diagram on the left illustrates a typical 5-brane configuration with an O5-plane (represented by the dashed line), which realizes a Spin($n$) gauge theory.  The diagram on the right shows a 5-brane web with an O5-plane where all D5-brane charges set to zero, simplifying the configuration for clarity and for ease of reference in later discussions.}
    \label{fig:simplifiedweb}
\end{figure}
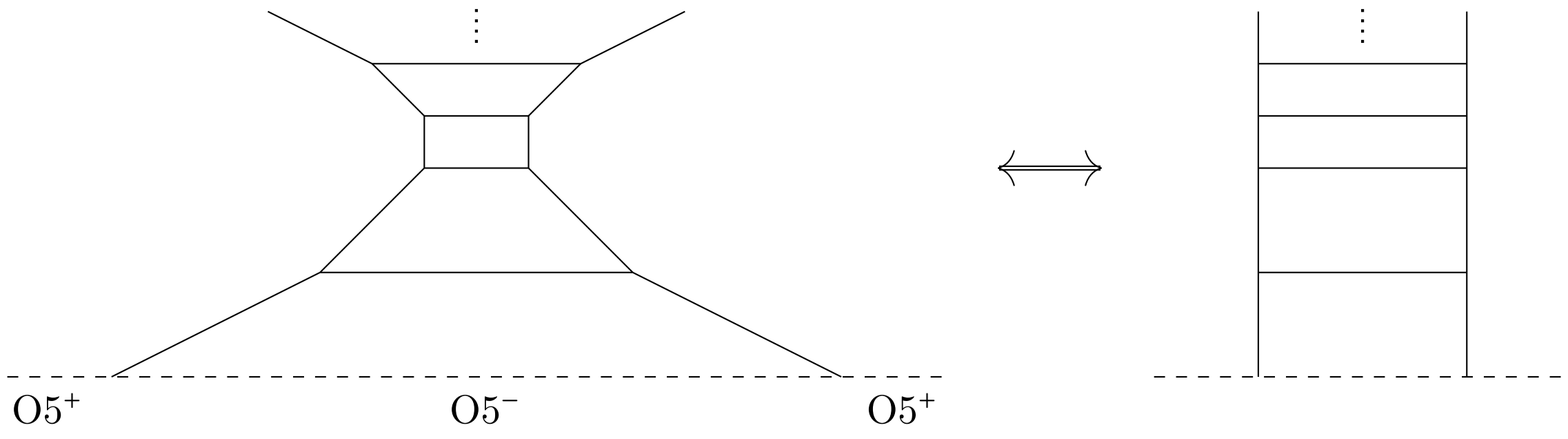

We now present how such 1-form symmetries can be realized on 5-brane webs with an O5-plane. For clarity and for ease of presentation, in this section the brane webs will be depicted in a schematic way, ignoring the bending of the 5-branes, as illustrated in Figure~\ref{fig:simplifiedweb}.

We begin by describing the procedures that will be used repeatedly to identify 1-form symmetries in 5-brane webs involving an O5-plane.

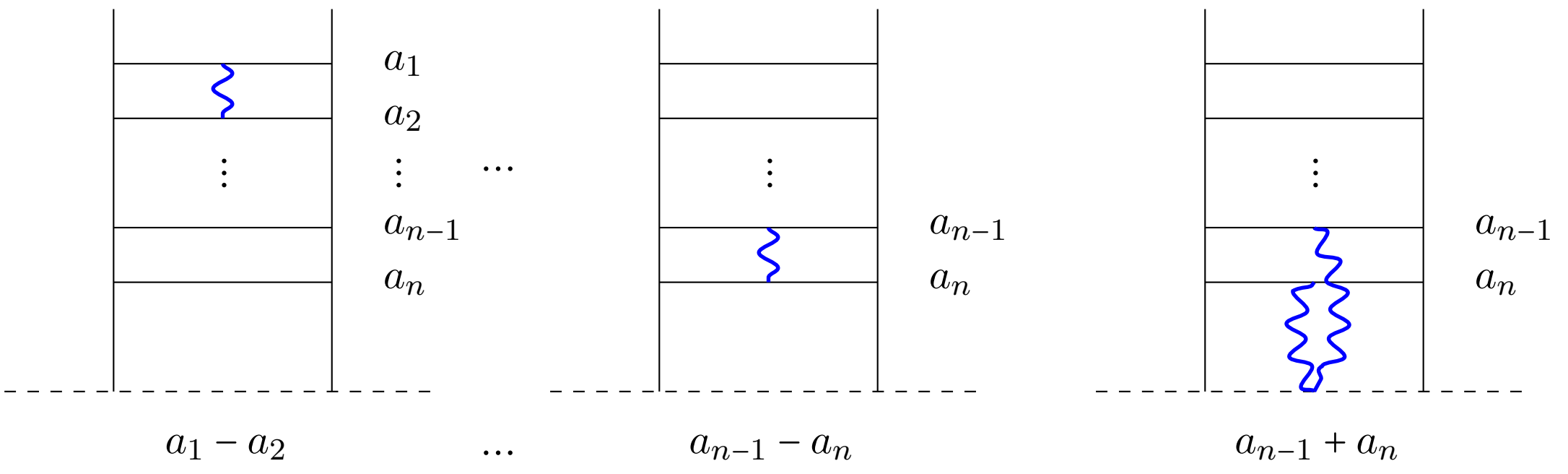
\begin{figure}[H]
    \centering
    \begin{tikzpicture}
\draw[dashed](0,0)--(4,0);
\draw(1,0)--(1,3.5);
\draw(3,0)--(3,3.5);
\draw(1,1)--(3,1);
\draw(1,1.5)--(3,1.5);
\node[label=right:{$a_{1}$}] at (3.2,3){};
\draw(1,2.5)--(3,2.5);
\node[label=right:{$a_{2}$}] at (3.2,2.5){};
\draw(1,3)--(3,3);
\node[label=right:{$\vdots$}] at (1.7,2){};
\node[label=right:{$\vdots$}] at (3.3,2){};
\draw[blue, line width=1.pt, decorate, decoration={snake, segment length=8pt}](2,3)--(2,2.5);
\node[label=right:{$a_{n-1}$}] at (3.2,1.5){};
\node[label=right:{$a_{n}$}] at (3.2,1){};
\node[label=right:{$a_{1}-a_2$}] at (1.2,-0.5){};
\node[label=right:{$\cdots$}] at (4.1,-0.6){};
\node[label=right:{$\cdots$}] at (4.1,2){};
\draw[dashed](5,0)--(9,0);
\draw(6,0)--(6,3.5);
\draw(8,0)--(8,3.5);
\draw(6,1)--(8,1);
\draw(6,1.5)--(8,1.5);
\node[label=right:{$\vdots$}] at (6.7,2){};
\draw(6,2.5)--(8,2.5);
\draw(6,3)--(8,3);
\node[label=right:{$a_{n-1}$}] at (8.2,1.5){};
\node[label=right:{$a_n$}] at (8.2,1){};
\draw[blue, line width=1.pt, decorate, decoration={snake, segment length=8pt}](7,1.5)--(7,1);
\node[label=right:{$a_{n-1}-a_n$}] at (6,-0.5){};

\draw[dashed](10,0)--(14,0);
\draw(11,0)--(11,3.5);
\draw(13,0)--(13,3.5);
\draw(11,1)--(13,1);
\draw(11,1.5)--(13,1.5);
\node[label=right:{$\vdots$}] at (11.7,2){};
\draw(11,2.5)--(13,2.5);
\draw(11,3)--(13,3);
\node[label=right:{$a_{n-1}$}] at (13.2,1.5){};
\node[label=right:{$a_n$}] at (13.2,1){};
\draw[blue, line width=1.pt, decorate, decoration={snake, segment length=8pt}]  (12,0) arc (30:-30:-1.) ;
\draw[blue, line width=1.pt,decorate, decoration={snake, segment length=8pt}] (12,1.5) arc (30:-30:1.5);  
\node[label=right:{$a_{n-1}+a_n$}] at (11,-0.5){};

\end{tikzpicture}
    \caption{Fundamental strings (blue wiggly lines) as the W-bosons corresponding to the simple roots.}
    \label{fig:W-bosons}
\end{figure}

\paragraph{W-bosons.}
The W-bosons of a $\mathrm{Spin}(2n)$ gauge theory are realized by fundamental strings connecting different color D5-branes. 
In particular, the simple roots, $a_1-a_2$, $\cdots$, $a_{n-1}-a_{n}$, and $a_{n-1}+a_n$, of $\mathrm{Spin}(2n)$ Lie algebra correspond to the fundamental strings connecting two nearby D5-branes, as given in Figure \ref{fig:W-bosons}. Other roots are generated by linear combinations of the simple roots, which are described by strings connecting different D5-branes.

\paragraph{D3-branes on the W-boson fundamental strings.}
As discussed in section \ref{sec:linesinweb}, $N$ Wilson lines are described by allocated $N$ D3-branes inside certain chambers of the Coulomb branch associated with the representation of the Wilson lines. For instance, a Wilson line in the vector representation corresponds to a D3-brane in the top chamber of the Coulomb branch, as the top chamber is associated with the fundamental weight. Likewise, a Wilson line in the second antisymmetric tensor representation corresponds to a D3-brane in the second chamber from the top. 

\begin{figure}[H]
    \centering
    \begin{tikzpicture}
\draw[dashed](0,0)--(4,0);
\draw(1,0)--(1,2.5);
\draw(3,0)--(3,2.5);
\draw(1,1)--(3,1);
\draw(1,2)--(3,2);
\draw[dotted, thick] (2,2.2)--(2,2.56);
    \node[cross, red,line width=1.5pt, scale=1.5]at(2,0.3){};

         \draw[<->, double] (4,1.2) -- (5,1.2);     
    \draw[dashed](5,0)--(9,0);
    \draw(6,0)--(6,2.5);
    \draw(8,0)--(8,2.5);
    \draw(6,1)--(8,1);
    \draw(6,2)--(8,2);
    \draw[dotted, thick] (7,2.2)--(7,2.56);
    \node[cross, red, line width=1.5pt, scale=1.5]at(6.8,0){};
    \node[cross, red,line width=1.5pt, scale=1.5]at(7.2,0){};

    \end{tikzpicture}
    \caption{Split of a D3-brane on an O5-plane. On the left, a (full) D3 is located in the bottom chamber. On the right, two half-D3's are located on an O5-plane, each of which can freely move along the O5-plane and also merge into a (full) D3 allowing to leave the O5-plane.}
    \label{fig:split}
\end{figure}
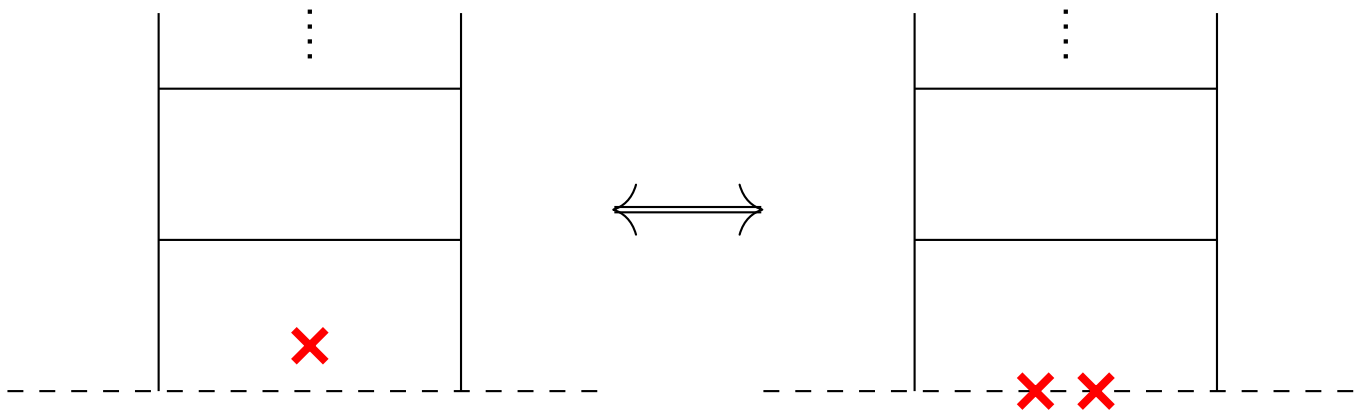

\paragraph{Split of a D3-brane into two half-D3-branes.}
It is important to note that a D3-brane located in the lower chamber of 
the Coulomb branch can reach the O5-plane and split, at no energy cost, into two half-D3-branes. The resulting half-D3-branes are confined to the 
O5-plane but can move independently along it, as illustrated in 
Figure~\ref{fig:split}. Conversely, two half-D3-branes confined to the 
O5-plane can merge into a full D3-brane. Since the resulting D3-brane is no longer confined to the O5-plane, it can move away from it.

\begin{figure}[H]
    \centering
    \begin{tikzpicture}
\draw[dashed](0,0)--(4,0);
\draw(1,0)--(1,3.5);
\draw(3,0)--(3,3.5);
\draw(1,1)--(3,1);
\draw(1,2.5)--(3,2.5);
\draw(1,3)--(3,3);
\draw[dotted, thick] (2,3.2)--(2,3.56);
 \draw[blue, line width=1.pt, decorate, decoration={snake, segment length=8pt}](2,1)--(2,2.5);
    \node[cross, red,line width=1.5pt, scale=1.5]at(2,2){};     

    \node[cross, red,line width=1.5pt, scale=1.5]at(2,1.3){};

         \draw[->, double] (4,2) -- (5,2);     

    \draw[dashed](5,0)--(9,0);
\draw(6,0)--(6,3.5);
\draw(8,0)--(8,3.5);
\draw(6,1)--(8,1);
\draw(6,2.5)--(8,2.5);
\draw(6,3)--(8,3);
\draw[dotted, thick] (7,3.2)--(7,3.56);
\draw[blue, line width=1.pt, decorate, decoration={snake, segment length=8pt}](7,0.99)--(7,1.29);
\draw[blue, line width=1.pt, decorate, decoration={snake, segment length=8pt}](7,2)--(7,2.5);
\node[cross, red,line width=1.5pt, scale=1.5]at(7,2){};
\node[cross, red,line width=1.5pt, scale=1.5]at(7,1.3){};
\draw[dotted, thick] (7,2)--(5,1.5);
\draw[blue, line width=1.pt, decorate, decoration={snake, segment length=8pt}](5.5,1.645)--(5.5,0.9);
\draw[dotted, thick] (7,1.3)--(5,0.8);
\node[label=left:{$x^{7,8,9}$}] at (5.4,0.7){};
        \draw[->, double] (9,2) -- (10,2); 
        \draw[dashed](10,0)--(14,0);
        \draw(11,0)--(11,3.5);
        \draw(13,0)--(13,3.5);
        \draw(11,1)--(13,1);
        \draw(11,2.5)--(13,2.5);
        \draw(11,3)--(13,3);
        \draw[dotted, thick] (12,3.2)--(12,3.56);
        
    \draw[blue, line width=1.pt, decorate, decoration={snake, segment length=8pt}](12,0.99)--(12,1.29);
    \draw[blue, line width=1.pt, decorate, decoration={snake, segment length=8pt}](12,2)--(12,2.5);
    \node[cross, red,line width=1.5pt, scale=1.5]at(12,2){};
    \node[cross, red,line width=1.5pt, scale=1.5]at(12,1.3){};
    \end{tikzpicture}
    \caption{Higgsing of a fundamental string suspended between two D3-branes.}
    \label{fig:HiggsingOfF1strings}
\end{figure}
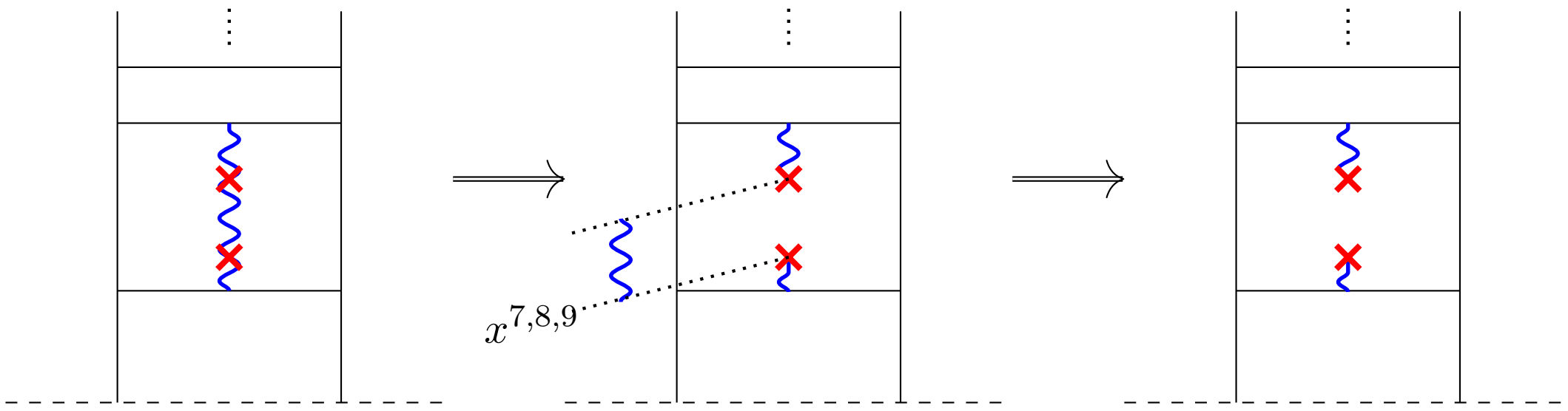

\paragraph{Higgsings.}
Just as a 5-brane stretched between two 7-brane of the same charge can be Higgsed away, a fundamental string stretched between two D3-branes can be Higgsed in an analogous manner. To illustrate this, consider two D3-branes lying in the Coulomb branch chamber of the Coulomb branch, as shown in Figure~\ref{fig:HiggsingOfF1strings}. In the left panel, the two D3-branes, indicated by red crosses, are positioned along a W-boson represented by a fundamental string stretched between two color D5-branes. The W-boson string then splits into three segments: two outer segments, each connecting a D5-brane to a D3-brane, and a middle segment suspended between the two D3-branes. Since the D3-branes extend along the $x^{7,8,9}$ directions, the middle string segment can be displaced away from the plane of the 5-brane web along these directions, as indicated by the dotted lines in the middle panel. Removing this segment through Higgsing yields the configuration shown in the right panel.

\begin{figure}[H]
    \centering
    \begin{tikzpicture}
\draw[dashed](0,0)--(4,0);
\draw(1,0)--(1,3.);
\draw(3,0)--(3,3.);
\draw(1,1)--(3,1);
\draw(1,2)--(3,2);
\draw(1,2.5)--(3,2.5);
\draw[dotted, thick] (2,3)--(2,2.6);
      
    \draw[blue, line width=1.pt,decorate, decoration={snake, segment length=8pt}](2,1.5)--(2,2);
    \draw[blue,line width=1.pt, decorate, decoration={snake, segment length=8pt}](2,1.5)--(2,0);
    \draw[blue, line width=1.pt,decorate, decoration={snake, segment length=8pt}](2,0)--(2.3,1);
    \node[cross, red, line width=1.5pt,scale=1.5]at(2,1.5){};      
    \node[cross, red,line width=1.5pt,scale=1.5]at(2,0){};

         \draw[->, double] (4,1.5) -- (5,1.5);     

    \draw[dashed](5,0)--(9,0);
\draw(6,0)--(6,3.);
\draw(8,0)--(8,3.);
\draw(6,1)--(8,1);
\draw(6,2)--(8,2);
\draw(6,2.5)--(8,2.5);

\draw[dotted, thick] (7,3)--(7,2.6);

\draw[blue,line width=1.pt, decorate, decoration={snake, segment length=8pt}](7,0)--(7.3,1);
\draw[blue, line width=1.pt,decorate, decoration={snake, segment length=8pt}](7,2)--(7,1.5);
\draw[dotted, thick] (7,0)--(5,-0.5);
\draw[blue, line width=1.pt,decorate, decoration={snake, segment length=8pt}](5.5,-0.40)--(5.5,1.12);
\node[cross, red,line width=1.5pt,scale=1.5]at(7,1.5){};
\node[cross, red,line width=1.5pt,scale=1.5]at(7,0){};

\draw[dotted, thick] (7,1.5)--(5,1);
        \draw[->, double] (9,1.5) -- (10,1.5); 
        \draw[dashed](10,0)--(14,0);
        \draw(11,0)--(11,3);
        \draw(13,0)--(13,3);
        \draw(11,1)--(13,1);
        \draw(11,2)--(13,2);
        \draw(11,2.5)--(13,2.5);
        \draw[dotted, thick] (12,3)--(12,2.6);    
    \draw[blue, line width=1.pt,decorate, decoration={snake, segment length=8pt}](12,2)--(12,1.5);
    \draw[blue, line width=1.pt,decorate, decoration={snake, segment length=8pt}](12,0)--(12,1);
        \node[cross, red,line width=1.5pt,scale=1.5]at(12,0){};
    \node[cross, red,line width=1.5pt,scale=1.5]at(12,1.5){};

    \end{tikzpicture}
    \caption{Higgsing associated with a half-D3-brane stuck on an O5-plane.}
    \label{fig:HiggsingonO5-plane}
\end{figure}
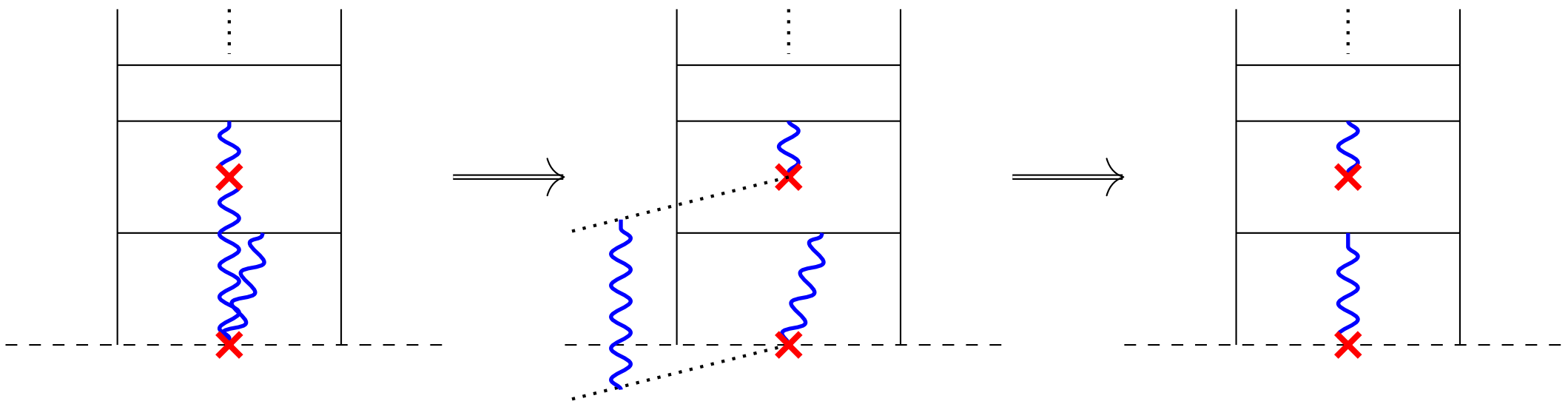

There are yet another two different types of Higgsing on a 5-brane configuration with an O5-plane. One type is the Higgsing associated a fundamental string stretched between a half-D3-brane stuck on an O5-plane and a full D3-brane. An example is shown in Figure~\ref{fig:HiggsingonO5-plane}.

\begin{figure}[H]
    \centering
    \begin{tikzpicture}
\draw[dashed](0,0)--(4,0);
\draw(1,0)--(1,3.);
\draw(3,0)--(3,3.);
\draw(1,1)--(3,1);
\draw(1,2)--(3,2);
\draw(1,2.5)--(3,2.5);
\draw[dotted, thick] (2,3)--(2,2.6);
     
    \draw[blue, decorate, line width=1.pt,decoration={snake, segment length=8pt}](2,1.5)--(2,2);
    \draw[blue, decorate, line width=1.pt,decoration={snake, segment length=8pt}]  (2,0) arc (30:-30:-1.54) ;
    \draw[blue, decorate, line width=1.pt, decoration={snake, segment length=8pt}] (2,1.5) arc (30:-30:1.54);
    \node[cross, red, line width=1.5pt,scale=1.5]at(2,1.5){};       %
         \draw[<->, double] (4,1.5) -- (5,1.5);     

\draw[dashed](5,0)--(9,0);
\draw(6,0)--(6,3);
\draw(8,0)--(8,3);
\draw(6,1)--(8,1);
\draw(6,2)--(8,2);
\draw(6,2.5)--(8,2.5);
\draw[brown](6,-3)--(6,0);
\draw[brown](8,-3)--(8,0);
\draw[brown](6,-1)--(8,-1);
\draw[brown](6,-2)--(8,-2);
\draw[brown](6,-2.5)--(8,-2.5);

\draw[dotted, thick] (7,3)--(7,2.6);
\draw[dotted, brown, thick] (7,-3)--(7,-2.6);

   \draw[blue, decorate, line width=1.pt, decoration={snake, segment length=8pt}](7,1.5)--(7,2);
   \draw[blue, decorate, line width=1.pt, decoration={snake, segment length=8pt}](7,-1.5)--(7,-2);
    \draw[blue, decorate, line width=1.pt,decoration={snake, segment length=8pt}](5.5,1.1)--(5.5,-1.4);
    \node[cross, red,line width=1.5pt,scale=1.5]at(7,1.5){};
\node[cross, red,line width=1.5pt,scale=1.5]at(7,-1.5){};

\draw[dotted, thick] (7,1.5)--(5,1);
\draw[dotted, thick] (7,-1)--(5,-1.5);
        \draw[->, double] (9,1.5) -- (10,1.5); 
        \draw[dashed](10,0)--(14,0);
        \draw(11,0)--(11,3);
        \draw(13,0)--(13,3);
        \draw(11,1)--(13,1);
        \draw(11,2)--(13,2);
        \draw(11,2.5)--(13,2.5);
        \draw[dotted, thick] (12,3)--(12,2.6);    
    \draw[blue, decorate, line width=1.pt,decoration={snake, segment length=8pt}](12,2)--(12,1.5);
    \node[cross, red,line width=1.5pt,scale=1.5]at(12,1.5){};
    
    \end{tikzpicture}
    \caption{A loop as a Higgsing through an O5-plane. On the middle figure, we introduced the mirror images due to the O5-plane, which is equivalent to the configuration based on the fundamental region that is depicted on the left.}
    \label{fig:HiggsingthroughO5-plane-loop}
\end{figure}
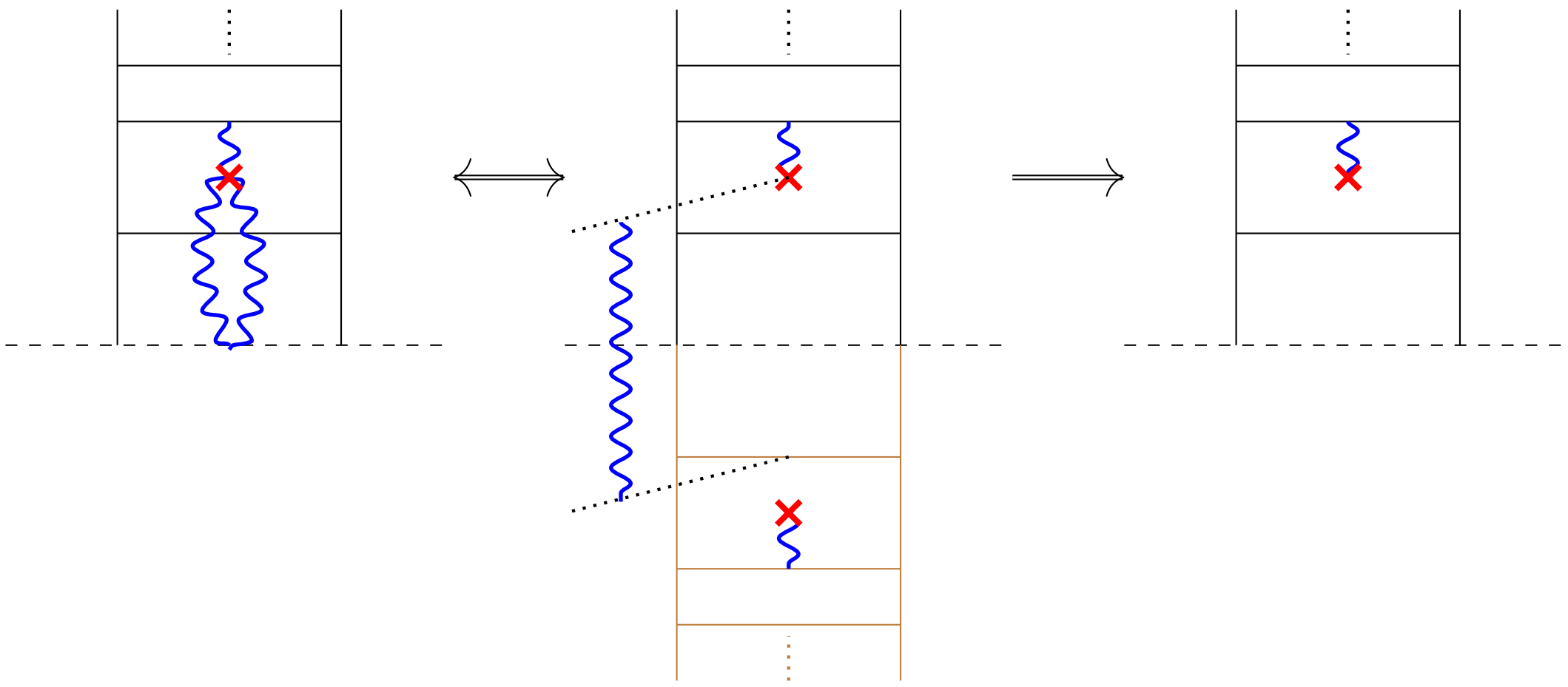

The other Higgsing is the configuration in which a fundamental string makes a loop through an O5-plane, as shown in Figure \ref{fig:HiggsingthroughO5-plane-loop}. This Higgsing can be understood as a string is stretched between a D3-brane and its mirror image through an O5-plane, which appears a loop in the fundamental region. 

\paragraph{Hanany-Witten moves of D3-branes.}
Let us consider a D3-brane move around crossing different chambers of Coulomb branch. By the Hanany-Witten transition, when a D3 moves across D5-branes, there is the creation/annihilation of a fundamental string. 
\begin{figure}[H]
    \centering
    \begin{tikzpicture}
\draw[dashed](0,0)--(4,0);
\draw(1,0)--(1,3.5);
\draw(3,0)--(3,3.5);
\draw(1,1)--(3,1);
\draw(1,2)--(3,2);
\draw(1,3)--(3,3);
\draw[dotted, thick] (2,3.2)--(2,3.56);
\draw[blue, decorate, line width=1.pt,decoration={snake, segment length=8pt}](2,1)--(2,0.3);
    \node[cross, red,line width=1.5pt,scale=1.5]at(2,0.3){};

         \draw[<->, double] (4,2) -- (5,2);     
    \draw[dashed](5,0)--(9,0);
    \draw(6,0)--(6,3.5);
    \draw(8,0)--(8,3.5);
    \draw(6,1)--(8,1);
    \draw(6,2)--(8,2);
    \draw(6,3)--(8,3);
    \draw[dotted, thick] (7,3.2)--(7,3.56);
    \node[cross, red,line width=1.5pt,scale=1.5]at(7,1.5){};
        \draw[<->, double] (9,2) -- (10,2);  
        \draw[dashed](10,0)--(14,0);
        \draw(11,0)--(11,3.5);
        \draw(13,0)--(13,3.5);
        \draw(11,1)--(13,1);
        \draw(11,2)--(13,2);
        \draw(11,3)--(13,3);
        \draw[dotted, thick] (12,3.2)--(12,3.56);
        \draw[blue, decorate, line width=1.pt,decoration={snake, segment length=8pt}](12,2)--(12,2.5);
        \node[cross, red,line width=1.5pt,scale=1.5]at(12,2.5){};
    \end{tikzpicture}
    \caption{Hanany-Witten moves of a D3-brane (represented by a red cross).}
    \label{fig:HWmoves}
\end{figure}
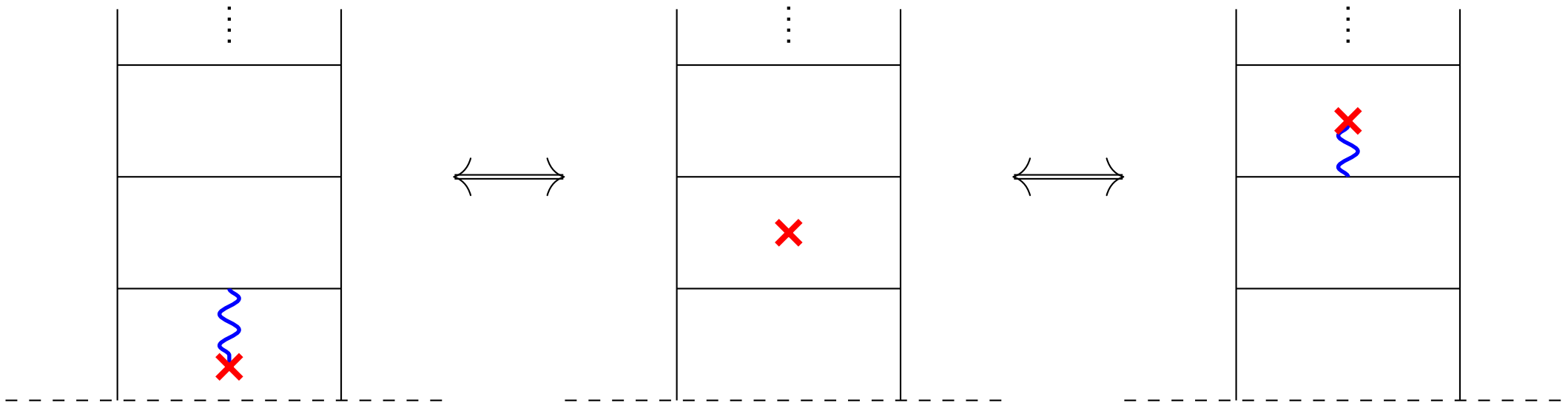

\subsection{\texorpdfstring{$\mathrm{Spin}(4n)$}{Spin(4n)}: \texorpdfstring{$\mathbb{Z}_2\times \mathbb{Z}_2$}{Z2xZ2} symmetry} 
As a representative example, we discuss 
$\mathrm{Spin}(4)$ which has $\mathbb{Z}_2\times \mathbb{Z}_2$ 1-form symmetry. First, recall the multiplication table for $\mathbb{Z}_2\times \mathbb{Z}_2$. As discussed, we take the elements for $\mathbb{Z}_2\times \mathbb{Z}_2$ as $Z =\{ 
(e_1,e_2), (\alpha,e_2), (e_1,\alpha), 
(\alpha,\alpha)
\}$ or $Z =\{ e, \alpha, \beta, 
\alpha \beta
\}$ and then the multiplication table for $\mathbb{Z}_2\times \mathbb{Z}_2$ is given by 
\begin{align}\label{table:Z2xZ2}
\begin{array}{c|cccc}
\mathbb{Z}_2\times \mathbb{Z}_2& e & \alpha &\beta &\alpha\beta\\
\hline
e &  e& \alpha &\beta &\alpha\beta\\
\alpha & \alpha &e & \alpha\beta &\beta\\
\beta & \beta & \alpha\beta &e& \alpha\\
\alpha\beta & \alpha\beta &\beta &\alpha &e
\end{array}    
\qquad {\rm or} \qquad
\begin{array}{c|cccc}
\mathbb{Z}_2\times \mathbb{Z}_2& (0,0) & (1,0) &(0,2) &(1,2)\\
\hline
(0,0) &  (0,0)& (1,0) &(0,2) &(1,2)\\
(1,0) & (1,0) &(0,0) & (1,2) &(0,2)\\
(0,2) & (0,2) & (1,2) &(0,0)& (1,0)\\
(1,2) & (1,2) &(0,2) &(1,0) &(0,0)
\end{array}    
\end{align}
where we used that the map $\phi:Z\to \mathcal{C}$ is a group isomorphism, as discussed in section \ref{sec:global form}:
    \begin{equation}
        \phi: 
\begin{array}{c}
(e_1,e_2) \leftrightarrow e \mapsto (0,0) \\
 (\alpha, e_2)\leftrightarrow \alpha \mapsto (1,0) \\
(e_1, \beta) \leftrightarrow\beta \mapsto (0,2) \\
(\alpha,\beta) \leftrightarrow \alpha\beta\mapsto (1,2)
\end{array}\ .
\end{equation}
Let us consider one $\mathbb{Z}_2$ is given as

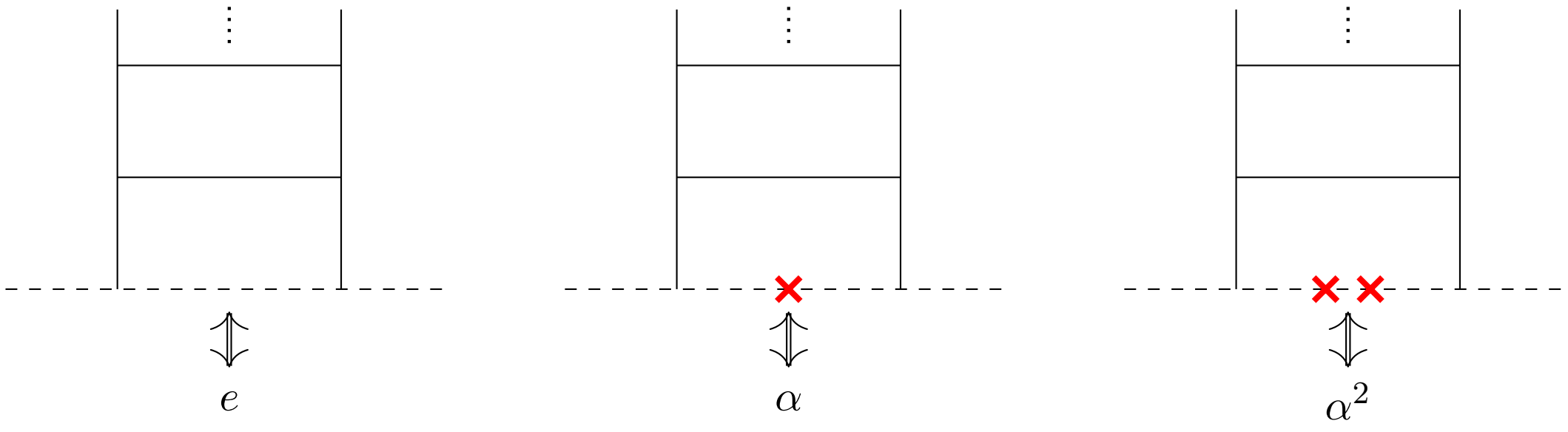
\begin{figure}[H]
    \centering
    \begin{tikzpicture}
\draw[dashed](5-10,0)--(9-10,0);
\draw(6-10,0)--(6-10,2.5);
\draw(8-10,0)--(8-10,2.5);
\draw(6-10,1)--(8-10,1);
\draw(6-10,2)--(8-10,2);
\draw[dotted, thick] (7-10,2.2)--(7-10,2.56);
\draw[<->,double] (-3,-0.2) -- (-3,-0.7);
\node[]at(-3,-1){$e$};

\draw[dashed](0,0)--(4,0);
\draw(1,0)--(1,2.5);
\draw(3,0)--(3,2.5);
\draw(1,1)--(3,1);
\draw(1,2)--(3,2);
\draw[dotted, thick] (2,2.2)--(2,2.56);
\node[cross, red,line width=1.5pt, scale=1.5]at(2,0){};
\node[]at(-3+5,-1){$\alpha$};
\draw[<->,double] (-3+5,-0.2) -- (-3+5,-0.7);
    %
    %
    \draw[dashed](5,0)--(9,0);
    \draw(6,0)--(6,2.5);
    \draw(8,0)--(8,2.5);
    \draw(6,1)--(8,1);
    \draw(6,2)--(8,2);
    \draw[dotted, thick] (7,2.2)--(7,2.56);
    \node[cross, red, line width=1.5pt, scale=1.5]at(6.8,0){};
    \node[cross, red,line width=1.5pt, scale=1.5]at(7.2,0){};
    \node[]at(-3+10,-1){$\alpha^2$};
    \draw[<->,double] (-3+10,-0.2) -- (-3+10,-0.7);
    \end{tikzpicture}
    \caption{An O5-plane realization of $\mathbb{Z}_2$ group representation. }
    \label{fig:Z2}
\end{figure}

\begin{figure}[H]
    \centering

    \begin{tikzpicture}[
        x=1cm,
        y=1cm,
        string/.style={
            blue,
            line width=1pt,
            decorate,
            decoration={snake, segment length=8pt, amplitude=1.8pt}
        },        
         oplane/.style={
        black,
        dashed,
        thick
        },
        brane cross/.style={
            cross,
            draw=red,
            line width=1.2pt,
            minimum size=6pt,
            inner sep=0pt
        }
    ]

    \newcommand{\ladder}[2]{%
        \draw (#1-1,#2)--(#1-1,#2+3.2);
        \draw (#1+1,#2)--(#1+1,#2+3.2);
        \draw (#1-1,#2+1)--(#1+1,#2+1);
        \draw (#1-1,#2+1.5)--(#1+1,#2+1.5);
        \draw (#1-1,#2+2.2)--(#1+1,#2+2.2);
        \draw (#1-1,#2+2.7)--(#1+1,#2+2.7);
        \draw[dotted,thick]
            (#1,#2+1.59)--(#1,#2+2.16);
        \draw[oplane]
            (#1-1.6,#2)--(#1+1.6,#2);
    }


    \node[font=\large] at (-3.05,5.35) {$\alpha^2$};
    \node[font=\Large] at (-2.30,5.35) {:};

    \ladder{0}{4}
    \node[label=right:{$a_1$}] at (0.9,6.7){};
    \node[label=right:{$a_{2}$}] at (0.9,6.2){};
    \node[label=right:{$a_{2n-1}$}] at (0.9,5.5){};
    \node[label=right:{$a_{2n}$}] at (0.9,5){};

    \node[brane cross] at (-0.22,4.03) {};
    \node[brane cross] at ( 0.22,4.03) {};

    \node[font=\Large] at (2.5,5.35) {$\Longleftrightarrow$};

    \ladder{5}{4}

    \node[brane cross] at (4.94,4.03) {};
    \node[brane cross] at (5.06,4.03) {};

    \node[font=\large] at (7.45,6.45) { $n~ W\text{-bosons:}$};
    \node[font=\large] at (7.45,6.05) { $a_{2i-1}+a_{2i}$};
    
    \node[font=\Large] at (7.45,5.35) {$\Longleftrightarrow$};

    \ladder{10}{4}

    \draw[string,cyan] (10,4.03)--(9.3,5);
    \draw[string,cyan] (10,4.03)--(9.5,5.5);
    \draw[string] (10,4.03)--(10.3,6.2);
    \draw[string] (10,4.03)--(10.8,6.7);
    \node[brane cross] at (10,4.03) {};
    \node[brane cross] at (10,4.03) {};

    \node[font=\small] at (0,3.55) {(a)};
    \node[font=\small] at (5,3.55) {(b)};
    \node[font=\small] at (10,3.55) {(c)};

    \node[font=\Large] at (11.5,3.5) {$\Updownarrow$};


    \node[font=\large] at (-3.05,1.35) {$1$};
    \node[font=\Large] at (-2.30,1.35) {:};

    \ladder{0}{0}

    \node[brane cross] at (0,3) {};

    \node[font=\Large] at (2.5,1.35) {$\Longleftrightarrow$};

    \ladder{5}{0}

    \draw[string] (5,1.75)--(5.2,2.7);
    \draw[string] (5,1.75)--(4.7,2.2);
    \node[brane cross] at (5,1.75) {};

    \node[font=\large] at (7.45,2.35) {$\mathrm{HW}$};
    \node[font=\Large] at (7.45,1.65) {$\Longleftrightarrow$};

    \ladder{10}{0}

    \draw[string,cyan] (10,0.5)--(9.3,1);
    \draw[string,cyan] (10,0.5)--(9.5,1.5);
    \draw[string] (10, 0.5)--(10.3,2.2);
    \draw[string] (10, 0.5)--(10.7,2.7);
    \node[brane cross] at (10,0.5) {};

    \node[font=\small] at (0,-0.45) {(f)};
    \node[font=\small] at (5,-0.45) {(e)};
    \node[font=\small] at (10,-0.45) {(d)};
    \end{tikzpicture}

    \caption{
        The relation between $\alpha^2$ and the identity configuration
        through $W$-boson and Hanany-Witten transitions. $\alpha^2=(\mathrm{a})\Leftrightarrow(\mathrm{b})\Leftrightarrow(\mathrm{c})\Leftrightarrow(\mathrm{d})\Leftrightarrow(\mathrm{e})\Leftrightarrow(\mathrm{f})=e.$
    }
    \label{fig:W-boson-1}    
\end{figure}
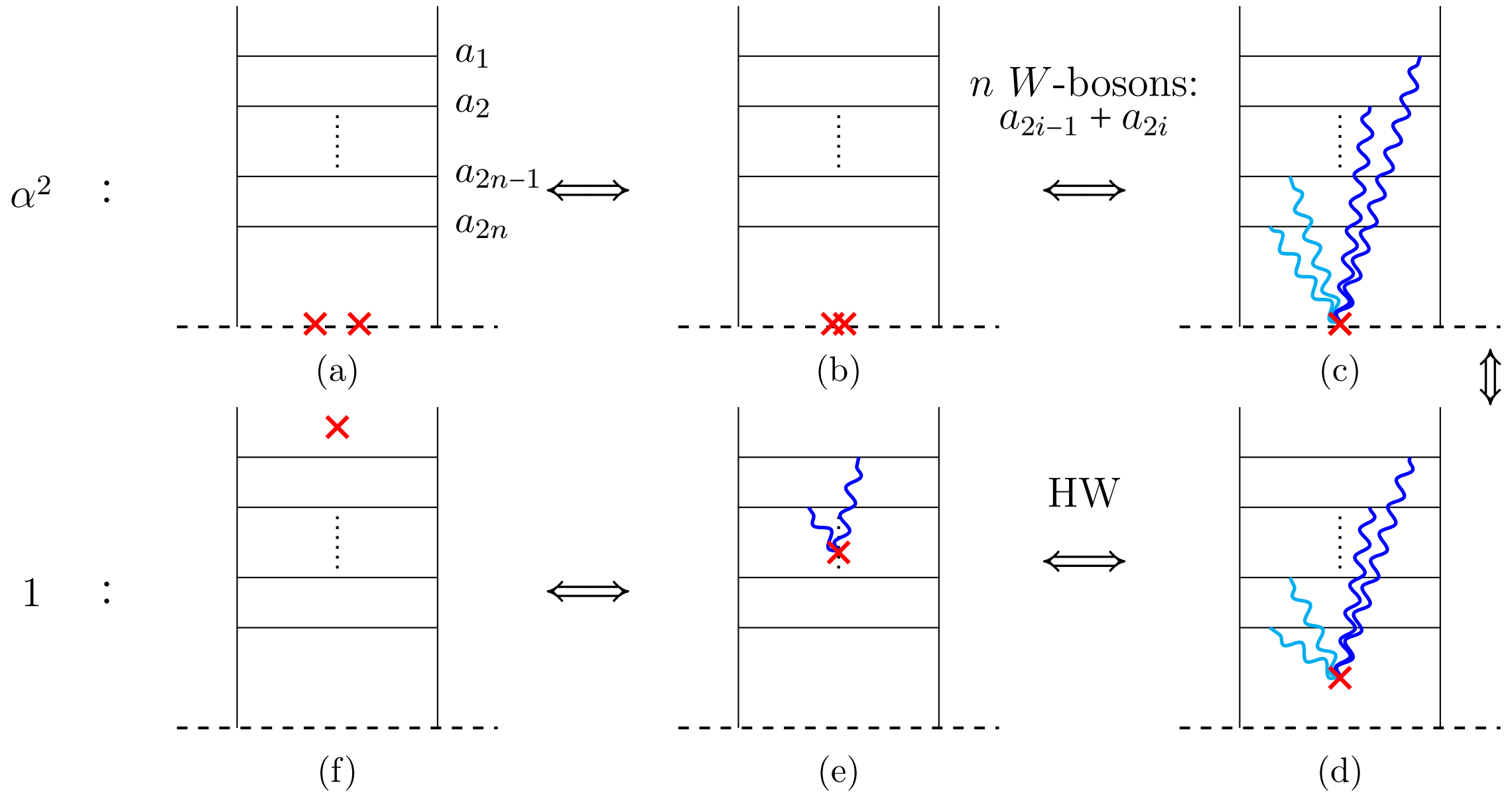

Here, in this brane realisation of $\mathbb{Z}_2$ group in Figure~\ref{fig:Z2}, the identity is denoted by $e$, which is the state where no D3-branes are present in the Coulomb branch chamber, a half-D3-brane stuck in an O5-plane is denoted by a (red) cross, corresponding to a $\mathbb{Z}_2$ element $\alpha$, and two half-D3-branes correspond to $\alpha^2$. One can see that the $\mathbb{Z}_2$ action $\alpha^2=\alpha\cdot\alpha=e$ from the brane with the help of the W-bosons, as illustrated in Figure~\ref{fig:W-boson-1}. 
The procedure described in Figure~\ref{fig:W-boson-1} is as follows: The web configuration given in Figure~\ref{fig:W-boson-1} has $2n$ colour D5-branes, which we denote by $a_i$. The top colour D5-brane is labelled by $a_1$ and the bottom colour D5-brane is labelled by $a_n$. With this label, it is easy to identify relevant W-bosons that play an important role in allocating the D3-brane out of the Coulomb branch. Figure~\ref{fig:W-boson-1}(a) corresponds to  $\alpha^2$ where there are two half-D3-branes on the O5-plane.
In Figure~\ref{fig:W-boson-1}(b), two half-D3-branes get closer and merge to form a full D3-brane. In Figure~\ref{fig:W-boson-1}(c), $n$ W-bosons associated with $a_{2i-1}+a_{2i}$ ($i=1,\cdots,n$) are attached to the full D3-brane on the O5-plane. In Figure~\ref{fig:W-boson-1}(d), the full D3-brane leaves the O5-plane. The procedure Figure~\ref{fig:W-boson-1}(e) illustrates Hanany-Witten moves so to put the D3-brane out of the Coulomb branch chamber. The final configuration in Figure~\ref{fig:W-boson-1}(f) shows that no D3-brane exists in Coulomb branch and hence it corresponds to the identity~$e$. Therefore, this shows that $\alpha^2=e$.

In a similar fashion, we can see the other $\mathbb{Z}_2$ which is independent of the above $\mathbb{Z}_2$:
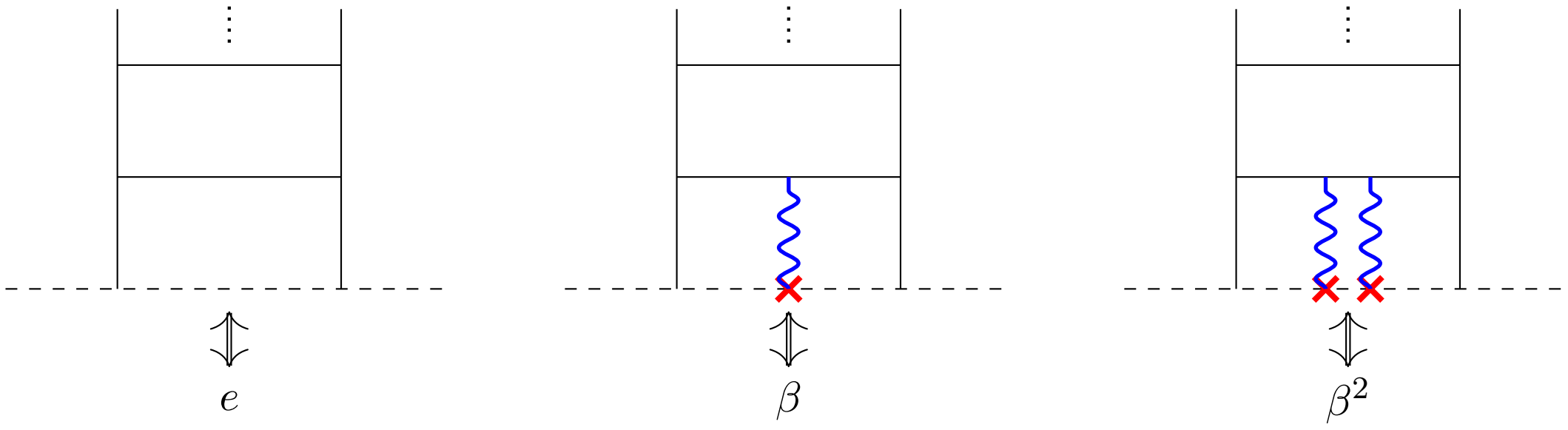
\begin{figure}[H]
    \centering
    \begin{tikzpicture}
\draw[dashed](5-10,0)--(9-10,0);
\draw(6-10,0)--(6-10,2.5);
\draw(8-10,0)--(8-10,2.5);
\draw(6-10,1)--(8-10,1);
\draw(6-10,2)--(8-10,2);
\draw[dotted, thick] (7-10,2.2)--(7-10,2.56);
\draw[<->,double] (-3,-0.2) -- (-3,-0.7);
\node[]at(-3,-1){$e$};

\draw[dashed](0,0)--(4,0);
\draw(1,0)--(1,2.5);
\draw(3,0)--(3,2.5);
\draw(1,1)--(3,1);
\draw(1,2)--(3,2);
\draw[dotted, thick] (2,2.2)--(2,2.56);
\node[cross, red,line width=1.5pt, scale=1.5]at(2,0){};
\node[]at(-3+5,-1){$\beta$};
\draw[<->,double] (-3+5,-0.2) -- (-3+5,-0.7);
\draw[blue, line width=1.pt,decorate, decoration={snake, segment length=8pt}](2,0)--(2,1);
    
    %
    %
    \draw[dashed](5,0)--(9,0);
    \draw(6,0)--(6,2.5);
    \draw(8,0)--(8,2.5);
    \draw(6,1)--(8,1);
    \draw(6,2)--(8,2);
    \draw[dotted, thick] (7,2.2)--(7,2.56);
    \node[cross, red, line width=1.5pt, scale=1.5]at(6.8,0){};
    \node[cross, red,line width=1.5pt, scale=1.5]at(7.2,0){};
    \node[]at(-3+10,-1){$\beta^2$};
    \draw[<->,double] (-3+10,-0.2) -- (-3+10,-0.7);
    \draw[blue, line width=1.pt,decorate, decoration={snake, segment length=8pt}](6.8,0)--(6.8,1);
    \draw[blue, line width=1.pt,decorate, decoration={snake, segment length=8pt}](7.2,0)--(7.2,1);
    \end{tikzpicture}
    \caption{Another O5-plane realization of $\mathbb{Z}_2$ group representation, with a fundamental string connected between a D5-brane and a half-D3-brane stuck in an O5-plane. }
    \label{fig:Z2-beta}
\end{figure}
\begin{figure}[H]
    \centering

    \begin{tikzpicture}[
        x=1cm,
        y=1cm,
        string/.style={
            blue,
            line width=1pt,
            decorate,
            decoration={snake, segment length=8pt, amplitude=1.8pt}
        },        
         oplane/.style={
        black,
        dashed,
        thick
        },
        brane cross/.style={
            cross,
            draw=red,
            line width=1.2pt,
            minimum size=6pt,
            inner sep=0pt
        }
    ]

 \newcommand{\ladder}[2]{%
        \draw (#1-1,#2)--(#1-1,#2+3.2);
        \draw (#1+1,#2)--(#1+1,#2+3.2);
        \draw (#1-1,#2+1)--(#1+1,#2+1);
        \draw (#1-1,#2+1.5)--(#1+1,#2+1.5);
        \draw (#1-1,#2+2.2)--(#1+1,#2+2.2);
        \draw (#1-1,#2+2.7)--(#1+1,#2+2.7);
        \draw[dotted,thick]
            (#1,#2+1.59)--(#1,#2+2.16);
        \draw[oplane]
            (#1-1.6,#2)--(#1+1.6,#2);
    }


    \node[font=\large] at (-3.05,5.35) {$\beta^2$};
    \node[font=\Large] at (-2.30,5.35) {:};

    \ladder{0}{4}
    \node[label=right:{$a_1$}] at (0.9,6.7){};
    \node[label=right:{$a_{2}$}] at (0.9,6.2){};
    \node[label=right:{$a_{2n-1}$}] at (0.9,5.5){};
    \node[label=right:{$a_{2n}$}] at (0.9,5){};

    \node[brane cross] at (-0.22,4.03) {};
    \node[brane cross] at ( 0.22,4.03) {};

    \draw[string] (-0.22,4.03)--(-0.22,5);
    \draw[string] ( 0.22,4.03)--( 0.22,5);

    \node[font=\Large] at (2.5,5.35) {$\Longleftrightarrow$};

    \ladder{5}{4}

    \node[brane cross] at (4.94,4.03) {};
    \node[brane cross] at (5.06,4.03) {};

    \draw[string] (4.94,4.03) -- (4.75,5);
    \draw[string] (5.06,4.03) -- (5.05,5);

    \node[font=\large] at (7.45,6.45) { $n~W\text{-bosons:}$};
    \node[font=\large] at (7.45,6.05) {\color{cyan}$a_{2i-1}+a_{2i},$};
    \node[font=\large] at (7.45,5.65) {\color{red}$a_{2n-1}-a_{2n}$};
    
    \node[font=\Large] at (7.45,5.35) {$\Longleftrightarrow$};

    \ladder{10}{4}

    \draw[string] (10,4.03)--(9.6,5);
    \draw[string] (9.9,5)-- (10,4.03);
    \draw[string,red] (9.9,5)--(9.8,5.5);
    \draw[string,cyan] (10,4.03)--(10.3,6.2);
    \draw[string,cyan] (10,4.03)--(10.8,6.7);

    \node[brane cross] at (10,4.03) {};

    \node[font=\small] at (0,3.45) {(a)};
    \node[font=\small] at (5,3.45) {(b)};
    \node[font=\small] at (10,3.45) {(c)};

   \node[font=\Large] at (11.5,3.5) {$\Updownarrow$};
        (11.5,3.9)--(11.5,3.25);

    \node[font=\large] at (-3.05,1.35) {$1$};
    \node[font=\Large] at (-2.30,1.35) {:};

    \ladder{0}{0}

    \node[brane cross] at (0,3) {};

    \node[font=\Large] at (2.5,1.35) {$\Longleftrightarrow$};

    \ladder{5}{0}

    \draw[string,cyan] (5,1.75)--(5.2,2.7);
    \draw[string,cyan] (5,1.75)--(4.7,2.2);
    \node[brane cross] at (5,1.75) {};

    \node[font=\large] at (7.45,2.35) {$\mathrm{HW}$};
    \node[font=\Large] at (7.45,1.65) {$\Longleftrightarrow$};

    \ladder{10}{0}

    \draw[string] (10,0.5)--(9.3,1);
    \draw[string] (10,0.5)--(9.5,1.5);
    \draw[string,cyan] (10, 0.5)--(10.3,2.2);
    \draw[string,cyan] (10, 0.5)--(10.7,2.7);
    \node[brane cross] at (10,0.5) {};

    \node[font=\small] at (0,-0.45) {(f)};
    \node[font=\small] at (5,-0.45) {(e)};
    \node[font=\small] at (10,-0.45) {(d)};
    \end{tikzpicture}

    \caption{
        The relation between $\beta^2$ and the identity configuration
        through $W$-boson and Hanany-Witten transitions.
    }
    \label{fig:W-boson-2}    
\end{figure}
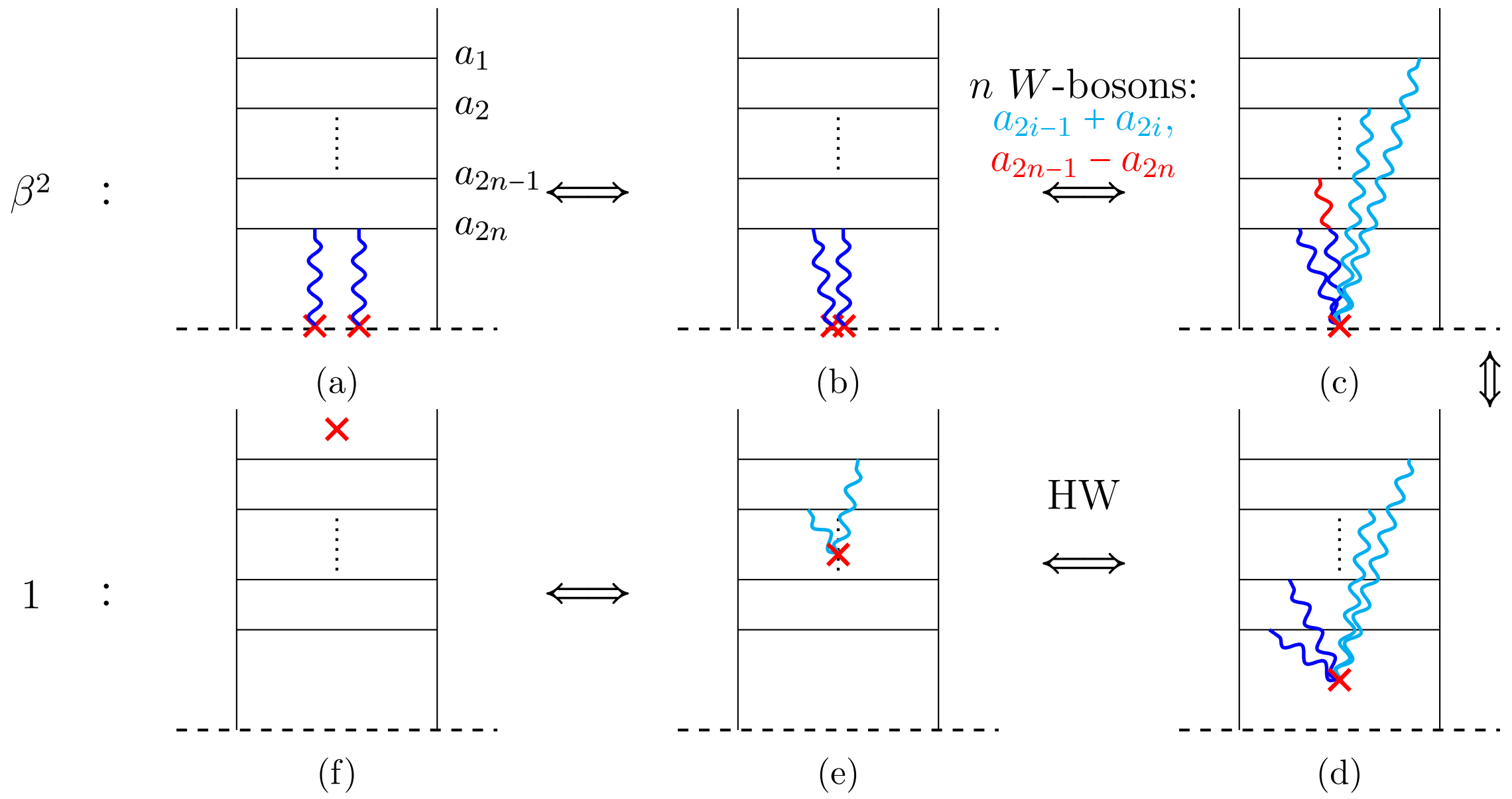
In a similar way, we can see $\beta^2=e$ as follows: Figure~\ref{fig:W-boson-2}(a) corresponds to $\beta^2$ where there are two half-D3-branes on the O5-plane, where a fundamental string is connected between the lower color D5-brane labelled $a_{2n}$ and this half-D3-brane.
In Figure~\ref{fig:W-boson-2}(b), two half-D3-branes get closer and merge to form a full D3-brane. In Figure~\ref{fig:W-boson-2}(c), $n$ W-bosons associated with $\color{cyan} a_{2i-1}+a_{2i}$ ($i=1,\cdots,n-1$) and $\color{red} a_{2n-1}-a_{2n}$ are attached to the full D3-brane on the O5-plane. Note that the W-boson $\color{red} a_{2n-1}-a_{2n}$ and one of the fundamental string connected to half-D3-brane (a blue string in Figure~\ref{fig:W-boson-2}(c)) are recombined together and then connected to each other to form a W-boson of $a_{2n-1}+a_{2n}$, so that the resulting configuration is the same as the one given in Figure~\ref{fig:W-boson-1}. The rest procedures described in Figure~\ref{fig:W-boson-2}(d),(e) and (f) are the same as the ones in Figure~\ref{fig:W-boson-1}(c), which thus shows $\beta^2=1$.

With these two distinct $\mathbb{Z}$s associated with $\alpha$ and $\beta$, let us consider how one can realize $\mathbb{Z}_2\times\mathbb{Z}_2$ symmetry from 5-brane webs with an O5-plane. To this end, we need to further consider mixture of $\alpha$ and $\beta$. First, $\alpha\beta$:
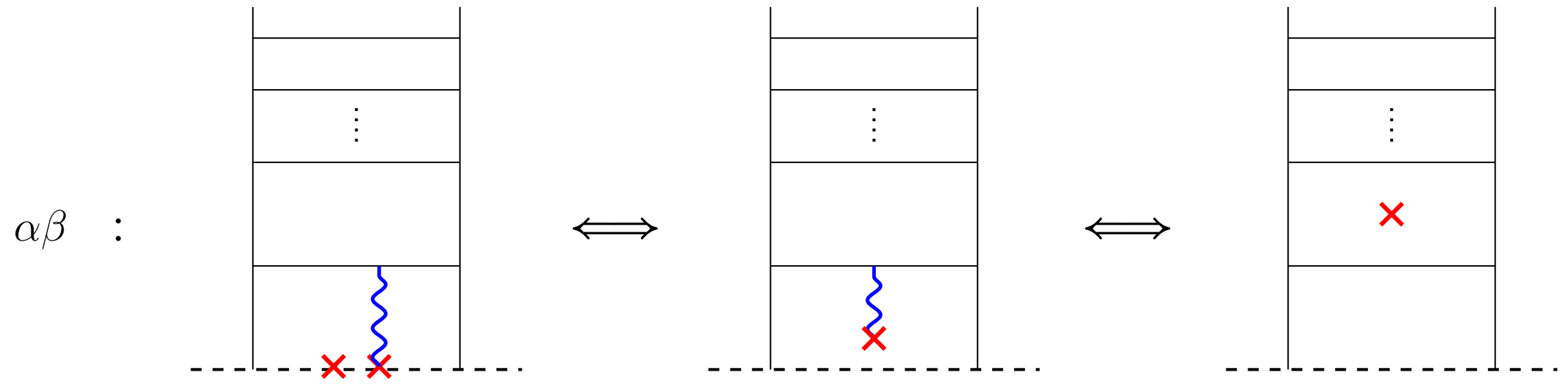
\begin{figure}[H]
    \centering

    \begin{tikzpicture}[
        x=1cm,
        y=1cm,
        string/.style={
            blue,
            line width=1pt,
            decorate,
            decoration={snake, segment length=8pt, amplitude=1.8pt}
        },        
         oplane/.style={
        black,
        dashed,
        thick
        },
        brane cross/.style={
            cross,
            draw=red,
            line width=1.2pt,
            minimum size=6pt,
            inner sep=0pt
        }
    ]

     \newcommand{\ladder}[2]{%
        \draw (#1-1,#2)--(#1-1,#2+3.5);
        \draw (#1+1,#2)--(#1+1,#2+3.5);
        \draw (#1-1,#2+1)--(#1+1,#2+1);
        \draw (#1-1,#2+2)--(#1+1,#2+2);
        \draw[dotted,thick]
            (#1,#2+2.2)--(#1,#2+2.56);
        \draw (#1-1,#2+2.7)--(#1+1,#2+2.7);
        \draw (#1-1,#2+3.2)--(#1+1,#2+3.2);
        \draw[oplane]
            (#1-1.6,#2)--(#1+1.6,#2);
    }


    \node[font=\large] at (-3.05,5.35) {$\alpha\beta$};
    \node[font=\Large] at (-2.30,5.35) {:};

    \ladder{0}{4}

    \node[brane cross] at (-0.22,4.03) {};
    \node[brane cross] at ( 0.22,4.03) {};

    \draw[string] ( 0.22,4.03)--( 0.22,5);

    \node[font=\Large] at (2.5,5.35) {$\Longleftrightarrow$};

    \ladder{5}{4}


    \draw[string] (5,4.3) -- (5,5);
    \node[brane cross] at (5.,4.3) {};

    
    \node[font=\Large] at (7.45,5.35) {$\Longleftrightarrow$};

    \ladder{10}{4}

    \node[brane cross] at (10,5.5) {};

        \end{tikzpicture}

    \caption{A brane realization for the element $\alpha\beta$ of $\mathbb{Z}_2\times \mathbb{Z}_2$.
    }
    \label{fig:ab}    
\end{figure}

Next, we show $(\alpha\beta)\alpha=\beta$:
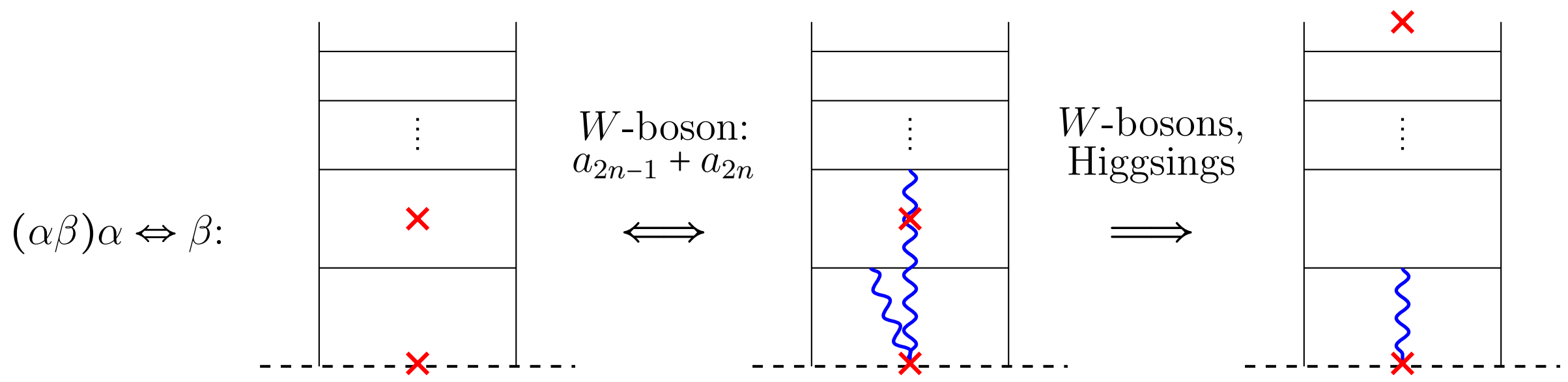
\begin{figure}[H]
    \centering

    \begin{tikzpicture}[
        x=1cm,
        y=1cm,
        string/.style={
            blue,
            line width=1pt,
            decorate,
            decoration={snake, segment length=8pt, amplitude=1.8pt}
        },        
         oplane/.style={
        black,
        dashed,
        thick
        },
        brane cross/.style={
            cross,
            draw=red,
            line width=1.2pt,
            minimum size=6pt,
            inner sep=0pt
        }
    ]

     \newcommand{\ladder}[2]{%
        \draw (#1-1,#2)--(#1-1,#2+3.5);
        \draw (#1+1,#2)--(#1+1,#2+3.5);
        \draw (#1-1,#2+1)--(#1+1,#2+1);
        \draw (#1-1,#2+2)--(#1+1,#2+2);
        \draw[dotted,thick]
            (#1,#2+2.2)--(#1,#2+2.56);
        \draw (#1-1,#2+2.7)--(#1+1,#2+2.7);
        \draw (#1-1,#2+3.2)--(#1+1,#2+3.2);
        \draw[oplane]
            (#1-1.6,#2)--(#1+1.6,#2);
    }


    \node[font=\large] at (-3.05,5.35) {$(\alpha\beta)\alpha
    \Leftrightarrow\beta$:};

    \ladder{0}{4}

    \node[brane cross] at (0,4.03) {};
    \node[brane cross] at (0,5.5) {};

    \node[font=\large] at (2.5,6.45) { $W\text{-boson:}$};
    \node[font=\large] at (2.5,6.05) { $a_{2n-1}+a_{2n}$};
    
    \node[font=\Large] at (2.5,5.35) {$\Longleftrightarrow$};

    \ladder{5}{4}

    \draw[string] (4.6,5) -- (5,4);
    \draw[string] (5,6) -- (5,4);
    \node[brane cross] at (5,4.03) {};
    \node[brane cross] at (5,5.5) {};

    \node[font=\large] at (7.45,6.45) { $W\text{-bosons,}$};
    \node[font=\large] at (7.45,6.05) {Higgsings};
        
    \node[font=\Large] at (7.45,5.35) {$\Longrightarrow$};

    \ladder{10}{4}
    \draw[string] (10,5) -- (10,4);
    \node[brane cross] at (10,4.03) {};
    \node[brane cross] at (10,7.5) {};
        \end{tikzpicture}

    \caption{A brane realization for the element 
        $(\alpha\beta)\alpha=\beta$ of $\mathbb{Z}_2\times \mathbb{Z}_2$.
    }
    \label{fig:aba=b}    
\end{figure}
In a similar way, one can see $(\alpha\beta)\beta=\alpha$:

\begin{figure}[H]
    \centering

    \begin{tikzpicture}[
        x=1cm,
        y=1cm,
        string/.style={
            blue,
            line width=1pt,
            decorate,
            decoration={snake, segment length=8pt, amplitude=1.8pt}
        },        
         oplane/.style={
        black,
        dashed,
        thick
        },
        brane cross/.style={
            cross,
            draw=red,
            line width=1.2pt,
            minimum size=6pt,
            inner sep=0pt
        }
    ]

     \newcommand{\ladder}[2]{%
        \draw (#1-1,#2)--(#1-1,#2+3.5);
        \draw (#1+1,#2)--(#1+1,#2+3.5);
        \draw (#1-1,#2+1)--(#1+1,#2+1);
        \draw (#1-1,#2+2)--(#1+1,#2+2);
        \draw[dotted,thick]
            (#1,#2+2.2)--(#1,#2+2.56);
        \draw (#1-1,#2+2.7)--(#1+1,#2+2.7);
        \draw (#1-1,#2+3.2)--(#1+1,#2+3.2);
        \draw[oplane]
            (#1-1.6,#2)--(#1+1.6,#2);
    }


    \node[font=\large] at (-3.05,5.35) {$(\alpha\beta)\beta
    \Leftrightarrow\alpha$:};

    \ladder{0}{4}

    \draw[string] (0,5) -- (0,4);
    \node[brane cross] at (0,4.03) {};
    \node[brane cross] at (0,5.5) {};

    \node[font=\large] at (2.5,6.45) { $W\text{-boson}:$};
    \node[font=\large] at (2.5,6.05) { $a_{2n-1}-a_{2n}$};
    \node[font=\Large] at (2.5,5.35) {$\Longleftrightarrow$};

    \ladder{5}{4}

    \draw[string] (5,6) -- (5,4);
    \node[brane cross] at (5,4.03) {};
    \node[brane cross] at (5,5.5) {};

    \node[font=\large] at (7.45,6.45) { $W\text{-bosons,}$};
    \node[font=\large] at (7.45,6.05) {Higgsings};
        
    \node[font=\Large] at (7.45,5.35) {$\Longrightarrow$};

    \ladder{10}{4}
    \node[brane cross] at (10,4.03) {};
    \node[brane cross] at (10,7.5) {};
        \end{tikzpicture}

    \caption{A brane realization for the element 
        $(\alpha\beta)\beta=\alpha$ of $\mathbb{Z}_2\times \mathbb{Z}_2$.
    }
    \label{fig:abb=a}    
\end{figure}
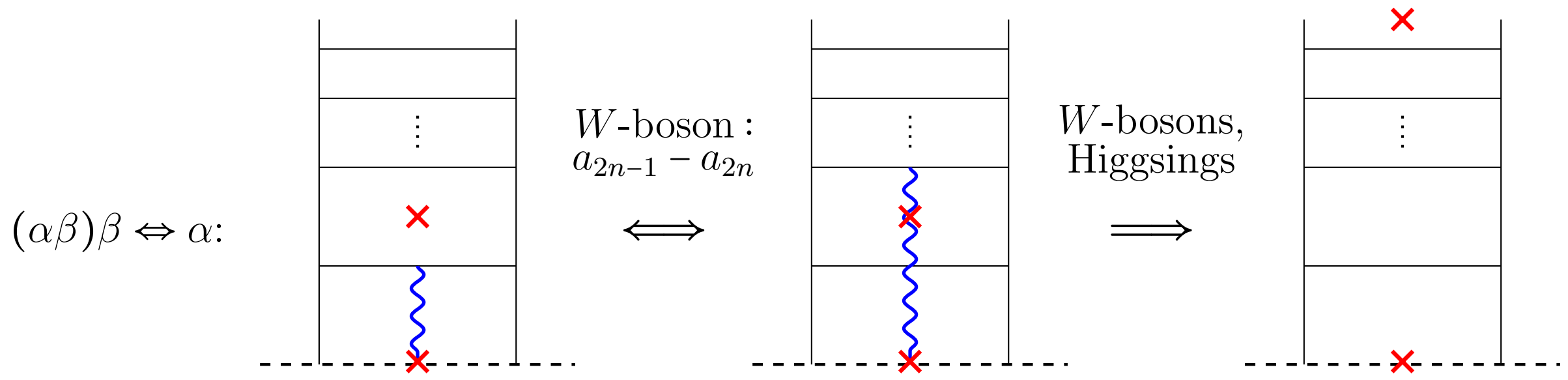
It follows that $(\alpha\beta)^2=1$ can be shown from the Higgsings discussed in Figure~\ref{fig:HiggsingOfF1strings} and Figure~\ref{fig:HiggsingonO5-plane}, and also by connecting the W-bosons.

Putting all possible operations, we can see that the multiplication table for $\mathbb{Z}_2\times\mathbb{Z}_2$ given in \eqref{table:Z2xZ2}, confirming that the proposal is indeed consistent with $\mathbb{Z}_2\times\mathbb{Z}_2$ 1-form symmetry. It is worth noting that in order to screen D3-branes or pulling it out of the Coulomb branch, the number of (color) D5-branes should be $2n$, yielding a $\mathrm{Spin}(4n)$ theory. This agrees with the 1-form symmetry of $\mathrm{Spin}(4n)$ theory being $\mathbb{Z}_2\times\mathbb{Z}_2$.

\subsection{\texorpdfstring{$\mathrm{Spin}(4n+2): \mathbb{Z}_4$}{Spin(4n+2):Z4} symmetry}

Spin($4n+2$) theories ($n\geq 1$) have a 1-form symmetry of $\mathbb{Z}_4$. The $\mathbb{Z}_4$ multiplication table takes the form 

\begin{align}\label{table:Z4}
\begin{array}{c|cccc}
\mathbb{Z}_4& e & \rho & \rho^2 & \rho^3\\
\hline
e &  e& \rho & \rho^2 & \rho^3\\
\rho & \rho & \rho^2 & \rho^3 & e\\
\rho^2 & \rho^2 & \rho^3 & e& \rho\\
\rho^3 & \rho^3 & e & \rho & \rho^2
\end{array}
\qquad{\rm or }\qquad 
\begin{array}{c|cccc}
\mathbb{Z}_4& (0,0) & (1,1) & (0,2) & (1,3)\\
\hline
(0,0) & (0,0) & (1,1) & (0,2) & (1,3)\\
(1,1) & (1,1) & (0,2) & (1,3) & (0,0)\\
(0,2) & (0,2) & (1,3) & (0,0) & (1,1)\\
(1,3) & (1,3) & (0,0) & (1,1) & (0,2)
\end{array}
\end{align}
Here, the group isomorphism 
is given as follows: $e\mapsto (0,0)$, $\rho\mapsto (1,1)$, $\rho^2\mapsto (0,2)$ and $\rho^3\mapsto (1,3)$.

Although the half-D3-branes confined to the O5-plane appear to admit two distinct configurations, denoted by \(\alpha\) and \(\beta\) in the previous subsection, these configurations are equivalent in the construction of the \(\mathbb{Z}_4\) symmetry. More precisely, they are related by
\begin{align}
    \beta=\alpha^3, \qquad
    \beta^2=\alpha^2, \qquad
    \beta^3=\alpha .
\end{align}
Consequently, either configuration may be chosen as a generator of the $\mathbb{Z}_4$ 1-form symmetry. In what follows, we adopt the $\alpha$ configuration and set $\rho=\alpha$.

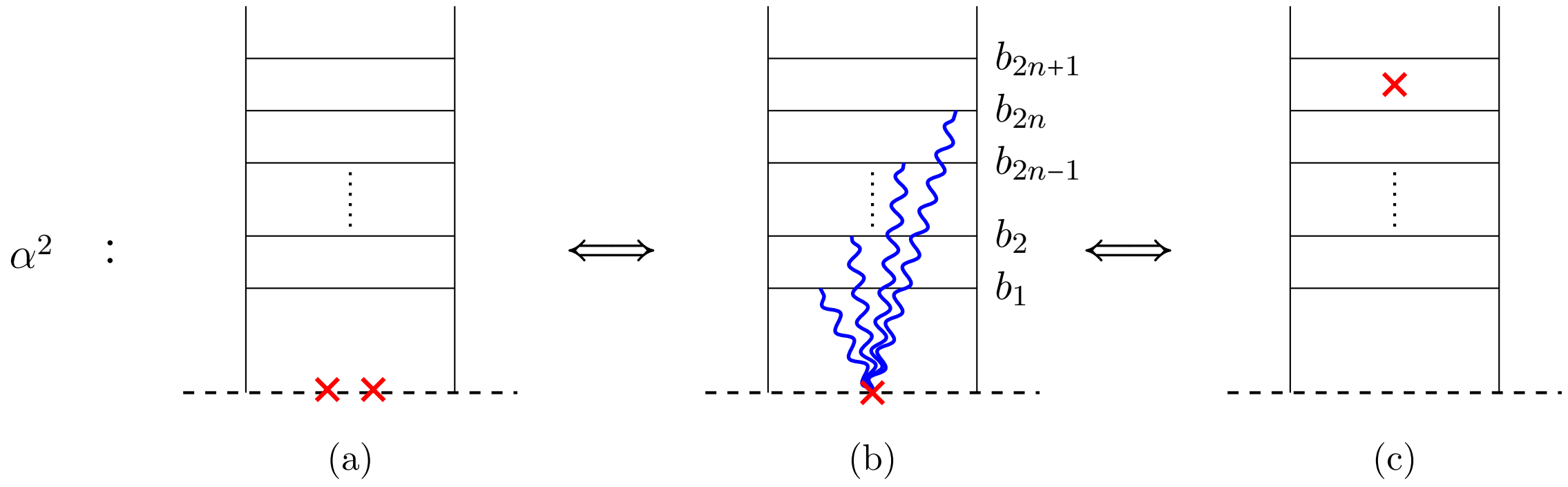
\begin{figure}[H]
    \centering
    \begin{tikzpicture}[
        x=1cm,
        y=1cm,
        string/.style={
            blue,
            line width=1pt,
            decorate,
            decoration={snake, segment length=8pt, amplitude=1.8pt}
        },        
         oplane/.style={
        black,
        dashed,
        thick
        },
        brane cross/.style={
            cross,
            draw=red,
            line width=1.2pt,
            minimum size=6pt,
            inner sep=0pt
        }
    ]

    \newcommand{\ladder}[2]{%
        \draw (#1-1,#2)--(#1-1,#2+3.7);
        \draw (#1+1,#2)--(#1+1,#2+3.7);
        \draw (#1-1,#2+1)--(#1+1,#2+1);
        \draw (#1-1,#2+1.5)--(#1+1,#2+1.5);
        \draw (#1-1,#2+2.2)--(#1+1,#2+2.2);
        \draw (#1-1,#2+2.7)--(#1+1,#2+2.7);
        \draw (#1-1,#2+3.2)--(#1+1,#2+3.2);
        \draw[dotted,thick]
            (#1,#2+1.59)--(#1,#2+2.16);
        \draw[oplane]
            (#1-1.6,#2)--(#1+1.6,#2);
    }


    \node[font=\large] at (-3.05,5.35) {$\alpha^2$};
    \node[font=\Large] at (-2.30,5.35) {:};

    \ladder{0}{4}
    
    \node[brane cross] at (-0.22,4.03) {};
    \node[brane cross] at ( 0.22,4.03) {};

    \node[font=\Large] at (2.5,5.35) {$\Longleftrightarrow$};

    \ladder{5}{4}
    \draw[string] (5,4.03)--(5-0.5,5);
    \draw[string] (5,4.03)--(5-0.2,5.5);
    \draw[string] (5,4.03)--(5.3,6.2);
    \draw[string] (5,4.03)--(5.8,6.7);
    
    \node[label=right:{$b_{2n+1}$}] at (5.9,7.2){};
    \node[label=right:{$b_{2n}$}] at (5.9,6.7){};
    \node[label=right:{$b_{2n-1}$}] at (5.9,6.2){};
    \node[label=right:{$b_{2}$}] at (5.9,5.5){};
    \node[label=right:{$b_{1}$}] at (5.9,5){};
    
    \node[brane cross] at (5,4) {};

    \node[font=\Large] at (7.45,5.35) {$\Longleftrightarrow$};

    \ladder{10}{4}
    \node[brane cross] at (10,6.95) {};

    \node[font=\small] at (0,3.35) {(a)};
    \node[font=\small] at (5,3.35) {(b)};
    \node[font=\small] at (10,3.35) {(c)};
  \end{tikzpicture}
  \vspace{-2em}
  
\caption{\textrm{(a)} $\alpha^2$ generator as two half-D3-branes on O5-plane. \textrm{(b)} two half-D3-branes merge to form a full D3 and W-bosons ($b_1+b_2, \cdots, b_{2n-1}+b_{2n}$) are attached to this D3. \textrm{(c)} The D3 being moved to the top Coulomb chamber by Hanany-Witten moves.}
    \label{fig:a^2-in-Z4}    
\end{figure}

Out of all the elements of the $\mathbb{Z}_4$ 1-form symmetry group, the following three $e$, $\alpha$, and $\alpha^2$ are realized as in Figure~\ref{fig:Z2}. However, note that there is a clear difference in treating $\alpha^2$ when constructing $\mathbb{Z}_2$ and $\mathbb{Z}_4$ symmetries. Recall that $\alpha^2=e$ in $\mathbb{Z}_2$ is due to the rank of $\mathrm{Spin}(4n)$ gauge group is \textit{even}. In other words, the W-bosons of the form $b_{2i-1}+b_{2i}$ ($i=1,\cdots, n)$ are attached to a full D3-brane on an O5-plane, whose resulting configuration is the same as the one created by Hanany-Witten moves when a D3-brane outside the Coulomb branch chambers is taken to the O5-plane. Therefore, the configuration with a full D3-brane connected to the W-bosons can be taken outside the Coulomb branch, as illustrated in Figure~\ref{fig:W-boson-1}. If the rank of $\mathrm{Spin}(4n+2)$ group is \textit{odd}, $\alpha^2$ is equivalent to a D3-brane moved to the top Coulomb chamber, and thus $\alpha^2\neq e$, as shown in Figure~\ref{fig:a^2-in-Z4}. 
Now, $\alpha^3$ then takes the following configuration:
\begin{figure}[H]
    \centering
    \begin{tikzpicture}[
        x=1cm,
        y=1cm,
        string/.style={
            blue,
            line width=1pt,
            decorate,
            decoration={snake, segment length=8pt, amplitude=1.8pt}
        },        
         oplane/.style={
        black,
        dashed,
        thick
        },
        brane cross/.style={
            cross,
            draw=red,
            line width=1.2pt,
            minimum size=6pt,
            inner sep=0pt
        }
    ]

    \newcommand{\ladder}[2]{%
        \draw (#1-1,#2)--(#1-1,#2+3.7);
        \draw (#1+1,#2)--(#1+1,#2+3.7);
        \draw (#1-1,#2+1)--(#1+1,#2+1);
        \draw (#1-1,#2+1.5)--(#1+1,#2+1.5);
        \draw (#1-1,#2+2.2)--(#1+1,#2+2.2);
        \draw (#1-1,#2+2.7)--(#1+1,#2+2.7);
        \draw (#1-1,#2+3.2)--(#1+1,#2+3.2);
        \draw[dotted,thick]
            (#1,#2+1.59)--(#1,#2+2.16);
        \draw[oplane]
            (#1-1.6,#2)--(#1+1.6,#2);
    }

    
    \node[font=\large] at (-3.05,5.35) {$\alpha^3$};
    \node[font=\Large] at (-2.30,5.35) {:};

    \ladder{0}{4}
    \node[label=right:{$b_{2n+1}$}] at (1.,7.2){};
    \node[label=right:{$b_{2n}$}] at (1.,6.7){};
    \node[label=right:{$\vdots$}] at (1,5.8){};
    \node[label=right:{$b_{1}$}] at (1,5){};
    
    \node[brane cross] at (0,4.03) {};
    \node[brane cross] at (0,6.95) {};

    \node[font=\Large] at (2.5,5.35) {$\Longleftrightarrow$};

    \ladder{5}{4}
    \draw[string] (5,4.03)--(5-0.2,5);
    \draw[string] (5,4.03)--(5.2,7.2);

     \node[label=right:{$b_{2n+1}$}] at (5.9,7.2){};
     \node[label=right:{$b_{2n}$}] at (5.9,6.7){};
    {};
    \node[label=right:{$\vdots$}] at (5.9,5.8){};
     \node[label=right:{$b_{1}$}] at (5.9,5){};
    
    \node[brane cross] at (5,4) {};
    \node[brane cross] at (5.2,6.95) {};

    \node[font=\Large] at (7.45,5.35) {$\Longleftrightarrow$};
    \node[font=\small] at (7.45,6.15) { Higgsing,};
    \node[font=\small] at (7.45,5.75) {HW};

    \ladder{10}{4}
    \draw[string] (10,4.03)--(10,5);
    \node[brane cross] at (10,4) {};
    \node[brane cross] at (10,7.45) {};

    \node[font=\small] at (0,3.35) {(a)};
    \node[font=\small] at (5,3.35) {(b)};
    \node[font=\small] at (10,3.35) {(c)};
  \end{tikzpicture}
  \vspace{-2em}
  
\caption{\textrm{(a)} $\alpha^3$ generator as a half-D3 on O5-plane and a full D3 in the top chamber. \textrm{(b)} A W-boson ($b_1+b_{2n+1}$) is connected to a half-D3 and a full D3. \textrm{(c)} The resulting configuration after Higgsing of F1 stretched between half-D3 and a full D3, followed by Hanany-Witten move taking a full D3 outside the Coulomb chamber.}
    \label{fig:a^3inZ4}    
\end{figure}
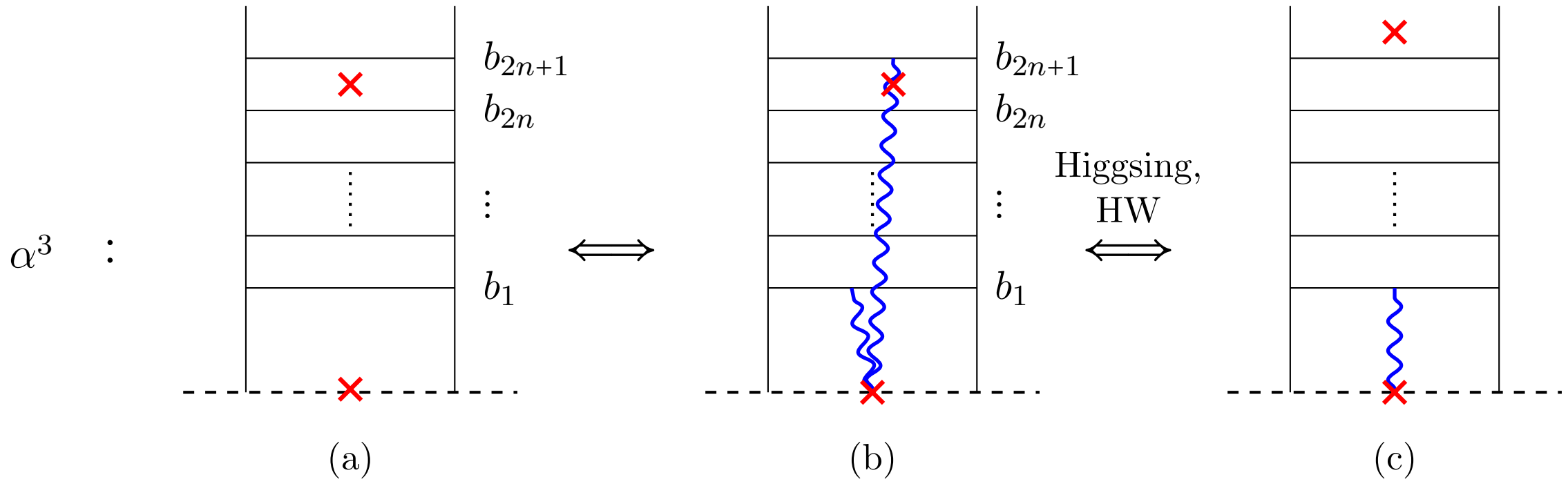
Then, it is straightforward to show that $\alpha^4=e$:
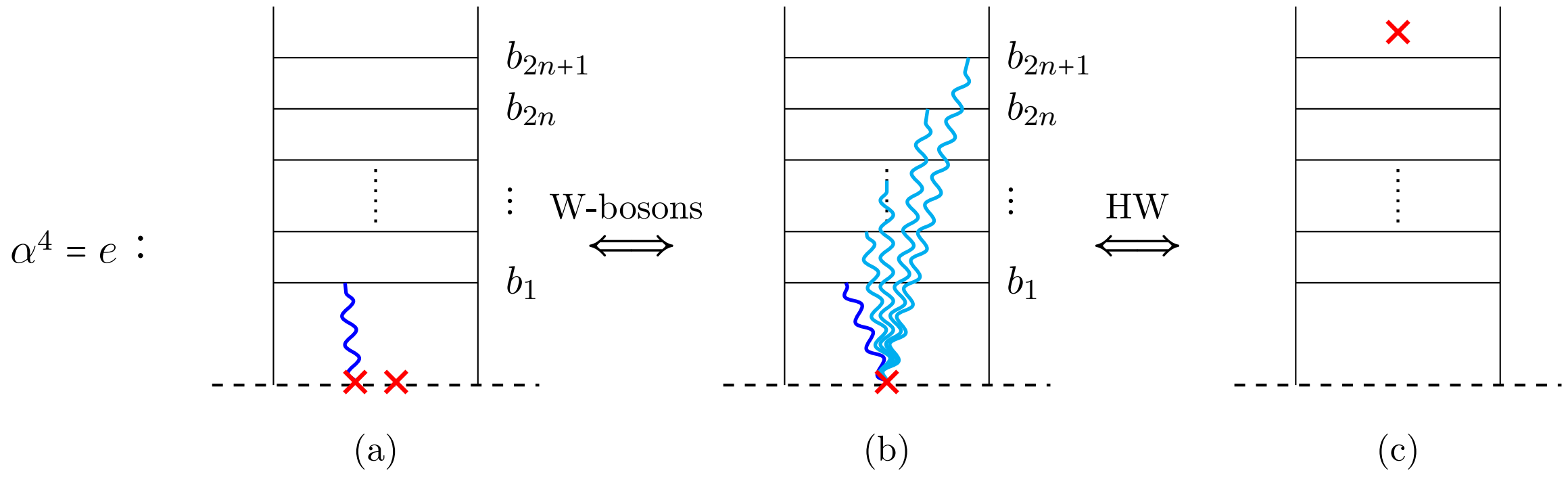
\begin{figure}[H]
    \centering
    \begin{tikzpicture}[
        x=1cm,
        y=1cm,
        string/.style={
            blue,
            line width=1pt,
            decorate,
            decoration={snake, segment length=8pt, amplitude=1.8pt}
        },        
         oplane/.style={
        black,
        dashed,
        thick
        },
        brane cross/.style={
            cross,
            draw=red,
            line width=1.2pt,
            minimum size=6pt,
            inner sep=0pt
        }
    ]

    \newcommand{\ladder}[2]{%
        \draw (#1-1,#2)--(#1-1,#2+3.7);
        \draw (#1+1,#2)--(#1+1,#2+3.7);
        \draw (#1-1,#2+1)--(#1+1,#2+1);
        \draw (#1-1,#2+1.5)--(#1+1,#2+1.5);
        \draw (#1-1,#2+2.2)--(#1+1,#2+2.2);
        \draw (#1-1,#2+2.7)--(#1+1,#2+2.7);
        \draw (#1-1,#2+3.2)--(#1+1,#2+3.2);
        \draw[dotted,thick]
            (#1,#2+1.59)--(#1,#2+2.16);
        \draw[oplane]
            (#1-1.6,#2)--(#1+1.6,#2);
    }

    
    \node[font=\large] at (-3.05,5.35) {$\alpha^4=e$};
    \node[font=\Large] at (-2.30,5.35) {:};

    \ladder{0}{4}
    \node[label=right:{$b_{2n+1}$}] at (1.,7.2){};
    \node[label=right:{$b_{2n}$}] at (1.,6.7){};
    \node[label=right:{$\vdots$}] at (1,5.8){};
    \node[label=right:{$b_{1}$}] at (1,5){};
    
    \draw[string] (-0.2,4.03)--(-0.3,5);
    \node[brane cross] at (-0.2,4.03) {};
    \node[brane cross] at (0.2,4.03) {};

    \node[font=\Large] at (2.5,5.35) {$\Longleftrightarrow$};
     \node[font=\small] at (2.45,5.75) {W-bosons};

    \ladder{5}{4}
    \draw[string] (5,4.03)--(5-0.4,5);
    \draw[string,cyan] (5,4.03)--(5-0.2,5.5);
    \draw[string,cyan] (5,4.03)--(5.,6);
    \draw[string,cyan] (5,4.03)--(5.4,6.7);
    \draw[string,cyan] (5,4.03)--(5.8,7.2);

     \node[label=right:{$b_{2n+1}$}] at (5.9,7.2){};
     \node[label=right:{$b_{2n}$}] at (5.9,6.7){};
    {};
    \node[label=right:{$\vdots$}] at (5.9,5.8){};
     \node[label=right:{$b_{1}$}] at (5.9,5){};
    
    \node[brane cross] at (5,4.03) {};

    \node[font=\Large] at (7.45,5.35) {$\Longleftrightarrow$};
    \node[font=\small] at (7.45,5.75) {HW};

    \ladder{10}{4}
    \node[brane cross] at (10,7.45) {};

    \node[font=\small] at (0,3.35) {(a)};
    \node[font=\small] at (5,3.35) {(b)};
    \node[font=\small] at (10,3.35) {(c)};
  \end{tikzpicture}

\caption{\textrm{(a)} $\alpha^4=\alpha^3\cdot\alpha$.~~\textrm{(b)} W-bosons ($b_2+b_{3}, \cdots, b_{2n}+b_{2n+1}$) are connected to a full D3 on O5-plane. \textrm{(c)} By Hanany-Witten, the D3 can be taken outside the Coulomb chamber, and hence $\alpha^4=e$.}
    \label{fig:a^4=e}    
\end{figure}

Next we take the $\beta$ configuration. It is straightforward to see that $\beta$ configuration is equivalent to the $\alpha$ configuration, as two configurations are related by
\begin{align}\label{eq:beta=alpha3}
    \beta=\alpha^3, \quad 
    \beta^2=\alpha^2 , \quad
    \beta^3= \alpha \ .
\end{align}

We note that it is also straightforward to see that the $n\ge 2$ cases are consistent with $\mathbb{Z}_4$ symmetry which can be seen in a similar way as the $n=1$ case by introducing proper W-bosons.

\paragraph{Special case: Spin(2)$\cong$ U(1):} The case $n=0$ is special and requires a separate discussion. In this instance Spin$(2)=\mathrm{U}(1)$, and the 1-form symmetry is actually $\mathrm{U}(1)$. That is, there are infinitely many line operators labelled by integers. In other words, the web should reflect that it is impossible to screen \emph{any} line operators. Indeed this can be seen immediately due to the boundary condition imposed by the O5-plane. Consider a Wilson line in the spinor representation of Spin$(2)$, equivalently one carrying unit charge under~$\mathrm{U}(1)$
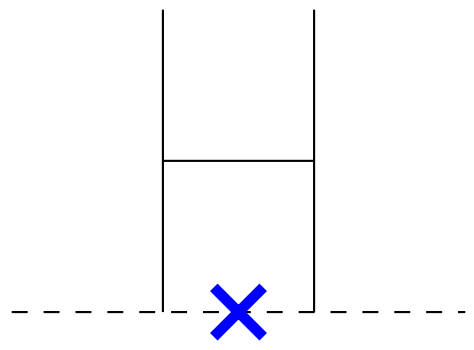
\begin{equation}
    \begin{tikzpicture}
        \draw(0,0)--(0,2) (1,0)--(1,2) (0,1)--(1,1);
        \draw[dashed](-1,0)--(2,0);
        \node[cross out, draw=blue, line width=2pt, minimum size=0.001cm] at (.5,0){};
    \end{tikzpicture}
\end{equation}
The only possible way to screen this D3-brane would be to have a number of them moved outside the Coulomb face by using W-bosons. But the lack of any W-bosons in abelian gauge theory forbids this. In other words, since the orientifold projects out the following would be W-boson
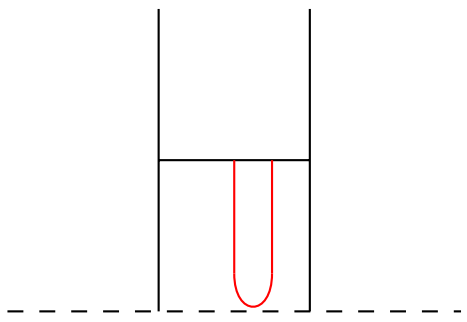
\begin{equation}
\begin{array}{c}
    \begin{tikzpicture}
        \draw(0,0)--(0,2) (1,0)--(1,2) (0,1)--(1,1);
        \draw[dashed](-1,0)--(2,0);
        \draw[red] (.5,.25)--(.5,1);
        \draw[red] (.75,.25)--(.75,1);
        \draw[red] (.75,0.25)to[out=-90,in=-90,loop,looseness=3](.5,0.25);
    \end{tikzpicture}
    \end{array}\;,
\end{equation}
one may insert infinitely many line operators at will, signalling a U(1) 1-form symmetry.

\subsection{\texorpdfstring{$\mathrm{Spin}(2n+1): \mathbb{Z}_2$ symmetry}{Spin(2n+1):Z2}}
The $\mathrm{Spin}(2n+1)$ theory possesses a $\mathbb{Z}_2$ 1-form symmetry, 
which can be identified using an argument analogous to those presented 
in the preceding subsections.
Before describing its 5-brane web realization, 
let us briefly introduce the relevant $\widetilde{\mathrm{O}5}$-planes. An $\widetilde{\mathrm{O}5}^{-}$-plane may be regarded as an 
$\mathrm{O}5^{-}$-plane together with a stuck half-D5-brane and the 
monodromy cut associated with a half-D7-brane. Similarly, an $\widetilde{\mathrm{O}5}^{+}$-plane consists of an $\mathrm{O}5^{+}$-plane accompanied by the monodromy cut of a half-D7-brane. In the 5-brane-web description, the $\mathrm{Spin}(2n+1)$ theory is obtained by Higgsing the $\mathrm{Spin}(2n+2)+1\mathbf{V}$ theory, and this  procedure naturally introduces an $\widetilde{\mathrm{O}5}$-plane. See \cite{Hayashi:2018bkd} for more explicit construction. 

Because the brane construction of the $\mathrm{Spin}(2n+1)$ theory starts from a 
configuration containing a hypermultiplet in the vector representation, 
the 1-form symmetry of the parent theory is explicitly reduced from 
either $\mathbb{Z}_4$ or $\mathbb{Z}_2 \times \mathbb{Z}_2$, depending on 
$N$, to $\mathbb{Z}_2$. Although the Higgsing procedure could, in 
principle, lead to an enhancement of the 1-form symmetry, we will show 
that no such enhancement occurs in the brane realization of the 
$\mathrm{Spin}(2n+1)$ theory. Consequently, the $\mathbb{Z}_2$ 1-form symmetry 
remains unchanged after Higgsing.

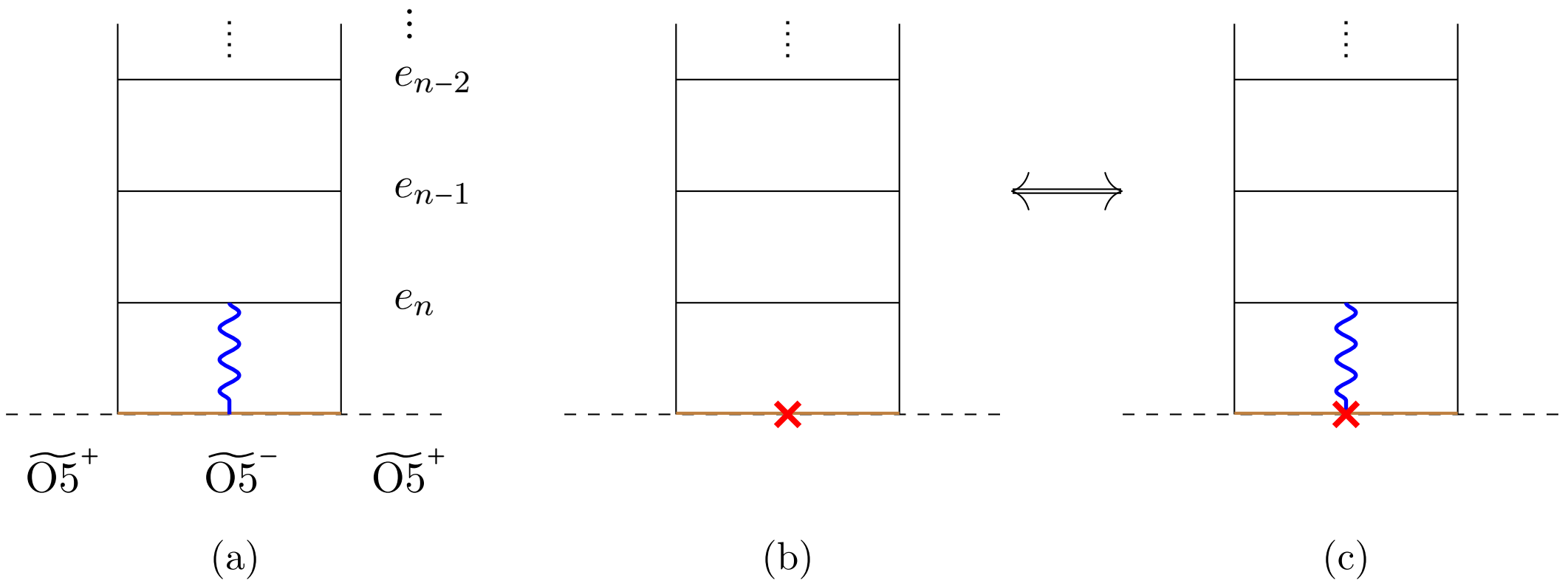
\begin{figure}[H]
    \centering
    \begin{tikzpicture}
\draw[dashed](0,0)--(4,0);
\draw(1,0)--(1,3.5);
\draw(3,0)--(3,3.5);
\draw(1,1)--(3,1);
\draw[brown, thick](1,0.01)--(3,0.01);
\node[label=right:{$e_{n-2}$}] at (3.2,3){};
\draw(1,2)--(3,2);
\node[label=right:{$e_{n-1}$}] at (3.2,2){};
\draw(1,3)--(3,3);
\node[label=right:{$e_n$}] at (3.2,1){};
\draw[dotted, thick] (2,3.2)--(2,3.56);
\node[label=right:{$\vdots$}] at (3.3,3.5){};
\draw[blue, decorate, line width=1.pt, decoration={snake, segment length=8pt}](2,1)--(2,0);
\node[label=right:{$\widetilde{\mathrm{O5}}^+$}] at (-0.1,-0.5){};
\node[label=right:{$\widetilde{\mathrm{O5}}^-$}] at (1.5,-0.5){};
\node[label=right:{$\widetilde{\mathrm{O5}}^+$}] at (3.0,-0.5){};

%
%
  %
    %
    \draw[dashed](5,0)--(9,0);
    \draw(6,0)--(6,3.5);
    \draw(8,0)--(8,3.5);
    \draw(6,1)--(8,1);
    \draw(6,2)--(8,2);
    \draw(6,3)--(8,3);
    \draw[brown, thick](6,0.01)--(8,0.01);
    \draw[dotted, thick] (7,3.2)--(7,3.56);
    \node[cross, red,line width=1.5pt,scale=1.5]at(7,0){};
        \draw[<->, double] (9,2) -- (10,2);  
        \draw[dashed](10,0)--(14,0);
        \draw(11,0)--(11,3.5);
        \draw(13,0)--(13,3.5);
        \draw(11,1)--(13,1);
        \draw(11,2)--(13,2);
        \draw[brown, thick](11,0.01)--(13,0.01);
        \draw(11,3)--(13,3);
        \draw[dotted, thick] (12,3.2)--(12,3.56);
        \draw[blue, decorate,,line width=1.pt, decoration={snake, segment length=8pt}](12,1)--(12,0);
        \node[cross, red,line width=1.5pt,scale=1.5]at(12,0){};
\node[font=\small] at  (2.05,-1.3){(a)};
    \node[font=\small] at (7,-1.3) {(b)};
    \node[font=\small] at (12,-1.3) {(c)};  
  \vspace{-2em}
\end{tikzpicture}
    \caption{$\mathbb{Z}_2$ 1-form symmetry configuration for the $\mathrm{Spin}(2n+1)$ theory.
{\rm (a)}    
    A W-boson string accounting for the short simple root $e_n$ of $\mathrm{Spin}(2n+1)$ Lie algebra, where an $\widetilde{\mathrm{O5}}^-$-plane possesses a half-D5-brane that is shaded in brown. When a half-D3-brane lies on $\widetilde{\rm O5}^-$-plane, the $\alpha$ configuration {\rm (b)} and the $\beta$ configuration {\rm (c)}, are not distinct but related by the W-boson associated with the simple root $e_n$. }
    \label{fig:W-boson-en}
\end{figure}

The Lie algebra of $\mathrm{Spin}(2n+1)$ is non-simply laced and contains a short 
simple root. Its brane realization is shown in 
Figure~\ref{fig:W-boson-en}\textrm{(a)}, where the short simple root is 
represented by the blue fundamental string stretching between the colour D5-brane labelled by $e_n$ and the $\widetilde{\mathrm{O}5}^{-}$-plane. 
The corresponding W-boson relates the $\alpha$ and $\beta$ configurations 
introduced in the preceding subsections: one configuration can be 
transformed into the other by adding the W-boson associated with the short 
root. Therefore, the configurations $\alpha$ and $\beta$, depicted in 
Figures~\ref{fig:W-boson-en}\textrm{(b)} and 
\ref{fig:W-boson-en}\textrm{(c)}, respectively, represent two equivalent brane configurations which give rise to  the same 1-form symmetry. It follows that the $\mathrm{Spin}(2n+1)$ theory has 
a $\mathbb{Z}_2$ 1-form symmetry.

\subsection{Spin gauge theory with spinor matter}

By introducing a hypermultiplet in spinor or conjugate spinor representation of $\mathrm{Spin}(2N)$ group the 1-form symmetry $\mathbb{Z}_2 \times \mathbb{Z}_2$ or $\mathbb{Z}_4$, depending on 
$N$, reduces to $\mathbb{Z}_2$ or trivial, respectively. In the following, we check this statement by specifying the strings used to screen the Wilson line operators.

\begin{figure}[H]
\centering
    \begin{tikzpicture}
\draw[dashed](-7,0)--(-1,0);
\draw (-4,1)--(-3,0)--(-2,1);
\fill (-6,0) circle (0.1cm); 
\draw[blue, decorate, decoration={snake, segment length=8pt}](-6,0)--(-3,0);
\draw (-6,0) node[above] {$(p,1)$};
\draw (-4.5,-0.2) node[below] {D-string};
\draw (-1.5,0) node[above] {O5};
\draw (-6,-1.1) node[right] {Allowed for odd $p$};
\draw (-6,-1.6) node[right] {Not allowed for even $p$};
\draw[dashed](1,0)--(7,0);
\draw (5,1)--(5,0);
\draw (5.1,1)--(5.1,0);
\fill (2,0) circle (0.1cm); 
\draw[blue, decorate, decoration={snake, segment length=8pt}](2,0)--(5,0);
\draw (2,0) node[above] {$(p,1)$};
\draw (3.5,-0.2) node[below] {D-string};
\draw (-1.5,0) node[above] {O5};
\draw (2,-1.1) node[right] {Allowed for even $p$};
\draw (2,-1.6) node[right] {Not allowed for odd $p$};
\end{tikzpicture}
\caption{Left: A D-string suspended between $(p,1)$ 7-brane and the intersection of $(1,1)$ 5-brane, $(1,-1)$ 5-brane, and an O5$^-$-plane. Such a D-string is allowed for odd $p$ and not allowed for even $p$. Right: A D-string suspended between $(p,1)$ 7-brane and the intersection of two coincident NS5-branes. Such a D-string is allowed for even $p$ and not allowed for odd $p$.}
\label{fig:conjecture}
\end{figure}
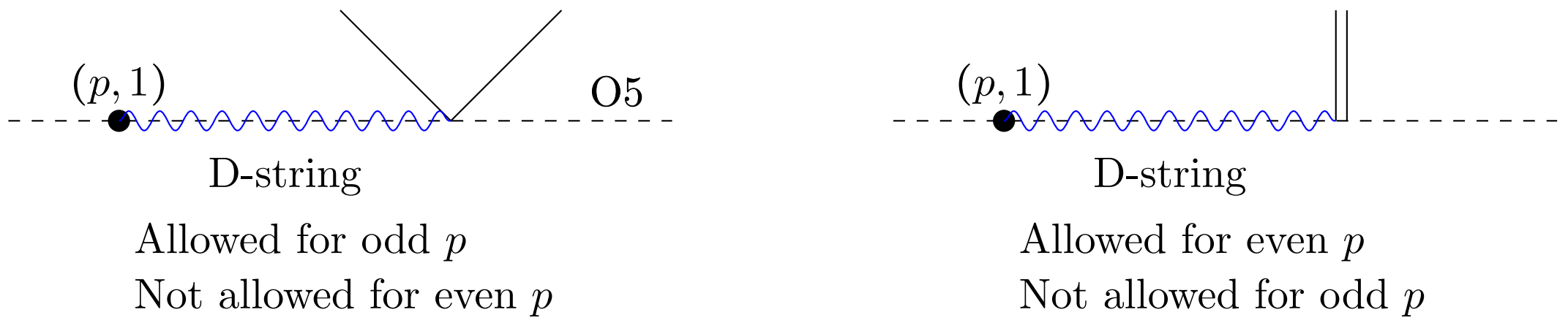

In this process, we encounter a problem whether a D-string on an O5$^-$-plane can be suspended between a $(p,1)$ 7-brane and a certain configuration of 5-branes. 
Without the O5$^-$-plane, the D-string can be attached only to (0,1) 7-branes and NS5-branes. It cannot be attached to 7-branes or 5-branes with $(p,q)$-charge other than (0,1). However, the situation would be different if the D-string is on the O5$^{-}$-plane.
We propose a conjecture that the D-string suspended between $(p,1)$ 7-brane and the 5-brane configuration depicted on
the left of Figure \ref{fig:conjecture} is allowed for odd $p$ and is not allowed for even $p$. On the contrary, we also proposed that the D-string suspended between $(p,1)$ 7-brane and the two coincident NS5-branes configuration depicted on the right of Figure \ref{fig:conjecture} is allowed for even $p$ and is not allowed for odd $p$.

Based on this conjecture, we discuss the screening of the Wilson line operators for the case of Spin(4), Spin(6), and Spin(8). We also provide consistency checks for the conjecture mentioned above by using the isomorphism $\mathrm{Spin}(4) = \mathrm{SU}(2) \times \mathrm{SU}(2)$, $\mathrm{Spin}(6) = \mathrm{SU}(4)$ and by comparing the with the analysis with the corresponding 5-brane web diagrams without an O5-plane. We also use triality of $\mathrm{Spin}(8)$ group for the consistency check.

\subsubsection{\texorpdfstring{$\mathrm{Spin}(4)+1\mathbf{S}$}{Spin4+1S}}
We start with the Spin(4) gauge theory with a spinor hypermultiplet, which is equivalent to $\SU(2)_1\times \SU(2)_2$ gauge theory with one fundamental hypermultiplet charged under $\SU(2)_2$. The lower indices 1,2 are put to distinguish two SU(2) groups. 
We use $b_1, b_2$ for Coulomb branch parameters of Spin(4) gauge theory, while $a^{(1)}$ and $a^{(2)}$ for $\SU(2)_1$ and $\SU(2)_2$, respectively. These are related by
\begin{align}\label{eq:map-Spin4-SU22}
a^{(1)} = \frac12( b_1 + b_2) \cr
a^{(2)} = \frac12( b_1 - b_2) 
\end{align}

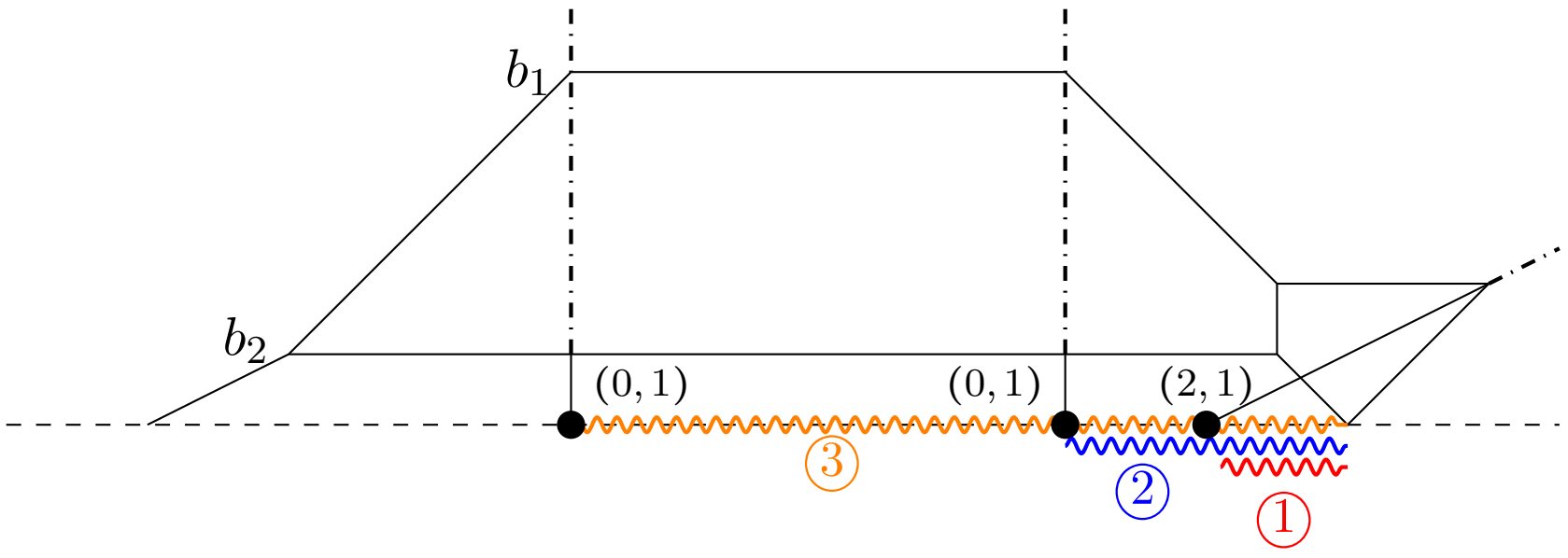
\begin{figure}[H]
\centering
\begin{tikzpicture}
\draw[dashed] (-6,0)--(5,0);
\draw(-5,0)--(-4,0.5)--(-2,2.5);
\draw(-2,2.5)--(1.5,2.5);
\draw (-2,2.5) node[left] {$b_1$};
\draw(-4,0.5)--(3,0.5);
\draw (-4,0.6) node[left] {$b_2$};
\draw(3,1)--(4.5,1);
\draw (1.5,2.5)--(3,1)--(3,0.5)--(3.5,0)--(4.5,1);
\draw[orange, thick, decorate, decoration={snake, segment length=4pt, amplitude=1.5pt}](-2,0)--(3.5,0);
\draw[blue, thick, decorate, decoration={snake, segment length=4pt, amplitude=1.5pt}](1.5,-0.15)--(3.5,-0.15);
\draw[red, thick, decorate, decoration={snake, segment length=4pt, amplitude=1.5pt}](2.6,-0.3)--(3.5,-0.3);
\fill (-2,0) circle (0.1); 
\draw (-2,0)--(-2,0.5);
\draw[dash dot, thick] (-2,0.5)--(-2,3);
\fill (1.5,0) circle (0.1); 
\draw (1.5,0)--(1.5,0.5);
\draw[dash dot, thick] (1.5,0.5)--(1.5,3);
\fill (2.5,0) circle (0.1); 
\draw(2.5,0)--(4.5,1);
\draw[dash dot, thick] (4.5,1)--(5,1.25);

\draw[red] (2.7,-0.7) node[right] {\textcircled{\small 1}};
\draw[blue] (1.7,-0.5) node[right] {\textcircled{\small 2}};
\draw[orange] (-0.5,-0.3) node[right] {\textcircled{\small 3}};

\draw (-1.5,0) node[above] {\scriptsize $(0,1)$};
\draw (1,0) node[above] {\scriptsize $(0,1)$};
\draw (2.5,0) node[above] {\scriptsize $(2,1)$};

\end{tikzpicture}
\caption{5-brane web diagram for Spin(4) + $1\mathbf{S}$ in the parameter region $b_2>0$. The relevant 7-branes are indicated by solid dots, with their associated monodromy cuts oriented to present their charges. 
 Three candidate D-strings are depicted in wiggly lines; One is a D-string attached to $(2,1)$ 7-brane, labelled \textcircled{\small 1}. The other is a D-string attached to the right $(0,1)$ 7-brane, labelled \textcircled{\small 2}. The last one is a D-string attached to the left $(0,1)$ 7-brane, labelled \textcircled{\small 3}. }
\label{fig:Spin4-1C-plus}
\end{figure}

\begin{figure}[H]
\centering
\begin{tikzpicture}
\draw[dashed] (-7.5,0)--(4,0);
\draw(-7,0)--(-6,0.5)--(-3,3.5);
\draw(-3,3.5)--(0.5,3.5);
\draw (-3,3.5) node[left] {$b_1$};
\draw(-6,0.5)--(3,0.5);
\draw (-6,0.6) node[left] {$- b_2$};
\draw(3,1)--(3.5,1);
\draw (0.5,3.5)--(3,1)--(3,0);
\draw (3,0.5)--(3.5,1);
\draw[magenta, thick, decorate, decoration={snake, segment length=4pt, amplitude=1.5pt}](1.5,-0.3)--(3,-0.3);
\draw[magenta] (2,-0.7) node[right] {\textcircled{\small 4}};
\draw[cyan, thick, decorate, decoration={snake, segment length=4pt, amplitude=1.5pt}](0.5,-0.15)--(3,-0.15);
\draw[cyan] (0.7,-0.5) node[right] {\textcircled{\small 5}};
\draw[brown, thick, decorate, decoration={snake, segment length=4pt, amplitude=1.5pt}](-3,0)--(3,0);
\draw[brown] (-1.5,-0.3) node[right] {\textcircled{\small 6}};
\fill (-3,0) circle (0.1); 
\draw (-3,0)--(-3,0.5);
\draw[dash dot, thick]  (-3,0.5)--(-3,4);
\draw (-2.5,0) node[above] {\scriptsize $(0,1)$};
\fill (0.5,0) circle (0.1); 
\draw (0.5,0)--(0.5,0.5);
\draw[dash dot, thick] (0.5,0.5)--(0.5,4);
\draw (0,0) node[above] {\scriptsize $(0,1)$};
\fill (1.5,0) circle (0.1); 
\draw(1.5,0)--(3,0.75);
\draw[dash dot, thick] (3,0.75)--(4,1.25);
\draw (1.3,0) node[above] {\scriptsize $(2,1)$};
\end{tikzpicture}
\caption{5-brane web diagram for Spin(4) + $1\mathbf{S}$ in the parameter region $b_2<0$. Three candidate D-strings are depicted; One is a D-string attached to (2,1) 7-brane, labelled \textcircled{\small 4}. The other is a D-string attached to the right (0,1) 7-brane, labelled \textcircled{\small 5}. The last one is a D-string attached to the left (0,1) 7-brane, labelled \textcircled{\small 6}.}
\label{fig:Spin4-1C-minus}
\end{figure}
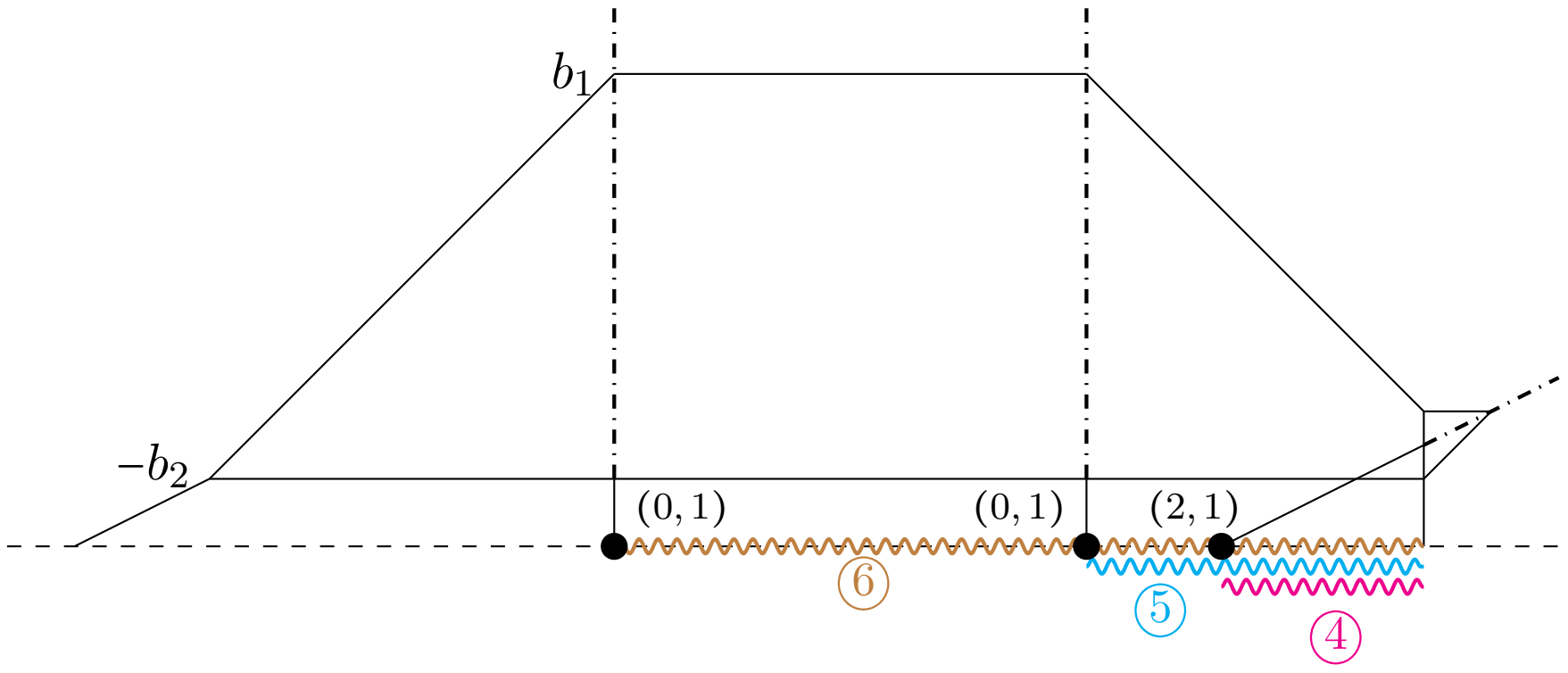

The 5-brane web diagram realizing this gauge theory is depicted in Figures \ref{fig:Spin4-1C-plus} and \ref{fig:Spin4-1C-minus} \cite{Zafrir:2015ftn}. The first figure corresponds to the parameter region $b_2>0$ and the second one corresponds to the parameter region $b_2<0$.

In order to consider the string configurations used for screening the Wilson line operators in the spinor representation and conjugate spinor representation, we consider six candidates of D-string configurations 
\textcircled{\small 1} -- \textcircled{\small 6}, 
which are given in Figures \ref{fig:Spin4-1C-plus} and \ref{fig:Spin4-1C-minus} \cite{Zafrir:2015ftn}. Among these, \textcircled{\small 4}, \textcircled{\small 5}, \textcircled{\small 6} are allowed configurations while \textcircled{\small 1}, \textcircled{\small 2}, \textcircled{\small 3} are not, according to our conjecture in Figure \ref{fig:conjecture}.
Computing the lengths of all these D-strings in terms of gauge theory parameters, they are given by
\begin{align}
\textrm{\textcircled{\small 1}}:& \,\, 
\frac12 (b_1+b_2)-m = a^{(1)}-m
\quad \text{(Not allowed)}
\cr
\textrm{\textcircled{\small 2}}:& \,\, 
\frac12 (b_1+b_2)+m = a^{(1)}+m
\quad \text{(Not allowed)}
\cr
\textrm{\textcircled{\small 3}}:& \,\, 
\frac12 (b_1+b_2)+m_0 +\frac12 m = a^{(1)}+m_0+\frac12 m
\quad \text{(Not allowed)}
\cr
\textrm{\textcircled{\small 4}}:& \,\, 
\frac12 (b_1-b_2)-m = a^{(2)}-m
\quad \fbox{\text{Allowed}}
\cr
\textrm{\textcircled{\small 5}}:& \,\, 
\frac12 (b_1-b_2)+m = a^{(2)}+m
\quad \fbox{\text{Allowed}}
\cr
\textrm{\textcircled{\small 6}}:& \,\, 
\frac12 (b_1-b_2)+m_0 +\frac12 m = a^{(2)}+m_0+\frac12 m
\quad \fbox{\text{Allowed}}
\end{align}
where we use the map \eqref{eq:map-Spin4-SU22}. 
These are obtained based on the parametrization discussed in Appendix \ref{app:parameterisation}. 

Strings with a length identical to the allowed configuration can also be found in the 5-brane web diagram with $\mathrm{SU}(2)_2$ gauge theory with one flavour as
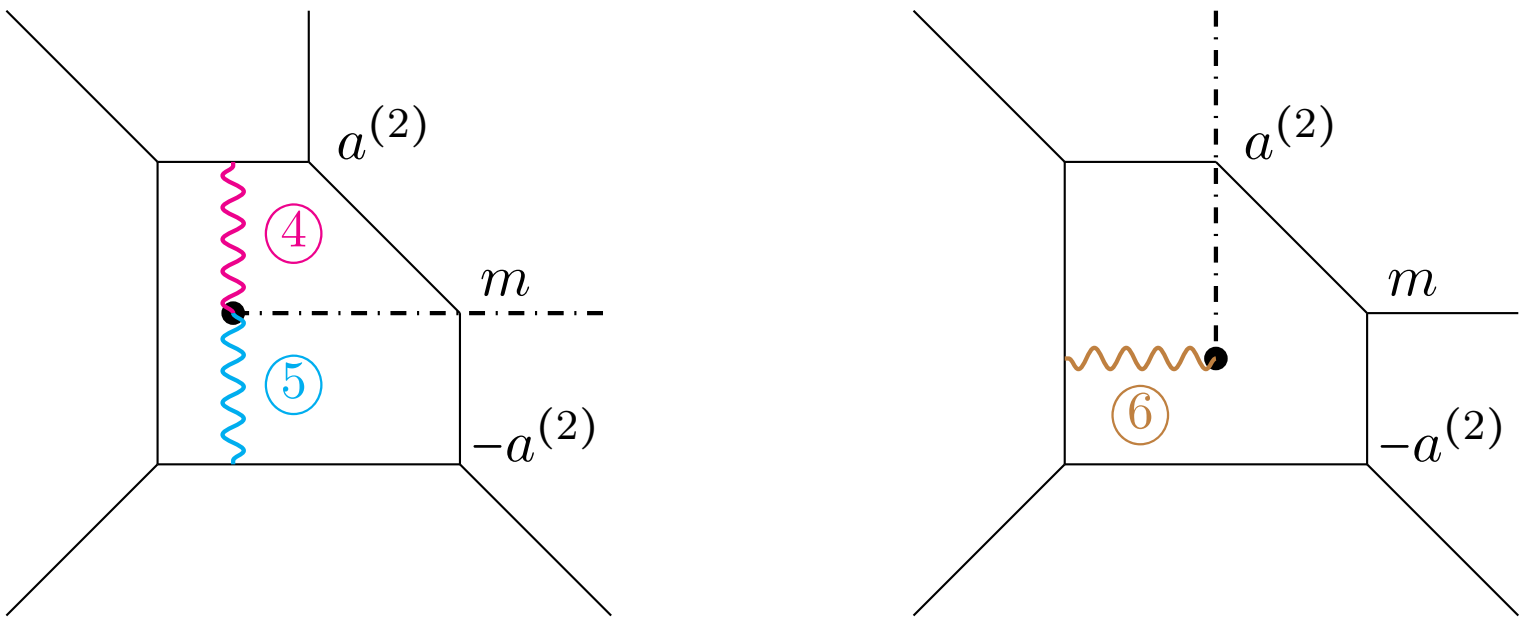
\begin{equation}\label{eq:web-SU2-2}
\centering
    \begin{tikzpicture}
    \begin{scope}[shift={(-3,0)}]
      \draw(-6,2)--(-4,2);
        \draw(-6,4)--(-5,4);
        \draw(-6,4)--(-6,2);
        \draw(-4,2)--(-4,3);
        \draw(-4,3)--(-5,4);
        \draw(-7,1)--(-6,2);
        \draw(-7,5)--(-6,4);
        \draw(-3,1)--(-4,2);
        \draw(-5,5)--(-5,4);
        \node[7brane] at (-5.5,3){};
        \draw[dash dot, thick](-5.5,3)--(-3,3);
        \draw[magenta, thick, decorate, decoration={snake, segment length=6pt, amplitude=2}](-5.5,3)--(-5.5,4);
        \draw[cyan, thick, decorate, decoration={snake, segment length=6pt, amplitude=2}](-5.5,3)--(-5.5,2);
        \node at (-5.1,3.5){\color{magenta}\textcircled{\small 4}};
        \node at (-5.1,2.5){\color{cyan}\textcircled{\small 5}};
        \node at (-4.5,4.2){$a^{(2)}$};
        \node at (-3.5,2.2){$-a^{(2)}$};
        \node at (-3.7,3.2){$m$};
    \end{scope}
    \begin{scope}[shift={(3,0)}]
      \draw(-6,2)--(-4,2);
        \draw(-6,4)--(-5,4);
        \draw(-6,4)--(-6,2);
        \draw(-4,2)--(-4,3);
        \draw(-4,3)--(-5,4);
        \draw(-7,1)--(-6,2);
        \draw(-7,5)--(-6,4);
        \draw(-3,1)--(-4,2);
        \draw(-4,3)--(-3,3);
        \node[7brane] at (-5,2.7){};
        \draw[dash dot, thick](-5,5)--(-5,2.7);
        \draw[brown, thick, decorate, decoration={snake, segment length=6pt, amplitude=2}](-5,2.7)--(-6,2.7);
        \node at (-5.5,2.3){\color{brown}\textcircled{\small 6}};
        \node at (-4.5,4.2){$a^{(2)}$};
        \node at (-3.5,2.2){$-a^{(2)}$};
        \node at (-3.7,3.2){$m$};
    \end{scope}
    \end{tikzpicture}
\end{equation}

On the contrary, strings with length identical to the configuration that are not allowed cannot be found in neither of the 5-brane web diagrams for $\mathrm{SU}(2)_1$ gauge theory nor $\mathrm{SU}(2)_2$ gauge theory. Since the length of \textcircled{\small 1}, \textcircled{\small 2}, \textcircled{\small 3} depend on $a^{(1)}$, the corresponding configuration should be found in the 5-brane web diagram for $\mathrm{SU}(2)_1$ gauge theory if they exist. However, the 5-brane web diagram for $\mathrm{SU}(2)_1$ gauge theory is given simply by 
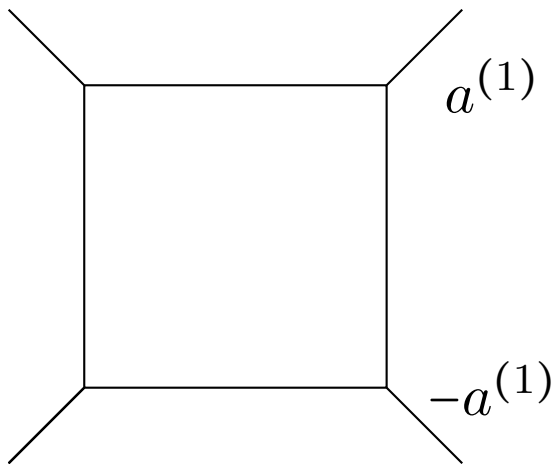
\begin{equation}
\centering
    \begin{tikzpicture}
    \draw(-1,-1)--(-1,1) -- (1,1) -- (1,-1) -- (-1,-1);
    \draw(-1,-1) -- (-1.5,-1.5);
    \draw(-1,1) -- (-1.5,1.5);
    \draw(-1,-1) -- (-1.5,-1.5);
    \draw(-1,-1) -- (-1.5,-1.5);
    \draw(1,1) -- (1.5,1.5);
    \draw(1,-1) -- (1.5,-1.5);
    \node at (1.7,1){$a^{(1)}$};
    \node at (1.7,-1){$-a^{(1)}$};
    \end{tikzpicture}       
\end{equation}
and obviously, such strings cannot be constructed in this 5-brane web. These observations provide non-trivial consistency checks for our conjecture.

Having these D-string configuration at hand, we discuss the screening of the Wilson lines. We consider the half-D3-brane configuration corresponding to $\alpha$ and $\beta$ that are discussed in Figure \ref{fig:Z2} and Figure \ref{fig:Z2-beta}. Considering the 5-brane web in the parameter region $b_2 < 0$ in Figure \ref{fig:Spin4-1C-minus}, we find that $\alpha$ can be screened by any of the allowed configurations of the D-strings \textcircled{\small 4}, \textcircled{\small 5}, \textcircled{\small 6} while $\beta$ cannot as shown below:
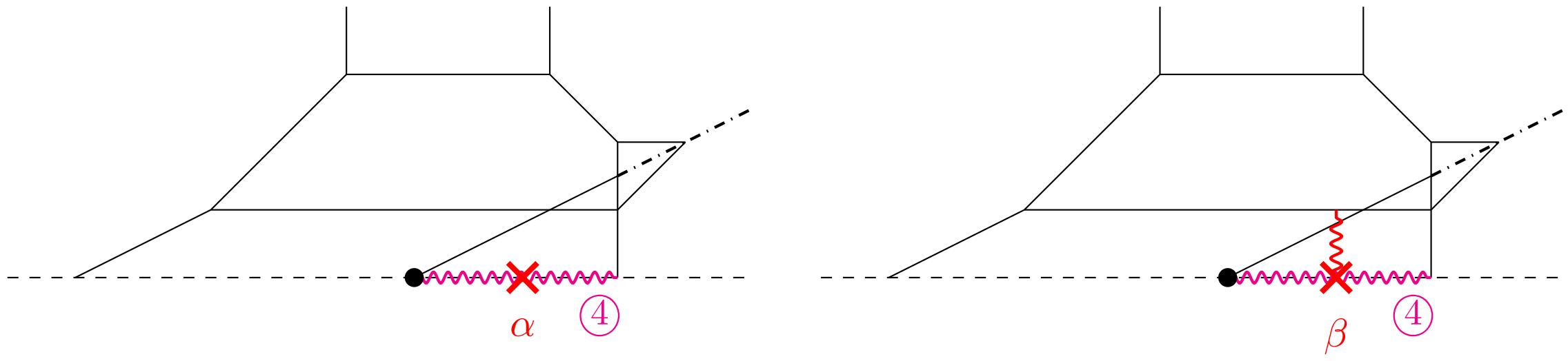
\begin{equation}
\centering
\begin{tikzpicture}
\begin{scope}[scale=1.3, shift={(-3,0)}]
\draw[dashed] (-1.5,0)--(4,0);
\draw(-1,0)--(0,0.5)--(,1.5)--(1,2);
\draw(1,1.5)--(2.5,1.5);
\draw(0,0.5)--(3,0.5);
\draw(3,1)--(3.5,1);
\draw (2.5,2)--(2.5,1.5)--(3,1)--(3,0);
\draw (3,0.5)--(3.5,1);
\draw[magenta, thick, decorate, decoration={snake, segment length=4pt, amplitude=1.5pt}](1.5,0)--(3,0);
\fill (1.5,0) circle (0.07); 
\draw(1.5,0)--(3,0.75);
\draw[dash dot, thick] (3,0.75)--(4,1.25);
\draw[magenta] (2.6,-0.3) node[right] {\textcircled{\small 4}};
\node[cross, red, line width=1.5pt,scale=2]at(2.3,0){};
\draw[red] (2.3,-0.2) node[below] {$\alpha$};
\end{scope}
\begin{scope}[scale=1.3, shift={(3,0)}]
\draw[dashed] (-1.5,0)--(4,0);
\draw(-1,0)--(0,0.5)--(,1.5)--(1,2);
\draw(1,1.5)--(2.5,1.5);
\draw(0,0.5)--(3,0.5);
\draw(3,1)--(3.5,1);
\draw (2.5,2)--(2.5,1.5)--(3,1)--(3,0);
\draw (3,0.5)--(3.5,1);
\draw[magenta, thick, decorate, decoration={snake, segment length=4pt, amplitude=1.5pt}](1.5,0)--(3,0);
\fill (1.5,0) circle (0.07); 
\draw(1.5,0)--(3,0.75);
\draw[dash dot, thick] (3,0.75)--(4,1.25);
\draw[magenta] (2.6,-0.3) node[right] {\textcircled{\small 4}};
\node[cross, red ,line width=1.5pt,scale=2]at(2.3,0){};
\draw[red] (2.3,-0.2) node[below] {$\beta$};
\draw [red, thick, decorate, decoration={snake, segment length=4pt, amplitude=1.5pt}] (2.3,0)--(2.3,0.5);
\end{scope}
\end{tikzpicture}
\end{equation}
where we draw only \textcircled{\small 4} as a representative for simplicity. 
We interpret that $\alpha$ corresponds to the Wilson line in the spinor representation and thus, can be screened due to the spinor matter in the theory. Thus, the 1-form symmetry generated by $\alpha$ reduces to trivial symmetry by introducing the spinor to the $\mathrm{Spin}(4)$ gauge theory. Meanwhile, $\beta$ is interpreted to correspond to the Wilson line in the conjugate spinor representation and thus, cannot be screened. Thus, the 1-form symmetry generated by $\beta$ still remains.

This situation can be understood in the equivalent $\mathrm{SU}(2)_1 \times \mathrm{SU}(2)_2$ description as follows. The Wilson line corresponding to $\alpha$ is in the fundamental representation of the $\mathrm{SU}(2)_2$ gauge theory and can be screened due to the flavour coupled to the $\mathrm{SU}(2)_2$. The Wilson line corresponding to $\beta$ is in the fundamental representation of the pure $\mathrm{SU}(2)_1$ gauge theory and, thus cannot be screened. 

This can be understood again also from the viewpoint of the 5-brane web as below:
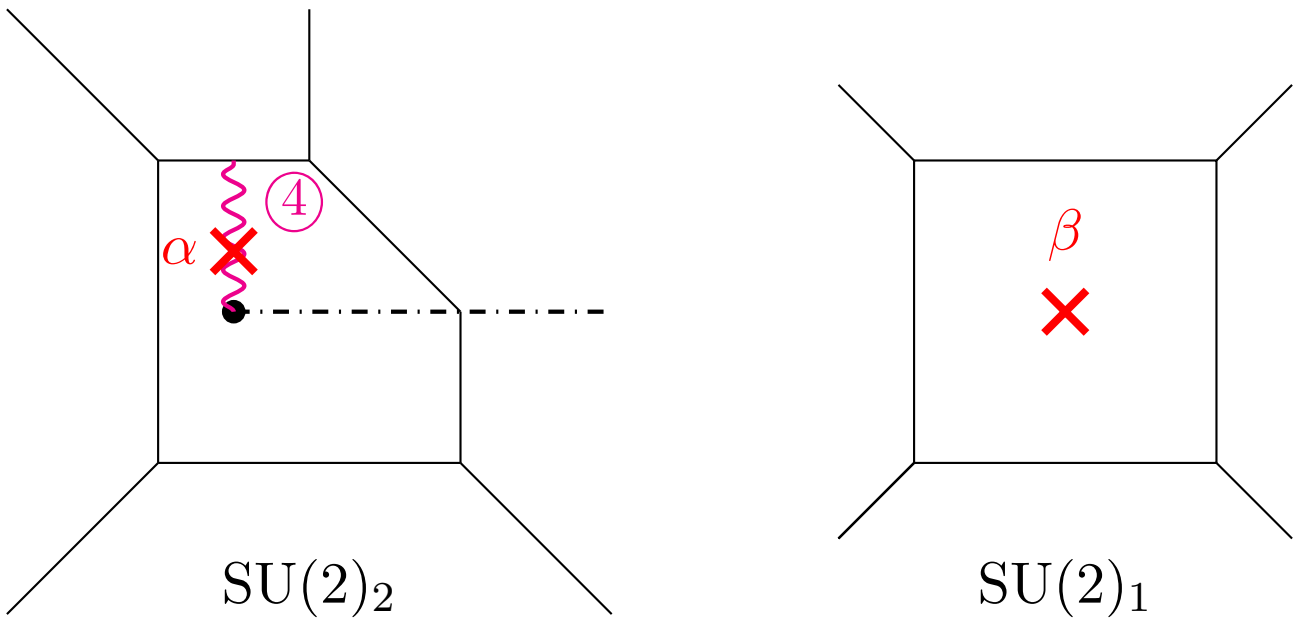
\begin{equation}
\centering
    \begin{tikzpicture}   
    \begin{scope}[shift={(0,-3)}]
      \draw(-6,2)--(-4,2);
        \draw(-6,4)--(-5,4);
        \draw(-6,4)--(-6,2);
        \draw(-4,2)--(-4,3);
        \draw(-4,3)--(-5,4);
        \draw(-7,1)--(-6,2);
        \draw(-7,5)--(-6,4);
        \draw(-3,1)--(-4,2);
        \draw(-5,5)--(-5,4);
        \node[7brane] at (-5.5,3){};
        \draw[dash dot, thick](-5.5,3)--(-3,3);
        \draw[magenta, thick, decorate, decoration={snake, segment length=6pt, amplitude=2}](-5.5,3)--(-5.5,4);
        \node at (-5.1,3.7){\color{magenta}\textcircled{\small 4}};
        \node[cross, red,line width=1.5pt,scale=2]at(-5.5,3.4){};
        \draw[red] (-5.6,3.4) node[left] {$\alpha$};
        \draw (-5,1.5) node[below] {SU(2)$_2$};
    \end{scope}
    \begin{scope}[shift={(0,0)}]
        \draw(-1,-1)--(-1,1) -- (1,1) -- (1,-1) -- (-1,-1);
        \draw(-1,-1) -- (-1.5,-1.5);
        \draw(-1,1) -- (-1.5,1.5);
        \draw(-1,-1) -- (-1.5,-1.5);
        \draw(-1,-1) -- (-1.5,-1.5);
        \draw(1,1) -- (1.5,1.5);
        \draw(1,-1) -- (1.5,-1.5);
        \node[cross, red,line width=1.5pt,scale=2]at(0,0){};
        \draw[red] (0,0.2) node[above] {$\beta$};
        \draw (0,-1.5) node[below] {SU(2)$_1$};
    \end{scope} 
    \end{tikzpicture}
\end{equation}
Here, $\alpha$ corresponds to the D3-brane inside the 5-brane web for $\mathrm{SU}(2)_2$ and can be screened by any of the string \textcircled{\small 4}, \textcircled{\small 5}, \textcircled{\small 6}, where we draw only \textcircled{\small 4} for simplicity. Meanwhile, $\beta$ corresponds to the D3-brane inside the 5-brane web for $\mathrm{SU}(2)_1$ and thus, cannot be screened.

In this way, we find the 1-form symmetry $\mathbb{Z}_2 \times \mathbb{Z}_2$ of the Spin(4) gauge theory reduces to $\mathbb{Z}_2$ when we introduce the spinor matter, from the viewpoint of the 5-brane web.

\subsubsection{\texorpdfstring{$\mathrm{Spin}(6)+1\mathbf{S}$}{Spin6+1S}}

We go on to the Spin(6) gauge theory with one hypermultiplet in the spinor representation with mass $m$. This theory is equivalent to the SU(4)$_\frac12$ gauge theory with one anti-fundamental hypermultiplet with mass $m$, or equivalently, one fundamental hypermultiplet with mass $-m$. We use $b_1, b_2, b_3$ for Coulomb branch parameters of Spin(6) gauge theory, while $a_1, a_2, a_3, a_4$ for SU(4) gauge theory, respectively. These are related by
\begin{align}\label{eq:map-Spin6-SU4}
a_1 &= \frac12 (b_1+b_2+b_3),
\cr
a_2 &= \frac12 (b_1-b_2-b_3),
\cr
a_3 &= \frac12 (-b_1+b_2-b_3),
\cr
a_4 &= \frac12 (-b_1-b_2+b_3).
\end{align}

We choose the Weyl chamber where 
\begin{align}
b_1 > b_2 > |b_3|.
\end{align}
When $b_3>0$, 5-brane web diagram is given as in Figure \ref{fig:SO6+1C-b3p}. 

\begin{figure}[H]
\centering 
\begin{tikzpicture}
\draw[dashed] (-6,0)--(5,0);
\draw(-5,0)--(-4,0.5)--(-3,1.5)--(-3,2);
\draw(-3,2)--(2.5,2);
\draw (-3.1,2) node[left] {$b_1$};
\draw(-3,1.5)--(2.5,1.5);
\draw (-3.1,1.5) node[left] {$b_2$};
\draw(-4,0.5)--(3,0.5);
\draw (-4,0.6) node[left] {$b_3$};
\draw(3,1)--(4.5,1);
\draw (2.5,2)--(2.5,1.5)--(3,1)--(3,0.5)--(3.5,0)--(4.5,1);
\draw[orange, thick, decorate, decoration={snake, segment length=4pt, amplitude=1.5pt}](-1,0)--(3.5,0);
\draw[blue, thick, decorate, decoration={snake, segment length=4pt, amplitude=1.5pt}](0.5,-0.15)--(3.5,-0.15);
\draw[red, thick, decorate, decoration={snake, segment length=4pt, amplitude=1.5pt}](2.6,-0.3)--(3.5,-0.3);
\fill (-1,0) circle (0.1); 
\draw (-1,0)--(-2.5,1.5);
\draw[dash dot, thick] (-2.5,1.5)--(-3.5,2.5);
\fill (0.5,0) circle (0.1); 
\draw (0.5,0)--(2,1.5);
\draw[dash dot, thick] (2,1.5)--(3,2.5);
\fill (2.5,0) circle (0.1); 
\draw(2.5,0)--(4.5,1);
\draw[dash dot, thick] (4.5,1)--(5,1.25);

\draw[red] (2.7,-0.7) node[right] {\textcircled{\small 1}};
\draw[blue] (1,-0.5) node[right] {\textcircled{\small 2}};
\draw[orange] (-0.5,-0.35) node[right] {\textcircled{\small 3}};

\draw (-0.8,0) node[above] {\scriptsize $(-1,1)$};
\draw (0.4,0) node[above] {\scriptsize $(1,1)$};
\draw (2.5,0) node[above] {\scriptsize $(2,1)$};

\end{tikzpicture}
\caption{5-brane web diagram for Spin(6) + $1\mathbf{S}$ in the parameter region $b_3>0$. Three candidate D-strings are depicted; One is a D-string attached to $(2,1)$ 7-brane, labelled \textcircled{\small 1}. The other is a D-string attached to the $(1,1)$ 7-brane, labelled \textcircled{\small 2}. The last one is a D-string attached to the $(-1,1)$ 7-brane, labelled \textcircled{\small 3}. }
\label{fig:SO6+1C-b3p}
\end{figure}

When $b_3<0$, 5-brane web diagram is given as shown in Figure \ref{fig:SO6+1C-b3m}. 
If we replace $b_3 \to - b_3$, this can also be interpreted as a 
Spin(6) gauge theory with one spinor representation in the $b_3>0$ region.

\begin{figure}[H]
\centering 
\begin{tikzpicture}
\draw[dashed] (-6.5,0)--(5,0);
\draw(-5,0)--(-4,0.5)--(-3,1.5)--(-3,2);
\draw(-3,2)--(2.5,2);
\draw (-3.1,2) node[left] {$b_1$};
\draw(-3,1.5)--(2.5,1.5);
\draw (-3.1,1.5) node[left] {$b_2$};
\draw(-4,0.5)--(3,0.5);
\draw (-4,0.6) node[left] {$-b_3$};
\draw(3,1)--(3.5,1);
\draw (2.5,2)--(2.5,1.5)--(3,1)--(3,0);
\draw (3,0.5)--(3.5,1);
\draw[magenta, thick, decorate, decoration={snake, segment length=4pt, amplitude=1.5pt}](1.5,-0.3)--(3,-0.3);
\draw[magenta] (2,-0.7) node[right] {\textcircled{\small 4}};
\draw[cyan, thick, decorate, decoration={snake, segment length=4pt, amplitude=1.5pt}](0.5,-0.15)--(3,-0.15);
\draw[cyan] (0.7,-0.5) node[right] {\textcircled{\small 5}};
\draw[brown, thick, decorate, decoration={snake, segment length=4pt, amplitude=1.5pt}](-1,0)--(3,0);
\draw[brown] (-0.5,-0.35) node[right] {\textcircled{\small 6}};
\fill (-1,0) circle (0.1); 
\draw (-1,0)--(-2.5,1.5);
\draw[dash dot, thick] (-2.5,1.5)--(-3.5,2.5);
\draw (-0.8,0) node[above] {\scriptsize $(-1,1)$};
\fill (0.5,0) circle (0.1); 
\draw (0.5,0)--(2,1.5);
\draw[dash dot, thick] (2,1.5)--(3,2.5);
\draw (0.4,0) node[above] {\scriptsize $(1,1)$};
\fill (1.5,0) circle (0.1); 
\draw(1.5,0)--(3,0.75);
\draw[dash dot, thick] (3,0.75)--(4,1.25);
\draw (1.5,0) node[above] {\scriptsize $(2,1)$};
\end{tikzpicture}
\caption{5-brane web diagram for Spin(6) + $1\mathbf{S}$ in the parameter region $b_3<0$. Three candidate D-strings are depicted; One is a D-string attached to $(2,1)$ 7-brane, labelled \textcircled{\small 4}. The other is a D-string attached to the $(1,1)$ 7-brane, labelled \textcircled{\small 5}. The last one is a D-string attached to the $(-1,1)$ 7-brane, labelled \textcircled{\small 6}. }
\label{fig:SO6+1C-b3m}
\end{figure}

In order to consider the string configurations used for screening the Wilson line operators in the spinor representation and the conjugate spinor representation, we again consider six candidates of D-string configurations 
\textcircled{\small 1} -- \textcircled{\small 6}, 
which are given in Figure \ref{fig:SO6+1C-b3p} and \ref{fig:SO6+1C-b3m}. This time, \textcircled{\small 2}, \textcircled{\small 3}, \textcircled{\small 4} among them are allowed configurations while \textcircled{\small 1}, \textcircled{\small 5}, \textcircled{\small 6} are not, according to our conjecture in Figure \ref{fig:conjecture}.
Computing the length of all these D-strings in terms of gauge theory parameters, they are given by
\begin{align}
\text{\textcircled{\small 1}}:& \,\, 
\frac12 (b_1+b_2+b_3)-m = a_1-m
\quad \text{(Not allowed)}
\cr
\text{\textcircled{\small 2}}:& \,\, 
\frac12 (b_1+b_2+b_3)+m = a_1+m
\quad \fbox{\text{Allowed}}
\cr
\text{\textcircled{\small 3}}:& \,\, 
\frac12 (b_1+b_2+b_3)+m_0 +\frac12 m = a_1+m_0+\frac12 m
\quad \fbox{\text{Allowed}}
\cr
\text{\textcircled{\small 4}}:& \,\, 
\frac12 (b_1+b_2-b_3)-m = -a_4-m
\quad \fbox{\text{Allowed}}
\cr
\text{\textcircled{\small 5}}:& \,\, 
\frac12 (b_1+b_2-b_3)+m = -a_4+m
\quad \text{(Not allowed)}
\cr
\text{\textcircled{\small 6}}:& \,\, 
\frac12 (b_1+b_2-b_3)+m_0 +\frac12 m = -a_4+m_0+\frac12 m
\quad \text{(Not allowed)}
\end{align}
where we use the map \eqref{eq:map-Spin6-SU4}.

Strings with length identical to the allowed configuration \textcircled{\small 2}, \textcircled{\small 3}, \textcircled{\small 4} can be found also in the 5-brane web diagram with $\mathrm{SU}(4)_{\frac12}$ gauge theory with one flavour as
\begin{figure}  [H]
\centering 
\begin{tikzpicture}
\draw (0,0) -- (0,0.5) -- (-0.5,1)--(-1.5,1.5)--(-2.5,2.5)--(-2.9,2.7);
\draw (3.5,2.5) -- (2,1) -- (2,0.5) -- (2.5,0) -- (2.9,-0.2);
\draw (-2.5,2.5) -- (3.5,2.5);
\draw (-2.5,2.4) node[left] {\small $a_1$};
\draw (-0.5,1) -- (2,1);
\draw (-0.5,1) node[left] {\small $a_2$};
\draw (0,0.5) -- (2,0.5);
\draw (0,0.5) node[left] {\small $a_3$};
\draw (0,0) -- (2.5,0);
\draw (0,0)node[left] {\small $a_4$};
\draw[blue, thick, decorate, decoration={snake, segment length=4pt, amplitude=1.5pt}](0.35,1.5)--(0.35,2.5);
\draw[magenta, thick, decorate, decoration={snake, segment length=4pt, amplitude=1.5pt}](0.35,1.5)--(0.35,0);
\draw[orange, thick, decorate, decoration={snake, segment length=4pt, amplitude=1.5pt}](1.65,1.6)--(1.65,2.5);
\draw [dash dot, thick](-2,1.5) -- (0.35,1.5);
\fill (0.35,1.5) circle (0.1); 
\draw (-2,1.5)node[left] {\small $-m$};
\draw [dash dot, thick] (-0.3,-0.3) -- (0.5,0.5);
\draw (0.5,0.5) -- (1.6,1.6);
\fill (1.6,1.6) circle (0.1); 
\draw [dash dot, thick] (1.5,1.5)--(3.9,2.7);
\fill (1.64,1.57) circle (0.1); 
\draw[blue] (0.4,2) node[left] {\textcircled{\small 2}};
\draw[orange] (1.6,2.1) node[right] {\textcircled{\small 3}};
\draw[magenta] (0.4,0.75) node[left] {\textcircled{\small 4}};
\draw [red](1.6,1.6) circle (0.3);
\draw[-latex,red,thick] (5.5,1.5) -- (2,1.6);
\begin{scope}[shift={(7,1.4)}]
\draw [red, thick](0,0) circle (1.5);
\draw[orange, thick, decorate, decoration={snake, segment length=4pt, amplitude=1.5pt}](0,0)--(0,1.4);
\draw[orange, thick, decorate, decoration={snake, segment length=4pt, amplitude=1.5pt}](0,0)--(-0.5,0.5);
\draw[orange, thick, decorate, decoration={snake, segment length=4pt, amplitude=1.5pt}](0,0)--(0.3,-0.6);
\fill (-0.5,0.5) circle (0.2);
\fill (0.3,-0.6) circle (0.2);
\draw [very thick](-0.5,0.5) -- (-1.4,-0.4);
\draw [dash dot, thick] (0.3,-0.6) -- (1.5,0);
\draw[orange] (0,0.5) node[right] {\textcircled{\small 3}};
\draw (-0.5,0.6) node[above] {\small $(1,1)$};
\draw (0.3,-0.7) node[below] {\small $(2,1)$};
\end{scope}
\end{tikzpicture}
\caption{5-brane web diagram for SU(4)$_\frac12$ gauge theory with one hypermultiplet in anti-fundamental representation. The figure inside the circle at the right side explains how the string junction including fundamental string can be terminated by a pair of the (1,1) 7-brane and the (2,1) 7-brane.}
\end{figure}
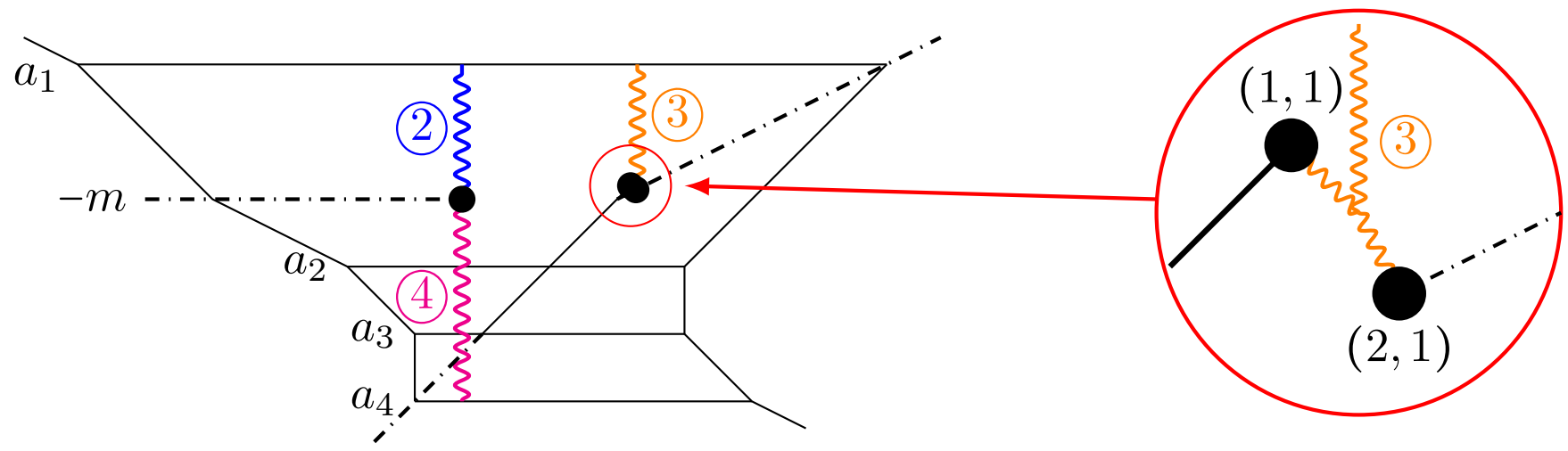
On the contrary, we cannot find strings with lengths identical to the configuration that is not allowed. These observations again provide non-trivial consistency checks for our conjecture.

Now, we discuss the screening of the Wilson lines using the 5-brane webs. In the 5-brane web in parameter region $b_3<0$, depicted in Figure \ref{fig:SO6+1C-b3m}, the allowed configuration is \textcircled{\small 4}. To be consistent with the previous convention, we associate the half D3-brane without a fundamental string as $\alpha$, while the half D3-brane with a fundamental string as $\beta$ in the parameter region $b_3<0$ as
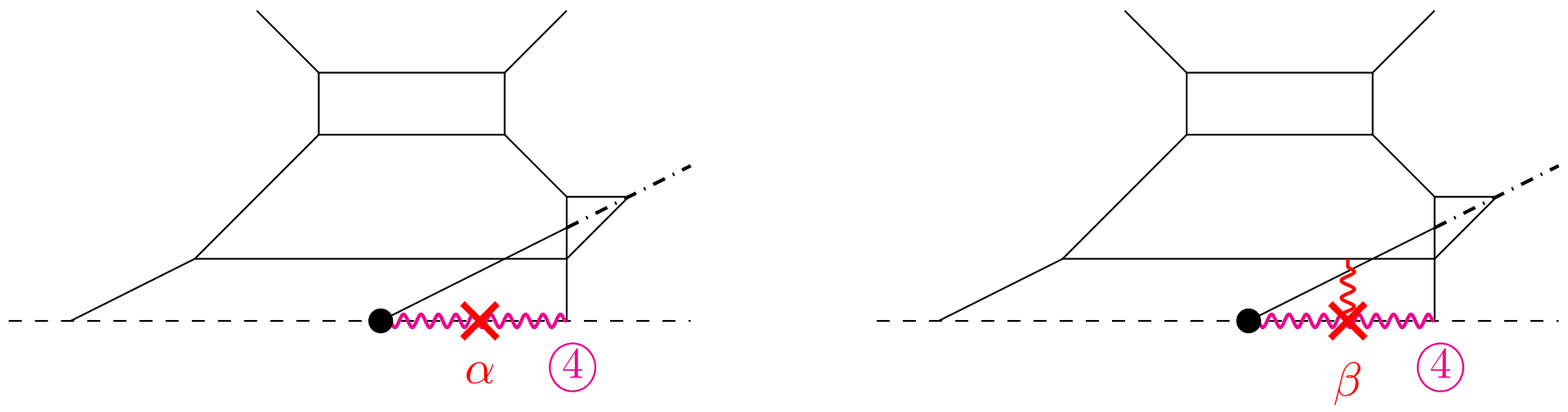
\begin{equation}
\centering 
\begin{tikzpicture}
\begin{scope}[shift={(-3.5,0)}]
\draw[dashed] (-1.5,0)--(4,0);
\draw(-1,0)--(0,0.5)--(1,1.5)--(1,2)--(0.5,2.5);
\draw(1,2)--(2.5,2);
\draw(1,1.5)--(2.5,1.5);
\draw(0,0.5)--(3,0.5);
\draw(3,1)--(3.5,1);
\draw (3,2.5)--(2.5,2)--(2.5,1.5)--(3,1)--(3,0);
\draw (3,0.5)--(3.5,1);
\draw[magenta, thick, decorate, decoration={snake, segment length=4pt, amplitude=1.5pt}](1.5,0)--(3,0);
\fill (1.5,0) circle (0.1); 
\draw(1.5,0)--(3,0.75);
\draw[dash dot, thick] (3,0.75)--(4,1.25);
\draw[magenta] (2.7,-0.4) node[right] {\textcircled{\small 4}};
\node[cross, red, line width=1.5pt,scale=2]at(2.3,0){};
\draw[red] (2.3,-0.2) node[below] {$\alpha$};
\end{scope}
\begin{scope}[shift={(3.5,0)}]
\draw[dashed] (-1.5,0)--(4,0);
\draw(-1,0)--(0,0.5)--(1,1.5)--(1,2)--(0.5,2.5);
\draw(1,2)--(2.5,2);
\draw(1,1.5)--(2.5,1.5);
\draw(0,0.5)--(3,0.5);
\draw(3,1)--(3.5,1);
\draw (3,2.5)--(2.5,2)--(2.5,1.5)--(3,1)--(3,0);
\draw (3,0.5)--(3.5,1);
\draw[magenta, thick, decorate, decoration={snake, segment length=4pt, amplitude=1.5pt}](1.5,0)--(3,0);
\fill (1.5,0) circle (0.1); 
\draw(1.5,0)--(3,0.75);
\draw[dash dot, thick] (3,0.75)--(4,1.25);
\draw[magenta] (2.7,-0.4) node[right] {\textcircled{\small 4}};
\node[cross, red, line width=1.5pt,scale=2]at(2.3,0){};
\draw[red] (2.3,-0.2) node[below] {$\beta$};
\draw[red, thick, decorate, decoration={snake, segment length=4pt, amplitude=1.5pt}] (2.3,0)--(2.3,0.5);
\end{scope}
\end{tikzpicture}
\end{equation}
Analogous to the previous case, the Wilson line corresponding to $\alpha$ can be screened by using this string \textcircled{\small 4}. Since $\beta = \alpha^3$ as discussed in \eqref{eq:beta=alpha3} or in Figure \ref{fig:a^3inZ4}, we can screen also $\beta$. This indicates that the $\mathbb{Z}_4$ 1-form symmetry, which is generated by $\alpha$, reduces to trivial. 

To see further consistency, it would still be interesting to consider the parameter region $b_3>0$ , where the allowed D-string is \textcircled{\small 2} and \textcircled{\small 3}, where we denote only \textcircled{\small 2} below for simplicity. 
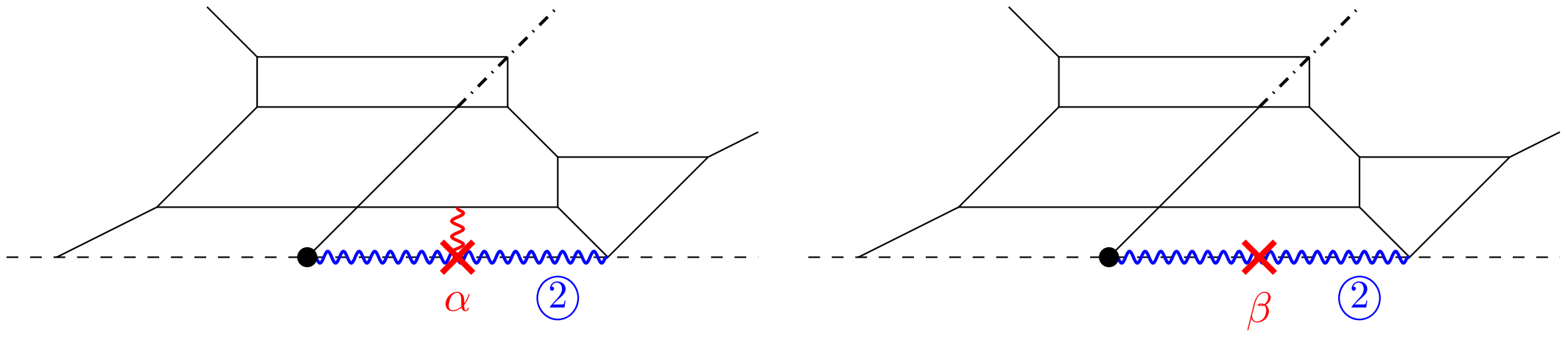
\begin{equation}
\centering 
\begin{tikzpicture}
\begin{scope}[scale=0.9, shift={(-4,0)}]
\draw[dashed] (-2.5,0)--(5,0);
\draw (-2,0)--(-1,0.5)--(0,1.5)--(0,2)--(-0.5,2.5);
\draw(0,2)--(2.5,2);
\draw(0,1.5)--(2.5,1.5);
\draw(-1,0.5)--(3,0.5);
\draw(3,1)--(4.5,1);
\draw (2.5,2)--(2.5,1.5)--(3,1)--(3,0.5)--(3.5,0)--(4.5,1);
\draw[blue, thick, decorate, decoration={snake, segment length=4pt, amplitude=1.5pt}](0.5,0)--(3.5,0);
\fill (0.5,0) circle (0.1); 
\draw (0.5,0)--(2,1.5);
\draw[dash dot, thick] (2,1.5)--(3,2.5);
\draw (4.5,1)--(5,1.25);
\draw[blue] (3,-0.1) node[below] {\textcircled{\small 2}};
\node[cross, red,line width=1.5pt,scale=2]at(2,0){};
\draw[red] (2,-0.2) node[below] {$\alpha$};
\draw[red, thick, decorate, decoration={snake, segment length=4pt, amplitude=1.5pt}] (2,0)--(2,0.5);
%
\end{scope}
\begin{scope}[scale=0.9, shift={(4,0)}]
\draw[dashed] (-2.5,0)--(5,0);
\draw (-2,0)--(-1,0.5)--(0,1.5)--(0,2)--(-0.5,2.5);
\draw(0,2)--(2.5,2);
\draw(0,1.5)--(2.5,1.5);
\draw(-1,0.5)--(3,0.5);
\draw(3,1)--(4.5,1);
\draw (2.5,2)--(2.5,1.5)--(3,1)--(3,0.5)--(3.5,0)--(4.5,1);
\draw[blue, thick, decorate, decoration={snake, segment length=4pt, amplitude=1.5pt}](0.5,0)--(3.5,0);
\fill (0.5,0) circle (0.1); 
\draw (0.5,0)--(2,1.5);
\draw[dash dot, thick] (2,1.5)--(3,2.5);
\draw (4.5,1)--(5,1.25);
\draw[blue] (3,-0.1) node[below] {\textcircled{\small 2}};
\node[cross, red,line width=1.5pt,scale=2]at(2,0){};
\draw[red] (2,-0.2) node[below] {$\beta$};
%
\end{scope}
\end{tikzpicture}
\end{equation}
We see from this figure that, with \textcircled{\small 2}, the Wilson line corresponding to the half D3-brane without fundamental string attached can be screened. However, we note that we associate such half D3-brane as $\beta$ instead of $\alpha$ in this parameter region $b_3>0$. Our claim is that, when we go from parameter region from $b_3<0$ to $b_3>0$, the string attached to $\beta$ is annihilated while new string is generated to $\alpha$, due to a variation of Hanany-Witten transition. Thus, our claim is that $\beta$ is screened by using \textcircled{\small 2}. 

This claim can be supported by comparing with the equivalent SU(4) gauge theory description. In this picture, $\alpha$ corresponds to the D3-brane inside the bottom room while $\beta$ corresponds to the D3-brane inside the top room, as depicted as follows:
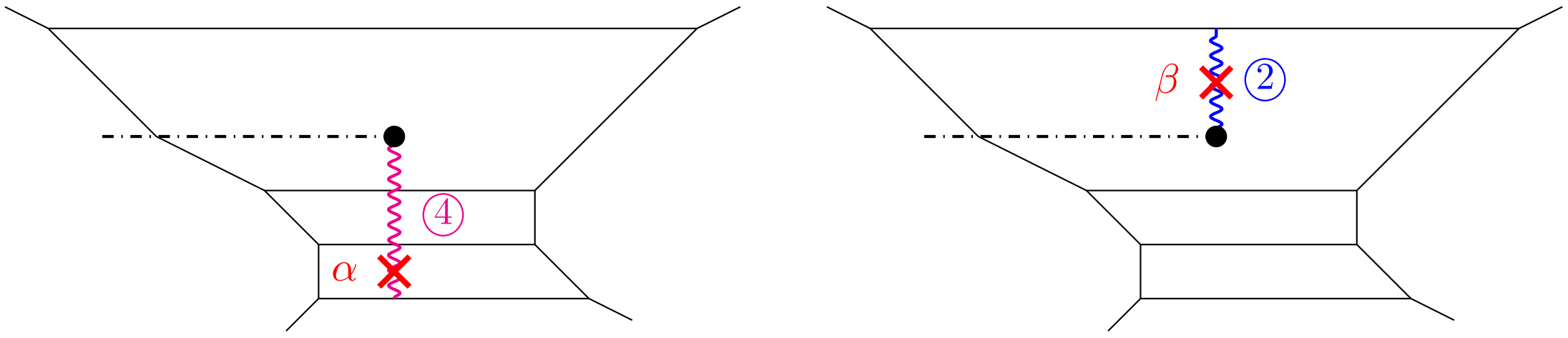
\begin{equation}
\centering 
\begin{tikzpicture}
\begin{scope}[shift={(-3.8,0)}]
\draw (-0.3,-0.3) -- (0,0) -- (0,0.5) -- (-0.5,1)--(-1.5,1.5)--(-2.5,2.5)--(-2.9,2.7);
\draw (3.9,2.7) -- (3.5,2.5) -- (2,1) -- (2,0.5) -- (2.5,0) -- (2.9,-0.2);
\draw (-2.5,2.5) -- (3.5,2.5);
\draw (-0.5,1) -- (2,1);
\draw (0,0.5) -- (2,0.5);
\draw (0,0) -- (2.5,0);
\draw[magenta, thick, decorate, decoration={snake, segment length=4pt, amplitude=1.5pt}](0.7,1.5)--(0.7,0);
\draw [dash dot, thick](-2,1.5) -- (0.7,1.5);
\fill (0.7,1.5) circle (0.1); 
\draw[magenta] (1.5,0.75) node[left] {\textcircled{\small 4}};
\node[cross, red,line width=1.5pt,scale=2]at(0.7,0.25){};
\draw[red] (0.5,0.25) node[left] {$\alpha$};
\end{scope}
\begin{scope}[shift={(3.8,0)}]
\draw (-0.3,-0.3) -- (0,0) -- (0,0.5) -- (-0.5,1)--(-1.5,1.5)--(-2.5,2.5)--(-2.9,2.7);
\draw (3.9,2.7) -- (3.5,2.5) -- (2,1) -- (2,0.5) -- (2.5,0) -- (2.9,-0.2);
\draw (-2.5,2.5) -- (3.5,2.5);
\draw (-0.5,1) -- (2,1);
\draw (0,0.5) -- (2,0.5);
\draw (0,0) -- (2.5,0);
\draw[blue, thick, decorate, decoration={snake, segment length=4pt, amplitude=1.5pt}](0.7,1.5)--(0.7,2.5);
\draw [dash dot, thick](-2,1.5) -- (0.7,1.5);
\fill (0.7,1.5) circle (0.1); 
\draw[blue] (1.5,2) node[left] {\textcircled{\small 2}};
\node[cross, red,line width=1.5pt,scale=2]at(0.7,2){};
\draw[red] (0.5,2) node[left] {$\beta$};
\end{scope}
\end{tikzpicture}
\end{equation}
With this figure, we see that $\alpha$ can be screened by the string \textcircled{\small 4} while $\beta$ can be screened by the string \textcircled{\small 2}, which is consistent with the analysis using the 5-brane web for Spin(6) gauge theory with one spinor. 

Thus, we find consistently from the view point of 5-brane webs that the 1-form symmetry $\mathbb{Z}_4$ of the Spin(6) gauge theory reduces to trivial when we introduce the spinor matter.

\subsubsection{\texorpdfstring{$\mathrm{Spin}(8) + 1\mathbf{S}$}{Spin8+1S}}
We finally discuss the Spin(8) gauge theory with one hypermultiplet in the spinor representation. Due to the triality of $\mathrm{Spin}(8)$ group, this theory is equivalent to the Spin(8) gauge theory with one hypermultiplet in the vector representation.

We use $b_1, b_2, b_3, b_4$ for Coulomb branch parameters of Spin(8) gauge theory with spinor, while $a_1, a_2, a_3, a_4$ for the Spin(8) gauge theory with vector. These are related by
\begin{align}\label{eq:map-Spin8-tirality}
a_1 &= \frac12 (b_1+b_2+b_3-b_4),
\cr
a_2 &= \frac12 (b_1+b_2-b_3+b_4),
\cr
a_3 &= \frac12 (b_1-b_2+b_3+b_4),
\cr
a_4 &= \frac12 (-b_1+b_2+b_3+b_4).
\end{align}
We choose the Weyl chamber where 
\begin{align}
b_1 > b_2 > b_3 > |b_4|.
\end{align}
is satisfied.

The 5-brane web diagram in the parameter region $b_4>0$ is given as in Figure \ref{fig:SO8+1C-b3p} while the 5-brane web diagram in the parameter region $b_4<0$ is given as in Figure \ref{fig:SO8+1C-b3n} 
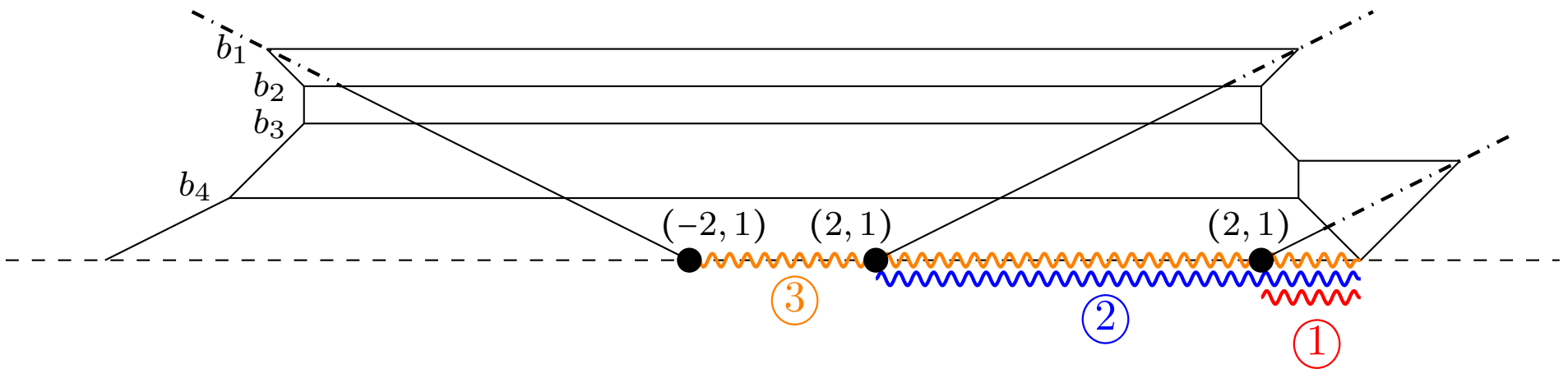
\begin{figure}[H]
\centering 
\begin{tikzpicture}
\draw[dashed] (-5.5,0)--(7,0);
\draw (-3.4,1.7)--(-3.1,1.4)--(-3.1,1.1)--(-3.7,0.5)--(-4.7,0);
\draw (-3.4,1.7)--(4.9,1.7);
\draw (-3.4,1.7) node[left] {\scriptsize $b_1$};
\draw (-3.1,1.4)--(4.6,1.4);
\draw (-3.1,1.4) node[left] {\scriptsize $b_2$};
\draw (-3.1,1.1)--(4.6,1.1);
\draw (-3.1,1.1) node[left] {\scriptsize $b_3$};
\draw (-3.7,0.5)--(4.9,0.5);
\draw (-3.7,0.6) node[left] {\scriptsize $b_4$};
\draw(4.9,0.8)--(6.2,0.8);
\draw (4.9,1.7)--(4.6,1.4)--(4.6,1.1)--(4.9,0.8)--(4.9,0.5)--(5.4,0)--(6.2,0.8);
\draw[orange, thick, decorate, decoration={snake, segment length=4pt, amplitude=1.5pt}](0,0)--(5.4,0);
\draw[blue, thick, decorate, decoration={snake, segment length=4pt, amplitude=1.5pt}](1.5,-0.15)--(5.4,-0.15);
\draw[red, thick, decorate, decoration={snake, segment length=4pt, amplitude=1.5pt}](4.6,-0.3)--(5.4,-0.3);
\draw[dash dot, thick] (-4,2) -- (-2.8,1.4);
\draw (-2.8,1.4)--(0,0);
\fill (0,0) circle (0.1); 
\fill (1.5,0) circle (0.1); 
\draw (1.5,0)--(4.3,1.4);
\draw[dash dot, thick] (4.3,1.4)--(5.5,2);
\fill (4.6,0) circle (0.1); 
\draw(4.6,0)--(5.1,0.25);
\draw[dash dot, thick] (5.1,0.25)--(6.6,1);

\draw[red] (4.7,-0.7) node[right] {\textcircled{\small 1}};
\draw[blue] (3,-0.5) node[right] {\textcircled{\small 2}};
\draw[orange] (0.5,-0.35) node[right] {\textcircled{\small 3}};

\draw (0.2,0) node[above] {\scriptsize $(-2,1)$};
\draw (1.3,0) node[above] {\scriptsize $(2,1)$};
\draw (4.5,0) node[above] {\scriptsize $(2,1)$};

\end{tikzpicture}
\caption{5-brane web diagram for Spin(8) + $1\mathbf{S}$ in the parameter region $b_4>0$. Three candidate D-strings are depicted. It turns out that none of them are allowed configuration.}
\label{fig:SO8+1C-b3p}
\end{figure}
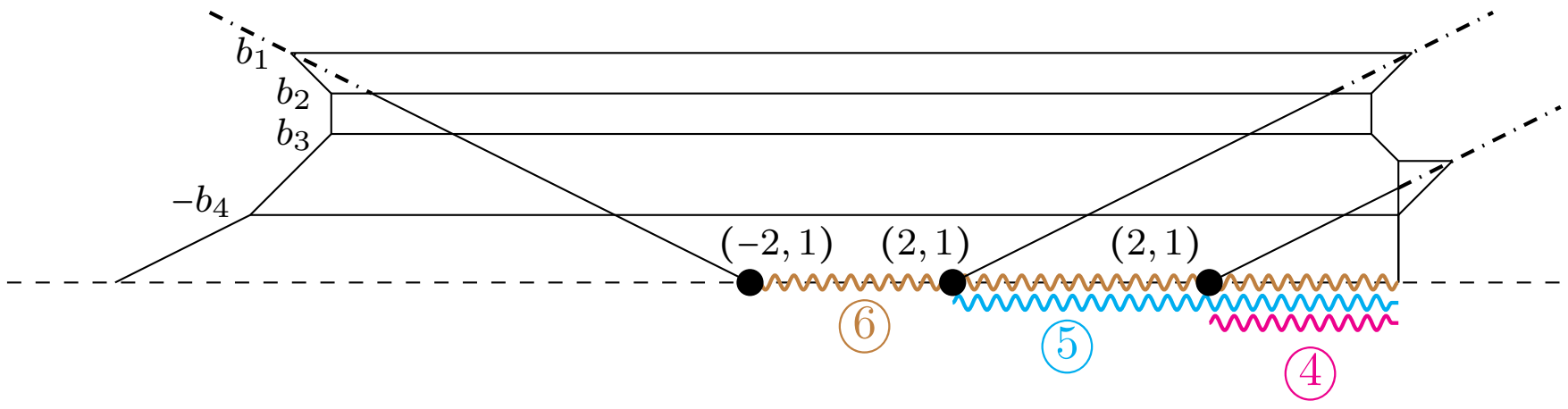
\begin{figure}[H]
\centering 
\begin{tikzpicture}
\draw[dashed] (-5.5,0)--(6,0);
\draw (-3.4,1.7)--(-3.1,1.4)--(-3.1,1.1)--(-3.7,0.5)--(-4.7,0);
\draw(-3.4,1.7)--(4.9,1.7);
\draw (-3.4,1.7) node[left] {\scriptsize $b_1$};
\draw(-3.1,1.4)--(4.6,1.4);
\draw (-3.1,1.4) node[left] {\scriptsize $b_2$};
\draw(-3.1,1.1)--(4.6,1.1);
\draw (-3.1,1.1) node[left] {\scriptsize $b_3$};
\draw(-3.7,0.5)--(4.8,0.5);
\draw (-3.7,0.6) node[left] {\scriptsize $-b_4$};
\draw(4.8,0.9)--(5.2,0.9);
\draw (4.9,1.7)--(4.6,1.4)--(4.6,1.1)--(4.8,0.9)--(4.8,0)--(4.8,0.5)--(5.2,0.9);
\draw[brown, thick, decorate, decoration={snake, segment length=4pt, amplitude=1.5pt}](0,0)--(4.8,0);
\draw[cyan, thick, decorate, decoration={snake, segment length=4pt, amplitude=1.5pt}](1.5,-0.15)--(4.8,-0.15);
\draw[magenta, thick, decorate, decoration={snake, segment length=4pt, amplitude=1.5pt}](3.4,-0.3)--(4.8,-0.3);
\draw[dash dot, thick] (-4,2) -- (-2.8,1.4);
\draw (-2.8,1.4)--(0,0);
\fill (0,0) circle (0.1); 
\fill (1.5,0) circle (0.1); 
\draw (1.5,0)--(4.3,1.4);
\draw[dash dot, thick] (4.3,1.4)--(5.5,2);
\fill (3.4,0) circle (0.1); 
\draw(3.4,0)--(4.8,0.7);
\draw[dash dot, thick] (4.8,0.7)--(6,1.3);

\draw[magenta] (3.8,-0.7) node[right] {\textcircled{\small 4}};
\draw[cyan] (2,-0.5) node[right] {\textcircled{\small 5}};
\draw[brown] (0.5,-0.35) node[right] {\textcircled{\small 6}};

\draw (0.2,0) node[above] {\scriptsize $(-2,1)$};
\draw (1.3,0) node[above] {\scriptsize $(2,1)$};
\draw (3,0) node[above] {\scriptsize $(2,1)$};

\end{tikzpicture}
\caption{5-brane web diagram for Spin(8) + $1\mathbf{S}$ in the parameter region $b_4<0$. Three candidate D-strings are depicted. }
\label{fig:SO8+1C-b3n}
\end{figure}

Analogous to the previous cases, we consider six candidates of D-string configurations \textcircled{\small 1} -- \textcircled{\small 6}, which are given in Figure \ref{fig:SO8+1C-b3p}  and Figure \ref{fig:SO8+1C-b3n}. 
The configuration \textcircled{\small 4}, \textcircled{\small 5}, \textcircled{\small 6} among them are allowed while \textcircled{\small 1}, \textcircled{\small 2}, \textcircled{\small 3} are not, according to our conjecture in Figure \ref{fig:conjecture}. The length of these D-strings are given by 
\begin{align}
\text{\textcircled{\small 1}}:& \,\, 
\frac12 (b_1+b_2+b_3+b_4)-m = \frac12 (a_1+a_2+a_3+a_4)-m
\quad \text{(Not allowed)}
\cr
\text{\textcircled{\small 2}}:& \,\, 
\frac12 (b_1+b_2+b_3+b_4)+m = \frac12 (a_1+a_2+a_3+a_4)+m
\quad \text{(Not allowed)}
\cr
\text{\textcircled{\small 3}}:& \,\, 
\frac12 (b_1+b_2+b_3+b_4)+m_0 = \frac12 (a_1+a_2+a_3+a_4)+m_0
\quad \text{(Not allowed)}
\cr
\text{\textcircled{\small 4}}:& \,\, 
\frac12 (b_1+b_2+b_3-b_4)-m = a_1-m
\quad \fbox{\text{Allowed}}
\cr
\text{\textcircled{\small 5}}:& \,\, 
\frac12 (b_1+b_2+b_3-b_4)+m = a_1+m
\quad \fbox{\text{Allowed}}
\cr
\text{\textcircled{\small 6}}:& \,\, 
\frac12 (b_1+b_2+b_3-b_4)+m_0 = a_1+m_0
\quad \fbox{\text{Allowed}}
\end{align}
where we use the map \eqref{eq:map-Spin8-tirality}.

Strings with lengths identical to the allowed configurations \textcircled{\small 4}, \textcircled{\small 5}, \textcircled{\small 6} can also be found in the 5-brane web diagram with $\mathrm{Spin}(8)$ gauge theory with one vector as

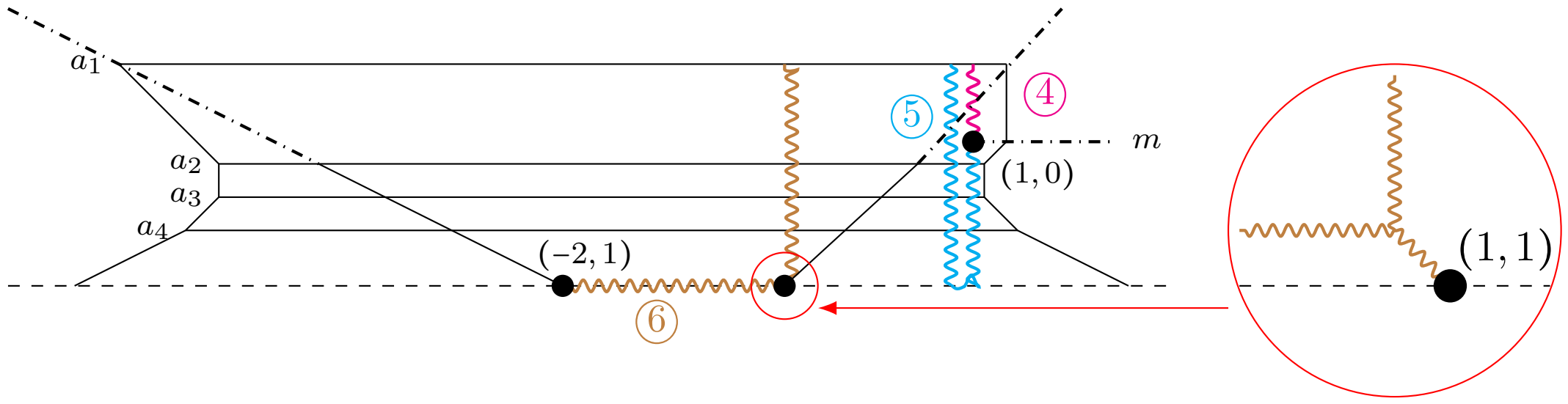
\begin{figure}[H]
\centering 
\begin{tikzpicture}
\draw[dashed] (-5,0)--(5.5,0);
\draw (-4,2)--(-3.1,1.1)--(-3.1,0.8)--(-3.4,0.5)--(-4.4,0);
\draw(-4,2)--(4,2);
\draw (-4,2) node[left] {\scriptsize $a_1$};
\draw(-3.1,1.1)--(3.8,1.1);
\draw (-3.1,1.1) node[left] {\scriptsize $a_2$};
\draw(-3.1,0.8)--(3.8,0.8);
\draw (-3.1,0.8) node[left] {\scriptsize $a_3$};
\draw(-3.4,0.5)--(4.1,0.5);
\draw (-3.4,0.5) node[left] {\scriptsize $a_4$};
\draw (4,2)--(4,1.3)--(3.8,1.1)--(3.8,0.8)--(4.1,0.5)--(5.1,0);
\draw[magenta, thick, decorate, decoration={snake, segment length=4pt, amplitude=1.5pt}](3.7,1.3)--(3.7,2);
\draw[magenta] (4,1.7) node[right] {\textcircled{\small 4}};
\draw[cyan, thick, decorate, decoration={snake, segment length=4pt, amplitude=1.5pt}](3.7,1.3)--(3.7,0)--(3.5,0);
\draw[cyan, thick, decorate, decoration={snake, segment length=4pt, amplitude=1.5pt}](3.5,0)--(3.5,2);
\draw[cyan] (2.8,1.5) node[right] {\textcircled{\small 5}};
\draw[brown, thick, decorate, decoration={snake, segment length=4pt, amplitude=1.5pt}](0,0)--(2,0)--(2,2);
\draw[brown] (0.5,-0.35) node[right] {\textcircled{\small 6}};
\draw[dash dot, thick] (-5,2.5) -- (-2.2,1.1);
\draw (-2.2,1.1)--(0,0);
\fill (0,0) circle (0.1); 
\draw (0.2,0) node[above] {\scriptsize $(-2,1)$};
\fill (2,0) circle (0.1); 
\draw (2,0)--(3.2,1.1);
\draw[dash dot, thick] (3.2,1.1)--(4.5,2.5);
\fill (3.7,1.3) circle (0.1); 
\draw[dash dot, thick] (3.7,1.3)--(5,1.3);
\draw (3.8,1) node[right] {\scriptsize $(1,0)$};
\draw (5,1.3) node[right] {\scriptsize $m$};
\draw[red,-Latex] (6,-0.2)--(2.3,-0.2);
\draw [red](2,0) circle (0.3);
\begin{scope}[shift={(7.5,0.5)}]
\draw [red](0,0) circle (1.5);
\draw[brown, thick, decorate, decoration={snake, segment length=4pt, amplitude=1.5pt}](0,0)--(0,1.4);
\draw[brown, thick, decorate, decoration={snake, segment length=4pt, amplitude=1.5pt}](0,0)--(0.5,-0.5);
\draw[brown, thick, decorate, decoration={snake, segment length=4pt, amplitude=1.5pt}](0,0)--(-1.4,0);
\fill (0.5,-0.5) circle (0.15);
\draw (1,-0.5) node[above] {$(1,1)$};
\draw[dashed] (-1.4,-0.5)--(1.4,-0.5);
\end{scope}
\end{tikzpicture}
\caption{5-brane web diagram for Spin(8) + $1\mathbf{V}$.}
\label{fig:SO8+1V}
\end{figure}

The configurations \textcircled{\small 4} and \textcircled{\small 5} correspond to typical BPS particles under the existence of a vector. These give further support for our conjecture in Figure \ref{fig:conjecture}. As for \textcircled{\small 6}, we assume that this is also an allowed configuration in order for our analysis to be consistent although it is not obvious whether this configuration is allowed a priori. 

Considering the half D3-branes corresponding to $\alpha$ and $\beta$ in the 5-brane web diagram in the parameter region $b_4<0$, depicted in Figure \ref{fig:SO8+1C-b3n}, we find that $\alpha$ can be screened by using \textcircled{\small 4}, \textcircled{\small 5} or \textcircled{\small 6} while $\beta$ cannot be screened. In the following, we draw $\alpha$, $\beta$ and only \textcircled{\small 4} for simplicity:
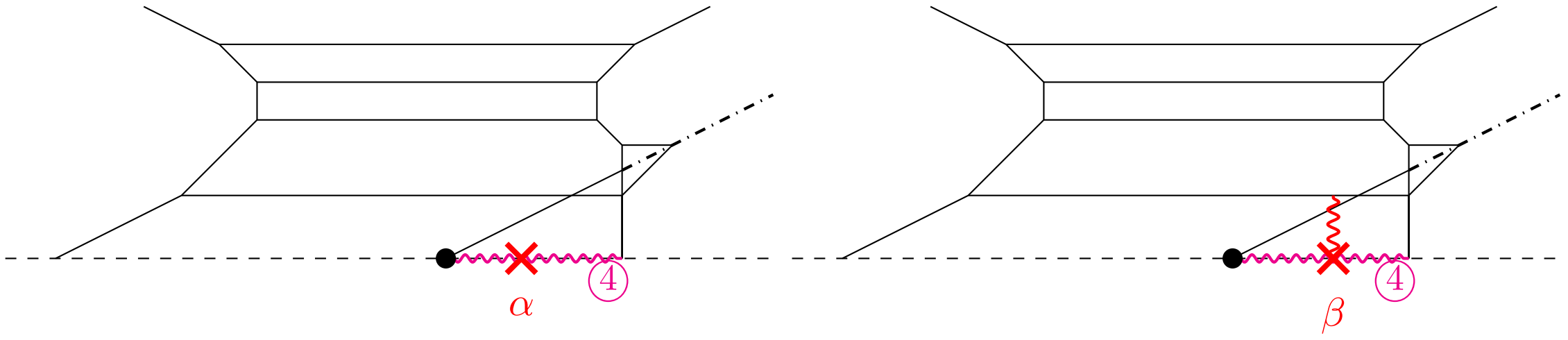
\begin{equation}
\centering 
\begin{tikzpicture}
\begin{scope}[shift={(-3.5,0)}, scale=1.2]
\draw[dashed] (-0.1,0)--(6,0);
\draw (1,2)--(1.6,1.7)--(1.9,1.4)--(1.9,1.1)--(1.3,0.5)--(0.3,0);
\draw(1.6,1.7)--(4.9,1.7);
\draw(1.9,1.4)--(4.6,1.4);
\draw(1.9,1.1)--(4.6,1.1);
\draw(1.3,0.5)--(4.8,0.5);
\draw(4.8,0.9)--(5.2,0.9);
\draw (5.5,2)--(4.9,1.7)--(4.6,1.4)--(4.6,1.1)--(4.8,0.9)--(4.8,0)--(4.8,0.5)--(5.2,0.9);
\draw[magenta, thick, decorate, decoration={snake, segment length=4pt, amplitude=1pt}](3.4,0)--(4.8,0);
\fill (3.4,0) circle (0.08); 
\draw(3.4,0)--(4.8,0.7);
\draw[dash dot, thick] (4.8,0.7)--(6,1.3);
\draw[magenta] (4.4,-0.2) node[right] {\textcircled{\small 4}};
\node[cross, red, line width=1.5pt,scale=2]at(4,0){};
\draw[red] (4,-0.2) node[below] {$\alpha$};
\end{scope}
\begin{scope}[shift={(4,0)}, scale=1.2]
\draw[dashed] (-0.1,0)--(6,0);
\draw (1,2)--(1.6,1.7)--(1.9,1.4)--(1.9,1.1)--(1.3,0.5)--(0.3,0);
\draw(1.6,1.7)--(4.9,1.7);
\draw(1.9,1.4)--(4.6,1.4);
\draw(1.9,1.1)--(4.6,1.1);
\draw(1.3,0.5)--(4.8,0.5);
\draw(4.8,0.9)--(5.2,0.9);
\draw (5.5,2)--(4.9,1.7)--(4.6,1.4)--(4.6,1.1)--(4.8,0.9)--(4.8,0)--(4.8,0.5)--(5.2,0.9);
\draw[magenta, thick, decorate, decoration={snake, segment length=4pt, amplitude=1pt}](3.4,0)--(4.8,0);
\fill (3.4,0) circle (0.08); 
\draw(3.4,0)--(4.8,0.7);
\draw[dash dot, thick] (4.8,0.7)--(6,1.3);
\draw[magenta] (4.4,-0.2) node[right] {\textcircled{\small 4}};
\node[cross, red, line width=1.5pt,scale=2]at(4.2,0){};
\draw[red] (4.2,-0.2) node[below] {$\beta$};
\draw[red, thick, decorate, decoration={snake, segment length=4pt, amplitude=1.5pt}] (4.2,0)--(4.2,0.5);
\end{scope}
\end{tikzpicture}
\end{equation}
Analogous to the pattern for the Spin(4) gauge theory with spinor, we see that the $\mathbb{Z}_2$ generated by $\alpha$ reduces to trivial symmetry while the $\mathbb{Z}_2$ generated by $\beta$ remains.

The corresponding screening can also be observed from the equivalent Spin(8) gauge theory with one vector. Due to the triality, the Wilson lines in the spinor representation and in the conjugate spinor representation of the Spin(8) + $1\mathbf{S}$ is mapped to the Wilson line in the vector representation and in the conjugate spinor representation of the Spin(8) + $1\mathbf{V}$, respectively. Thus, $\alpha$ in this picture is given by the D3-brane in the top room and can be screened by the string \textcircled{\small 4} while $\beta$ is the half D3-brane on O5-plane with a fundamental string attached and cannot be screened. The half D3-brane without a string corresponds to $\alpha \beta$ in this notation. 
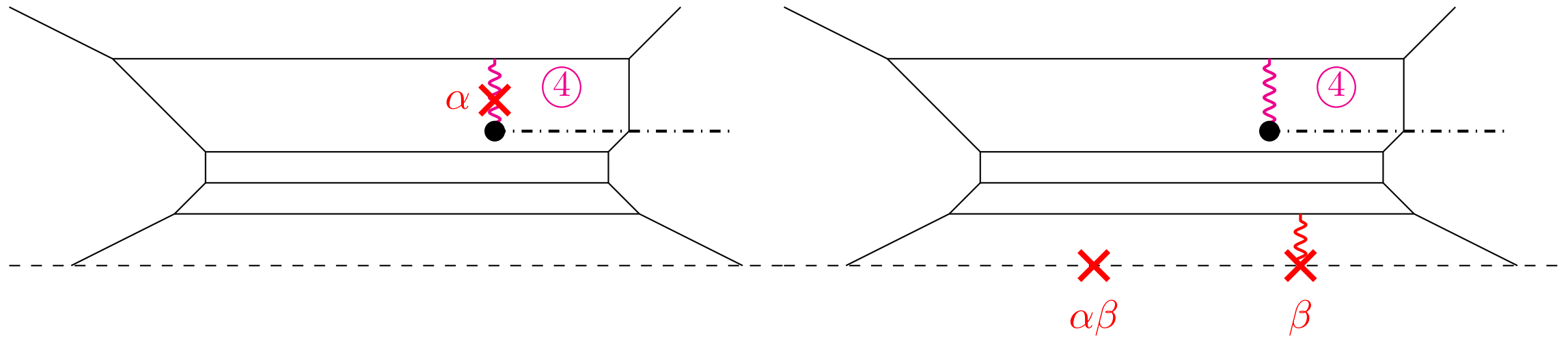
\begin{equation}
\centering 
\begin{tikzpicture}
\begin{scope}[shift={(-3.5,0)}]
\draw[dashed] (-2,0)--(5.5,0);
\draw (-2,2.5)--(-1,2)--(-0.1,1.1)--(-0.1,0.8)--(-0.4,0.5)--(-1.4,0);
\draw(-1,2)--(4,2);
\draw(-0.1,1.1)--(3.8,1.1);
\draw(-0.1,0.8)--(3.8,0.8);
\draw(-0.4,0.5)--(4.1,0.5);
\draw (4.5,2.5)--(4,2)--(4,1.3)--(3.8,1.1)--(3.8,0.8)--(4.1,0.5)--(5.1,0);
\draw[magenta, thick, decorate, decoration={snake, segment length=4pt, amplitude=1.5pt}](2.7,1.3)--(2.7,2);
\draw[magenta] (3,1.7) node[right] {\textcircled{\small 4}};
\fill (2.7,1.3) circle (0.1); 
\draw[dash dot, thick] (2.7,1.3)--(5,1.3);
\node[cross, red, line width=1.5pt,scale=2]at(2.7,1.6){};
\draw[red] (2.6,1.6) node[left] {$\alpha$};
\end{scope}
\begin{scope}[shift={(4,0)}]
\draw[dashed] (-2,0)--(5.5,0);
\draw (-2,2.5)--(-1,2)--(-0.1,1.1)--(-0.1,0.8)--(-0.4,0.5)--(-1.4,0);
\draw(-1,2)--(4,2);
\draw(-0.1,1.1)--(3.8,1.1);
\draw(-0.1,0.8)--(3.8,0.8);
\draw(-0.4,0.5)--(4.1,0.5);
\draw (4.5,2.5)--(4,2)--(4,1.3)--(3.8,1.1)--(3.8,0.8)--(4.1,0.5)--(5.1,0);
\draw[magenta, thick, decorate, decoration={snake, segment length=4pt, amplitude=1.5pt}](2.7,1.3)--(2.7,2);
\draw[magenta] (3,1.7) node[right] {\textcircled{\small 4}};
\fill (2.7,1.3) circle (0.1); 
\draw[dash dot, thick] (2.7,1.3)--(5,1.3);
\node[cross, red, line width=1.5pt,scale=2]at(3,0){};
\draw[red] (3,-0.2) node[below] {$\beta$};
\draw[red, thick, decorate, decoration={snake, segment length=4pt, amplitude=1.5pt}] (3,0)--(3,0.5);
\node[cross, red, line width=1.5pt,scale=2]at(1,0){};
\draw[red] (1,-0.2) node[below] {$\alpha \beta$};
\end{scope}
\end{tikzpicture}
\end{equation}
Since $\alpha$ is identified as an identity $e$ due to the screening, we can also identify $\alpha \beta$ as $\beta$. In other words, the $\mathbb{Z}_2$ generated by $\beta$ and the $\mathbb{Z}_2$ generated by $\alpha' := \alpha\beta$ are identified with each other. 
Thus, we can interpret that the $\mathbb{Z}_2 \times \mathbb{Z}_2$ 1-form symmetry reduces to their diagonal $\mathbb{Z}_2$ in this picture. This is consistent with the expected result. 

In this way, we find that the 1-form symmetry $\mathbb{Z}_2 \times \mathbb{Z}_2$ of the Spin(8) gauge theory reduces to $\mathbb{Z}_2$ when we introduce the spinor matter from the viewpoint of the 5-brane web in a consistent manner.

\vspace{0.5cm}

\section{Conclusion}\label{sec:conclusion}

In this work, we developed a brane-based approach for determining the 1-form symmetries of five-dimensional theories engineered by 5-brane webs. In the brane picture, screening is implemented by fragmenting and reconnecting strings, Higgsing segments along the directions shared by D3- and 7-branes, and subsequently removing the remaining D3-branes through Hanany-Witten transitions. This provides a microscopic interpretation of the equivalence relation among line operators whose Pontryagin dual defines the 1-form symmetry group.

For ordinary 5-brane webs, our approach agrees with the Smith normal form prescription based on the asymptotic 7-brane charges. In this sense, the screening construction provides a physical interpretation of the arithmetic information encoded by the Smith normal form. Moreover, our approach is in agreement for Lagrangian theories, with the expected 1-form symmetries based on gauge theoretic techniques. 
At the same time, because the construction is formulated directly in terms of branes and strings, it admits extensions to systems for which a purely asymptotic-charge prescription, or Lagrangian formulation is not immediately available.

We first applied these ideas to webs containing orientifold 7-planes. For an $O7^-$-plane, resolving the orientifold into a pair of 7-branes allows its contribution to be incorporated into the usual Smith normal form computation. This reproduces the expected 1-form symmetries of $\mathrm{SU}(N)$ theories with antisymmetric matter and of pure $\mathrm{Sp}(N)_\theta$ theories. For an $O7^+$-plane, which does not admit such a resolution, we proposed assigning the effective charge $[2,0]$, motivated by its monodromy. Including this charge in the Smith normal form correctly reproduces the gauge-theory results for the $\mathrm{SU}(N)$ theories with symmetric matter and for the $SO(N)$ theories considered here. The same prescription also gives predictions for non-Lagrangian theories: for the higher-rank generalisations of the Bhardwaj theory, it yields a trivial 1-form symmetry for even rank and a $\mathbb{Z}_2$ symmetry for odd rank.

The novel microscopic prescription of screening, was applied in particular to 5-brane webs with O5-planes, and those with O7-planes. 

A central lesson of the O5 analysis is that asymptotic charge data alone do not, in general, determine the spectrum of unscreened lines. Webs with identical boundary behaviour can distinguish different global structures through the local configuration of branes near the orientifold. The 1-form symmetry is therefore sensitive to information near the origin of the web, including the possible half-D3-brane configurations and the local rules governing their screening. This observation refines the usual expectation that the global variants of a theory should be completely encoded by boundary data at infinity.

Several questions remain open. It would be useful to derive both the effective $[2,0]$ prescription for O7$^+$ planes and the parity rule for D-strings on O5$^-$ planes from a more fundamental worldsheet or geometric analysis. Extending the construction to magnetic and dyonic extended operators may also make higher-form symmetries and their anomalies directly visible in brane webs. More broadly, applying the screening approach to additional non-Lagrangian theories could provide a systematic way to determine their global structures and associated symmetry topological field theories.

It would be desirable to analyse Hanany-Witten systems for field theories in other dimensions, and extend the analysis presented here, to determine the 1-form symmetries of brane constructions for field theories across spacetime dimensions.


\paragraph{Acknowledgement.} The authors thank Siddharth Dwivedi and Hirotaka Hayashi for collaboration at an initial stage of this project. It is a pleasure to thank Adi Armoni, and Shigeki Sugimoto for discussions related to \cite{Armoni:2022xhy} which sparked the initial idea for this project. We also thank Jin Chen, Hee-Cheol Kim, Kimyeong Lee, Marcus Sperling, Sinan Moura Soysüren, Fabio Marino, Francesca Pretto for discussions. MA is supported in part
by an INFN fellowship under the “Iniziativa Specifica” 24 ST\&FI. F.C. is supported by the Italian Ministry of Universities and Research (MUR) through the grant “StringGeom” (grant CUP code no. I33c25000420006). S.K. is supported by National Natural Science Foundation of China grant No. 12250610188 and also by the Brain Pool Program funded by the Ministry of Science and ICT through the National Research Foundation of Korea (RS-2025-25457100). F.Y. is supported by start-up research grant RK21121 from Huzhou Normal University.

\appendix
\section{Pontryagin duality}
\label{app:pontryagin}

Let $G$ be a locally compact abelian group. Its \emph{Pontryagin dual} is the group
\begin{equation}
    \widehat{G}
    =
    \operatorname{Hom}_{\mathrm{cont}}(G,U(1))
\end{equation}
of continuous homomorphisms from $G$ to $\mathrm{U}(1)$. The elements of $\widehat{G}$ are called the \emph{characters} of $G$. If the group operation on $G$ is written additively, they satisfy
\begin{equation}
    \chi(g_1+g_2)=\chi(g_1)\chi(g_2).
\end{equation}
The group operation on $\widehat{G}$ is pointwise multiplication. The Pontryagin duality theorem states that the natural evaluation map
\begin{equation}
    G\longrightarrow\widehat{\widehat{G}},
    \qquad
    g\longmapsto\bigl(\chi\mapsto\chi(g)\bigr),
\end{equation}
is an isomorphism of topological groups. Thus, a locally compact abelian group can be canonically recovered from its double dual.

Some standard examples of Pontryagin duals are
\begin{equation}
    \widehat{\mathbb{Z}}\simeq U(1),
    \qquad
    \widehat{U(1)}\simeq\mathbb{Z},
    \qquad
    \widehat{\mathbb{R}}\simeq\mathbb{R}.
\end{equation}
The characters of $\mathrm{U}(1)$ are $z\mapsto z^n$, with $n\in\mathbb{Z}$, while the characters of $\mathbb{R}$ take the form
\begin{equation}
    \chi_\xi(x)=e^{2\pi i\xi x}.
\end{equation}
More generally, the Pontryagin dual of a compact abelian group is discrete, while the dual of a discrete abelian group is compact. If $G$ is finite and abelian, then $\widehat{G}$ is noncanonically isomorphic to $G$.

\section{\texorpdfstring{Spin($N$) parametrisations}{Spin(N) parametrisations}} \label{app:parameterisation}
In this appendix, we discuss the parameterisation of spinor matter in Spin($N$) gauge theories engineered by 5-brane webs in the presence of an O5-plane, as used in section~\ref{sec:withO5}. We explicitly work out two examples that admit equivalent SU gauge theory descriptions: $\Spin(4)+1\mathbf{S}~\simeq~\SU(2)\times\SU(2)-[1]$ and $\Spin(6)+1\mathbf{S}~\simeq~\SU(4)+1\mathbf{F}$. As the parameterisation of an $\SU$-gauge theory with a hypermultiplet in the fundamental representation is well established, 
we determine the appropriate parameterisation for the 5-brane webs for the Spin gauge theories by comparing the areas of the compact faces, which gives the tension of monopoles, with the derivatives of the prepotentials of their corresponding SU descriptions. We further test this parameterisation in various flopped phases and find consistent agreement.  

\subsection{Spin(4) parameterisation} 

\subsubsection*{Pure Spin(4)}
Since $\mathfrak{so}(4)\simeq\mathfrak{su}(2)\oplus \mathfrak{su}(2)$, we expect the gauge theory prepotential to be of the form
\begin{equation}
    \mathcal{F}_{E_1\oplus E_1}=m_0(a_1^2+a_2^2)+\frac{4}{3}(a_1^3+a_2^3)\;.
\end{equation}
Consider the following parameterisation
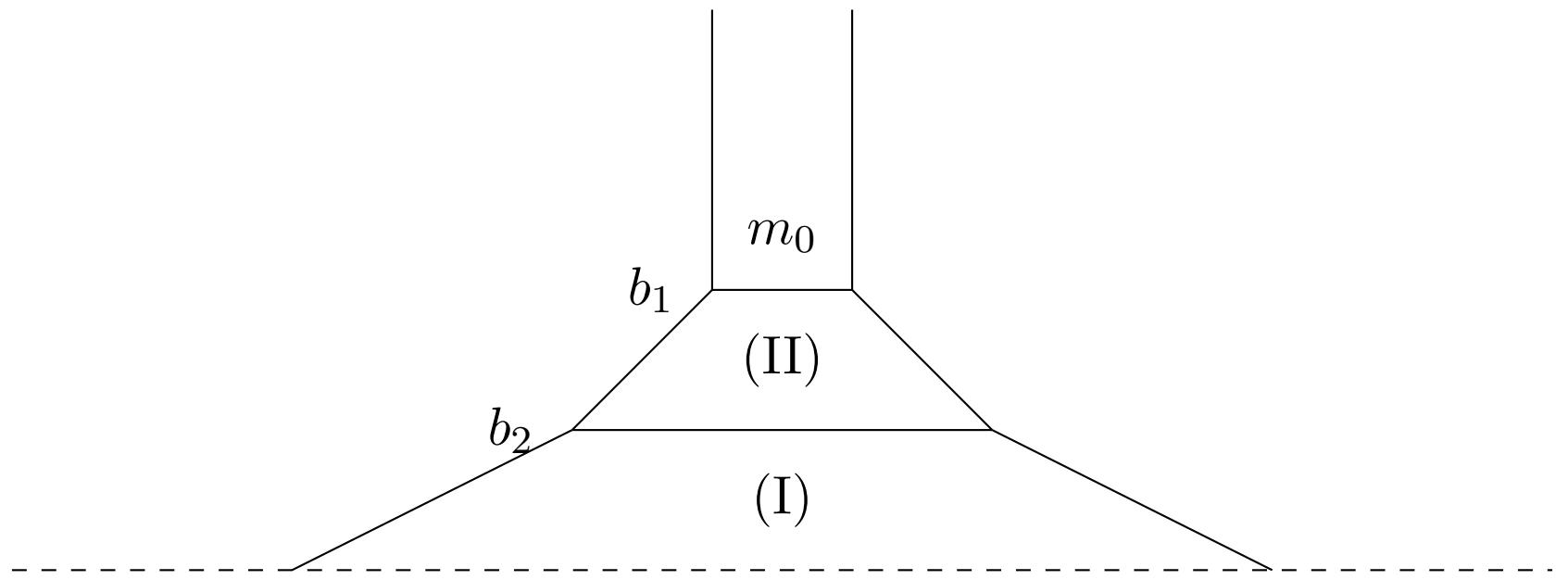
\begin{equation}
    \begin{tikzpicture}
        \draw(0,0)--(2,1);
        \draw(2,1)--(3,2);
        \draw(3,4)--(3,2);
        \draw(5,1)--(2,1);
        \draw(4,2)--(3,2);
        \draw(4,4)--(4,2);
        \draw(5,1)--(4,2);
        \draw(5,1)--(7,0);
        \draw[dashed](-2,0)--(9,0);
        \node[label=left:{$b_2$}] at (2,1){};
        \node[label=left:{$b_1$}] at (3,2){};
        \node[label=above:{$m_0$}] at (3.5,2){};
        \node at (3.5,1.5){(II)};
        \node at (3.5,.5){(I)};
    \end{tikzpicture}
\end{equation}
Computing the area of the two faces in this parametrisation yields
\begin{align}
    T_\mathrm{(II)}&=(b_1-b_2)(m_0+b_1-b_2)\;,\\
    T_{\mathrm{(I)}}&=b_2(m_0+2b_1) .
\end{align}
It is straightforward to check that, upon identifying
\begin{equation}
    b_1=a_1+a_2\;,\qquad b_2=a_1-a_2\;,
\end{equation}
the parametrisation above agrees with the gauge theory prepotential
\begin{equation}
    \frac{\partial\mathcal{F}_{E_1\oplus E_1}}{\partial a_1}=2T_{\mathrm{(I)}}+T_\mathrm{(II)}\;,\qquad \frac{\partial\mathcal{F}_{E_1\oplus E_1}}{\partial a_2}=T_\mathrm{(II)}\;.
\end{equation}

\subsubsection*{Spin(4)+1S}
In the presence of a single spinor hypermultiplet, we expect the gauge theory prepotential to be that of the $E_1\oplus E_2$ theory
\begin{equation}
    \mathcal{F}_{E_1\oplus E_2}=(m_0-\frac{m}{2})a_1^2+m_0a_2^2+\frac{4}{3}(a_1^3+a_2^3)-\frac{1}{12}|a_2\pm m|^3\;.
\end{equation}
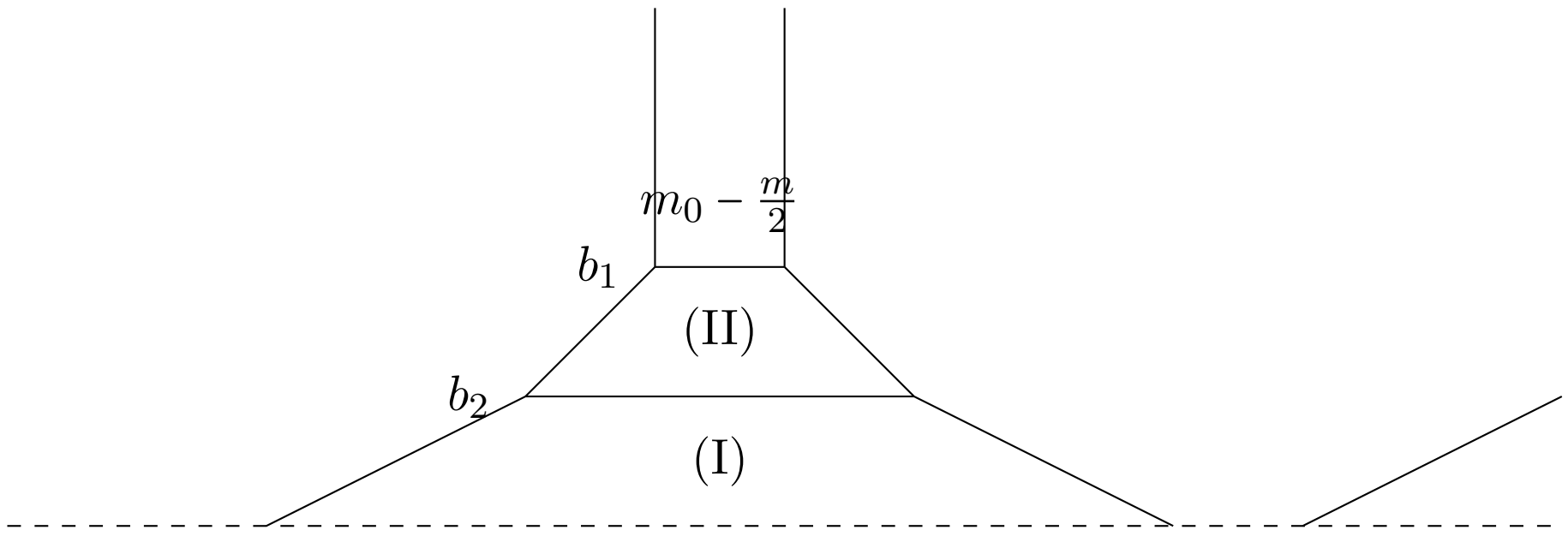
\begin{equation}
    \begin{tikzpicture}
        \draw(0,0)--(2,1);
        \draw(2,1)--(3,2);
        \draw(3,4)--(3,2);
        \draw(5,1)--(2,1);
        \draw(4,2)--(3,2);
        \draw(4,4)--(4,2);
        \draw(5,1)--(4,2);
        \draw(5,1)--(7,0);
        \draw[dashed](-2,0)--(10,0);
        \node[label=left:{$b_2$}] at (2,1){};
        \node[label=left:{$b_1$}] at (3,2){};
        \node[label=above:{$m_0-\frac{m}{2}$}] at (3.5,2){};
        \node at (3.5,1.5){(II)};
        \node at (3.5,.5){(I)};
        \draw(8,0)--(10,1);
    \end{tikzpicture}
\end{equation}
This parametrisation is consistent with the gauge theory prepotential
\begin{equation}
    \frac{\partial\mathcal{F}_{E_1\oplus E_2}}{\partial a_1}=2T_{\mathrm{(I)}}+T_\mathrm{(II)}\;,\qquad \frac{\partial\mathcal{F}_{E_1\oplus E_2}}{\partial a_2}=T_\mathrm{(II)}\;.
\end{equation}

\paragraph{}
\paragraph{Spin(4)+1S after one flop.} It turns out that the phase after one flop does not give new information.

\paragraph{}
\paragraph{Spin(4)+1S after two flops.}
Let us now go to a different phase on the Coulomb branch with prepotential
\begin{equation}
   \mathcal{F}= m_0 \left(a_1^2+a_2^2\right)+\frac{4}{3} \left(a _1^3+a_2^3\right)-\frac{1}{12} \left(a_2-m \right){}^3-\frac{1}{12} \left(m +a_2\right){}^3-\frac{1}{2} m  a_1^2\;.
\end{equation}
The corresponding 5-brane web is given in~\eqref{eq:spin4+1S-2flop}, where $x=\frac{1}{2}(b_1+b_2)-m$.
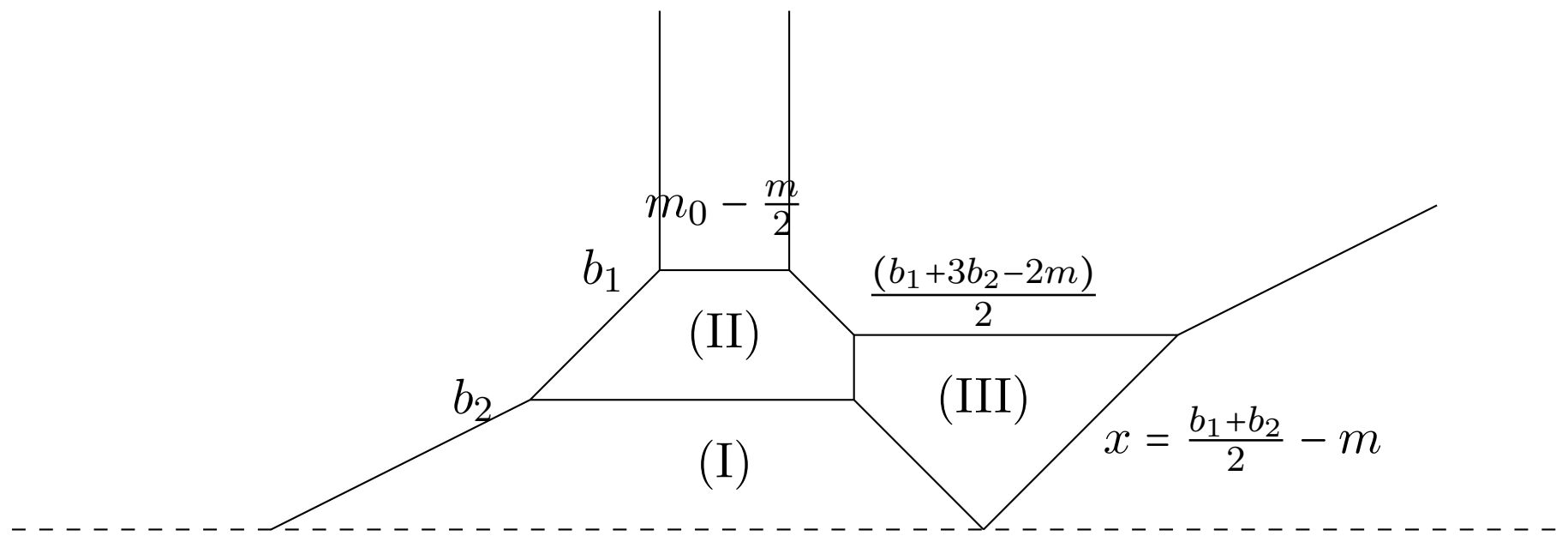
\begin{equation}
    \begin{tikzpicture}
        \draw(0,0)--(2,1);
        \draw(2,1)--(3,2);
        \draw(3,4)--(3,2);
        \draw(4.5,1)--(2,1);
        \draw(4,2)--(3,2);
        \draw(4,4)--(4,2);
        \draw(4.5,1.5)--(4,2);
        \draw(4.5,1.5)--(4.5,1);
        \draw(5.5,0)--(4.5,1);
        \draw(5.5,0)--(7,1.5);
        \draw(4.5,1.5)--(7,1.5);
        \draw(9,2.5)--(7,1.5);
        \draw[dashed](-2,0)--(10,0);
        \node[label=left:{$b_2$}] at (2,1){};
        \node[label=left:{$b_1$}] at (3,2){};
        \node[label=above:{$m_0-\frac{m}{2}$}] at (3.5,2){};
        \node[label=above:{$\frac{ \left(b_1+3 b_2-2 m\right)}{2}$}] at (5.5,1.3){};
        \node at (5.5,1){(III)};
        \node at (3.5,1.5){(II)};
        \node at (7.5,0.7){$x= \frac{b_1+b_2}{2}-m$};
        \node at (3.5,.5){(I)};
    \end{tikzpicture}
    \label{eq:spin4+1S-2flop}
\end{equation}\;
from which one obtains the following tensions
\begin{equation}
\begin{split}
    T_{\mathrm{(I)}}&=\frac{1}{2} b_2 \left(3 b_1+2 m_0+m \right)\\
    T_\mathrm{(II)}&=\frac{1}{8} \left(\left(b_1-b_2\right) \left(7 b_1-7 b_2+8 m_0\right)-4 m ^2\right)\\
    T_\mathrm{(III)}&=\frac{1}{8} \left(-4 \left(b_1+3 b_2\right) m +b_1^2+6 b_2 b_1+b_2^2+4 m ^2\right)
\end{split}
\end{equation}
It is straightforward to verify the consistency of the 5-brane parameterisation with the gauge theory prepotential
\begin{equation}
    \frac{\partial\mathcal{F}}{\partial a_1}=2T_{\mathrm{(I)}}+T_\mathrm{(II)}+T_\mathrm{(III)}\;,\qquad \frac{\partial\mathcal{F}}{\partial a_2}=T_\mathrm{(II)}\;.
\end{equation}

\subsection{Spin(6) parameterisation}
\subsubsection*{Pure Spin(6)}
The expected gauge theory prepotential is
\begin{equation}
    \mathcal{F}_{SU(4)_0}=\frac{m_0}{2}\sum_{i=1}^4 a_i^2+\frac{1}{6}\sum_{i<j}^4(a_i-a_j)^3\ ,
\end{equation}
where $m_0$ is the instanton mass for $\mathrm{SU}(4)_{0}$. The corresponding 5-brane web is given in \eqref{eq:spin6_pure}.

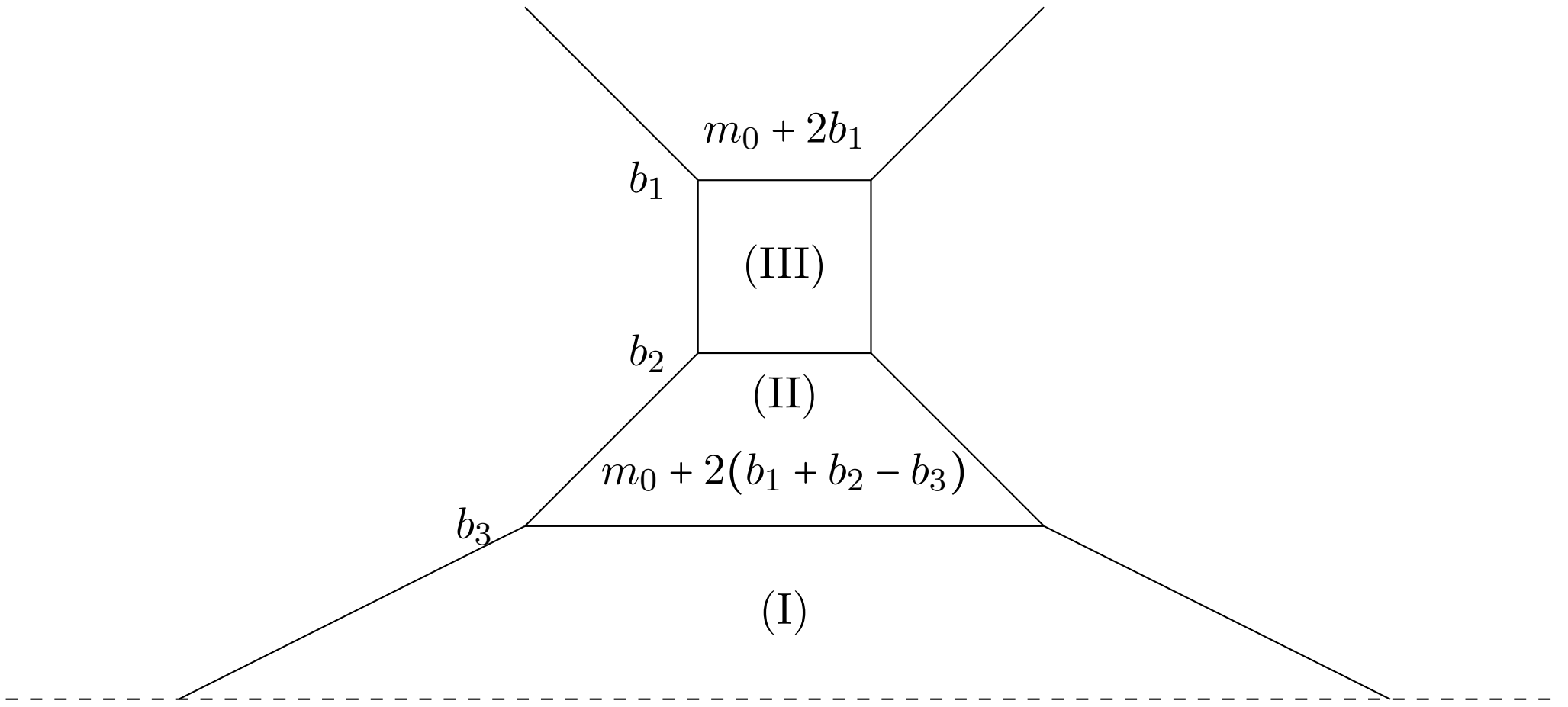
\begin{equation}
    \begin{tikzpicture}[scale =1.5]
        \draw(0,0)--(2,1);
        \draw(2,1)--(3,2);
        \draw(3,3)--(3,2);
        \draw(3,3)--(2,4);
        \draw(5,1)--(2,1);
        \draw(4,2)--(3,2);
        \draw(4,3)--(4,2);
        \draw(4,3)--(5,4);
        \draw(4,3)--(3,3);
        \draw(5,1)--(4,2);
        \draw(5,1)--(7,0);
        \draw[dashed](-1,0)--(8,0);
        \node[label=left:{$b_3$}] at (2,1){};
        \node[label=left:{$b_2$}] at (3,2){};
        \node[label=left:{$b_1$}] at (3,3){};
        \node[label=above:{$m_0+2b_1$}] at (3.5,3){};
        \node[label=above:{$m_0+2(b_1+b_2-b_3)$}] at (3.5,1){};
        \node at (3.5,2.5){(III)};
        \node at (3.5,1.75){(II)};
        \node at (3.5,.5){(I)};
    \end{tikzpicture}
    \label{eq:spin6_pure}
\end{equation}
From the diagram, we read the following area/tensions
\begin{align}
    T_{\mathrm{(I)}}&=b_3(m_0+2(b_1+b_2))\\
    T_\mathrm{(II)}&=(b_2-b_3)(m_0+2b_1+b_2-b_3)\\
    T_\mathrm{(III)}&=(b_1-b_2)(m_0+2b_1)\ .
\end{align}
Upon making the substitutions
\begin{equation}\label{eq:SO6-SU4map}
\begin{split}
    b_1&=\phi_2,\;\quad b_2=\phi_1-\phi_2+\phi_3\;,\quad b_3=\phi_1-\phi_3,\\
    a_1&=\phi_1\;,\quad a_2=\phi_2-\phi_1\;,\quad a_3=\phi_3-\phi_2\;,\quad a_4=-\phi_3\;,
\end{split}
\end{equation}
one finds that the parameterisation of the web is consistent with the gauge theory prepotential
\begin{equation}
    \frac{\partial\mathcal{F}_{SU(4)_0}}{\partial\phi_1}=2T_{\mathrm{(I)}}+T_\mathrm{(II)}\;,\qquad \frac{\partial\mathcal{F}_{SU(4)_0}}{\partial\phi_2}=T_\mathrm{(III)}\;,\qquad \frac{\partial\mathcal{F}_{SU(4)_0}}{\partial\phi_3}=T_\mathrm{(II)}\;.
\end{equation}
It follows from \eqref{eq:SO6-SU4map} that 
\begin{equation}
    \frac{b_1+b_2+b_3}{2}=\phi_1=a_1\ .
\end{equation}
\subsubsection*{Spin(6)+1S}
The expected gauge theory prepotential is
\begin{equation}
    \mathcal{F}_{SU(4)_{\frac{1}{2}}+1\mathbf{F}}=m_0\sum_{i=1}^4 a_i^2+\frac{1}{6}\sum_{i<j}^4(a_i-a_j)^3-\frac{1}{12}\sum_{i=1}^4\left( a_i+m \right)^3+\frac{1}{12}\sum_{i=1}^4a_i^3\ ,
\end{equation}
where $m_0$ is the instanton mass for $\mathrm{SU}(4)_{\frac{1}{2}}+1\mathbf{F}$. The corresponding 5-brane web is given in \eqref{eq:spin6_1s}.

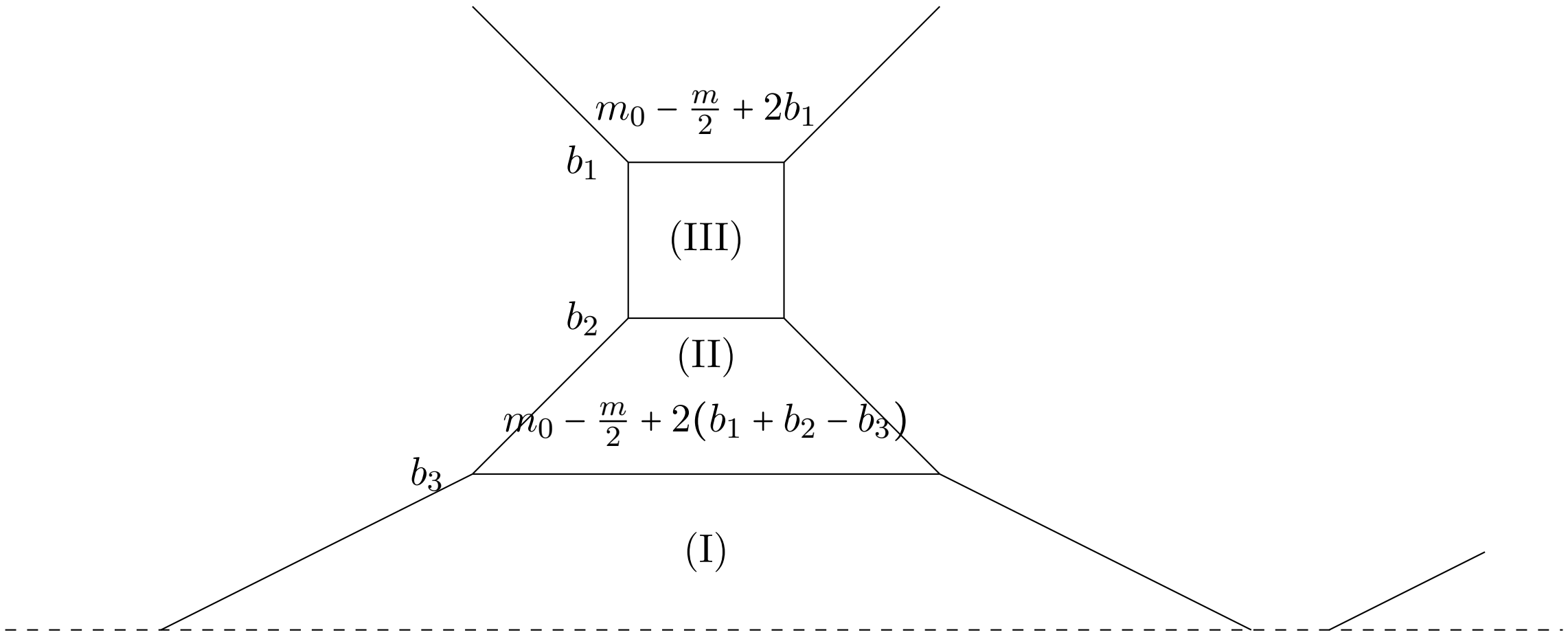
\begin{equation}
    \begin{tikzpicture}[scale =1.5]
        \draw(0,0)--(2,1);
        \draw(2,1)--(3,2);
        \draw(3,3)--(3,2);
        \draw(3,3)--(2,4);
        \draw(5,1)--(2,1);
        \draw(4,2)--(3,2);
        \draw(4,3)--(4,2);
        \draw(4,3)--(5,4);
        \draw(4,3)--(3,3);
        \draw(5,1)--(4,2);
        \draw(5,1)--(7,0);
        \draw(7.5,0)--(8.5,.5);
        \draw[dashed](-1,0)--(9,0);
        \node[label=left:{$b_3$}] at (2,1){};
        \node[label=left:{$b_2$}] at (3,2){};
        \node[label=left:{$b_1$}] at (3,3){};
        \node[label=above:{$m_0-\frac{m}{2}+2b_1$}] at (3.5,3){};
        \node[label=above:{$m_0-\frac{m}{2}+2(b_1+b_2-b_3)$}] at (3.5,1){};
        \node at (3.5,2.5){(III)};
        \node at (3.5,1.75){(II)};
        \node at (3.5,.5){(I)};
    \end{tikzpicture}
    \label{eq:spin6_1s}
\end{equation}
From the diagram, we read the following area/tensions
\begin{align}
    T_{\mathrm{(I)}}&=\frac{1}{2}b_3(2m_0-m+4(b_1+b_2))\\
    T_\mathrm{(II)}&=\frac{1}{2}(b_2-b_3)(2m_0-m+4b_1+2b_2-2b_3)\\
    T_\mathrm{(III)}&=(b_1-b_2)(m_0-\frac{m}{2}+2b_1)\ .
\end{align}
It is straightforward to check that these tensions are consistent with the gauge theory prepotential
\begin{equation}
    \frac{\partial\mathcal{F}_{SU(4)_\frac{1}{2}+1\textbf{F}}}{\partial\phi_1}=2T_{\mathrm{(I)}}+T_\mathrm{(II)}\;,\quad \frac{\partial\mathcal{F}_{SU(4)_\frac{1}{2}+1\textbf{F}}}{\partial\phi_2}=T_\mathrm{(III)}\;,\quad \frac{\partial\mathcal{F}_{SU(4)_\frac{1}{2}+1\textbf{F}}}{\partial\phi_3}=T_\mathrm{(II)}\;.
\end{equation}

\paragraph{}
\paragraph{Spin(6)+1S after one flop.} It also turns out that no new information arises from the first flop.

\paragraph{}
\paragraph{Spin(6)+1S after two flops.}
The expected gauge theory prepotential is
\begin{equation}
    \mathcal{F}_{SU(4)_{\frac{1}{2}}+1\mathbf{F}}=m_0\sum_{i=1}^4 a_i^2+\frac{1}{6}\sum_{i<j}^4(a_i-a_j)^3\!-\!\frac{1}{12}\left(\sum_{i=1}^3( a_i+m)^3-( a_4+m)^3\right)+\frac{1}{12}\sum_{i=1}^4a_i^3.
\end{equation}
The corresponding 5-brane web is given in \eqref{eq:spin6_1s_1flop}.

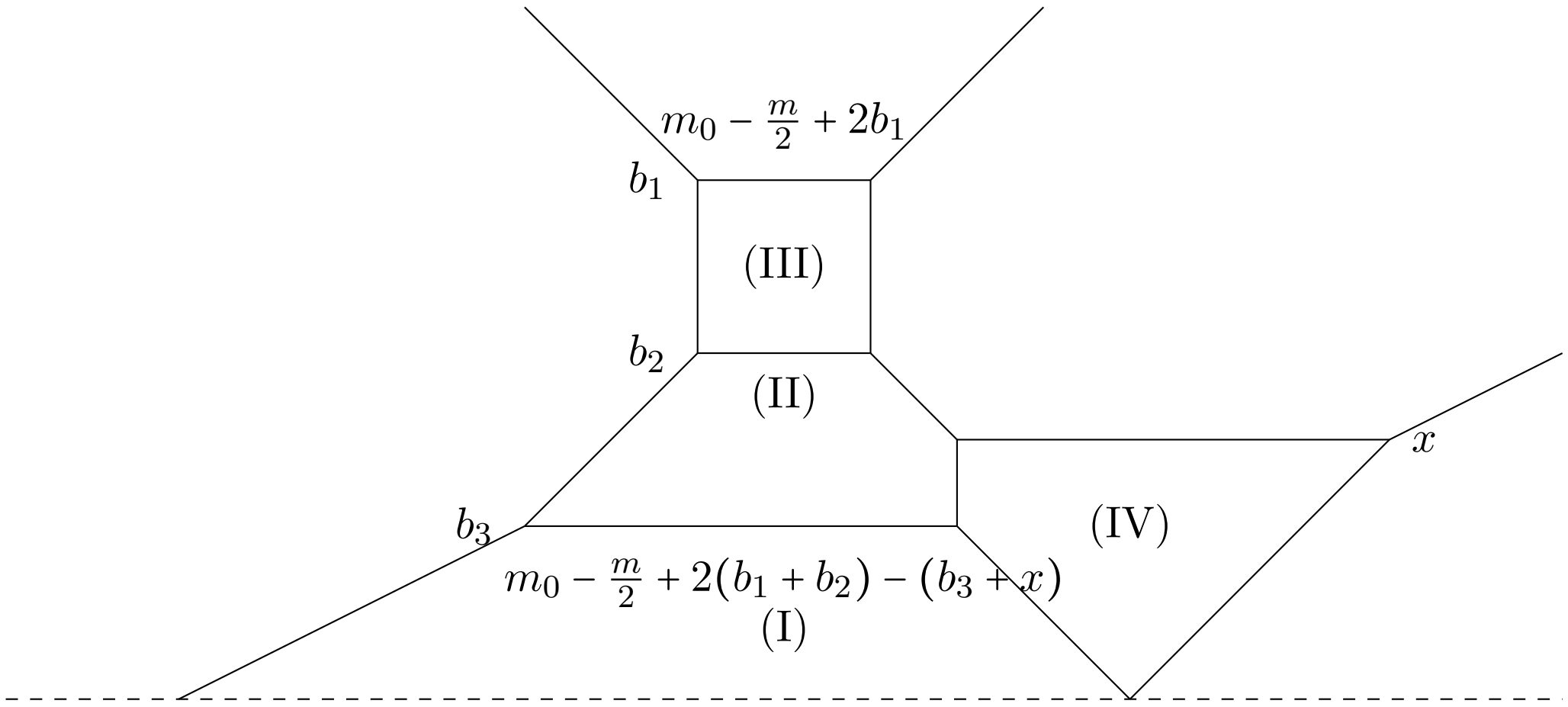
\begin{equation}
    \begin{tikzpicture}[scale =1.5]
        \draw(0,0)--(2,1);
        \draw(2,1)--(3,2);
        \draw(3,3)--(3,2);
        \draw(3,3)--(2,4);
        \draw(4.5,1)--(2,1);
        \draw(4,2)--(3,2);
        \draw(4,3)--(4,2);
        \draw(4,3)--(5,4);
        \draw(4,3)--(3,3);
        \draw(4.5,1.5)--(4,2);
        \draw(4.5,1.5)--(4.5,1);
        \draw(4.5,1.5)--(7,1.5);
        \draw(8,2)--(7,1.5);
        \draw(4.5,1)--(5.5,0);
        \draw(7,1.5)--(5.5,0);
        \draw[dashed](-1,0)--(8,0);
        \node[label=left:{$b_3$}] at (2,1){};
        \node[label=left:{$b_2$}] at (3,2){};
        \node[label=left:{$b_1$}] at (3,3){};
        \node[label=above:{$m_0-\frac{m}{2}+2b_1$}] at (3.5,3){};
        \node[label=below:{$m_0-\frac{m}{2}+2(b_1+b_2)-(b_3+x)$}] at (3.5,1){};
        \node[label=above:{$x$}] at (7.2,1.25){};
        \node at (3.5,2.5){(III)};
        \node at (3.5,1.75){(II)};
        \node at (3.5,.4){(I)};
        \node at (5.5,1){(IV)};
    \end{tikzpicture}
\label{eq:spin6_1s_1flop}
\end{equation}
From the diagram, we read the following area/tensions
\begin{align}
    T_{\mathrm{(I)}}&=\frac{1}{2} b_3 (4 b_1 + 4 b_2 + b_3 + 2 m_0 - m - 2 x )\\
    T_\mathrm{(II)}&=(b_2 - b_3) (2 b_1 +b_2-b_3+ m_0 - \frac{m}{2}) - \frac12 (b_3 - x)^2\\
    T_\mathrm{(III)}&=(b_1-b_2)(m_0-\frac{m}{2}+2b_1)\\
    T_\mathrm{(IV)}&=x^2-\frac{1}{2}(x-b_3)^2 \ .
\end{align}
A consistency check with the gauge theory prepotential reads:

\begin{align}
&\frac{\partial\mathcal{F}_{SU(4)_\frac{1}{2}+1\textbf{F}}}{\partial\phi_1}=2T_{\mathrm{(I)}}+T_\mathrm{(II)}+T_\mathrm{(IV)}=(b_2 + b_3)\left( 2 b_1 + b_2 + b_3 + m_0 -\frac12 m \right)
    \;, \\ &\frac{\partial\mathcal{F}_{SU(4)_\frac{1}{2}+1\textbf{F}}}{\partial\phi_2}=T_\mathrm{(III)}\;,\\
    & \frac{\partial\mathcal{F}_{SU(4)_\frac{1}{2}+1\textbf{F}}}{\partial\phi_3}=T_\mathrm{(II)}.
\end{align}

We find the following two solutions:
\begin{align}
x = \frac12 (b_1+b_2+b_3-2m), \qquad
\frac12 (-b_1 - b_2 + 3b_3 +2 m)\ ,
\end{align}
where the solution consistent with other chambers (flop) is  
\begin{align}
    x = \frac12 (b_1+b_2+b_3-2m) 
    = a_1-m. 
\end{align}

\paragraph{}
\paragraph{Spin(6)+1S after three flops.} The expected gauge theory prepotential is
\begin{equation}
    \mathcal{F}_{SU(4)_{\frac{1}{2}}+1\mathbf{F}}=m_0\sum_{i=1}^4 a_i^2+\frac{1}{6}\sum_{i<j}^4(a_i-a_j)^3-\frac{1}{12}\left( \sum_{i=1}^2(a_i+m)^3-\sum_{i=3}^4(a_i+m)^3 \right)+\frac{1}{12}\sum_{i=1}^4a_i^3.
\end{equation}
The corresponding 5-brane web is given in \eqref{eq:spin6_1s_2flop}.

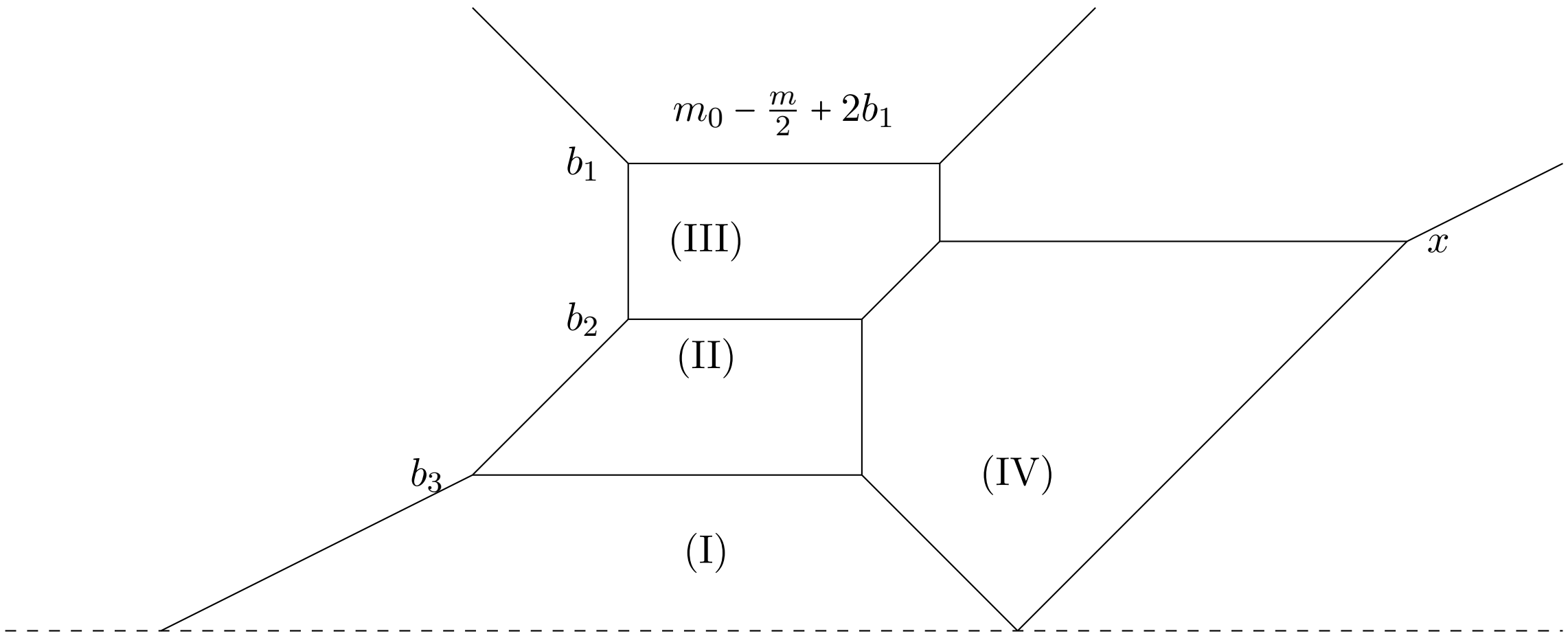
\begin{equation}
    \begin{tikzpicture}[scale =1.5]
        \draw(0,0)--(2,1);
        \draw(2,1)--(3,2);
        \draw(3,3)--(3,2);
        \draw(3,3)--(2,4);
        \draw(4.5,1)--(2,1);
        \draw(4.5,2)--(3,2);
        \draw(5,3)--(5,2.5);
        \draw(5,2.5)--(4.5,2);
        \draw(5,3)--(6,4);
        \draw(5,3)--(3,3);
        \draw(4.5,2)--(4.5,1);
        \draw(5,2.5)--(8,2.5);
        \draw(9,3)--(8,2.5);
        \draw(4.5,1)--(5.5,0);
        \draw(8,2.5)--(5.5,0);
        \draw[dashed](-1,0)--(9,0);
        \node[label=left:{$b_3$}] at (2,1){};
        \node[label=left:{$b_2$}] at (3,2){};
        \node[label=left:{$b_1$}] at (3,3){};
        \node[label=above:{$m_0-\frac{m}{2}+2b_1$}] at (4,3){};
        \node[label=above:{$x$}] at (8.2,2.25){};
        \node[label=below:{
        }] at (3.5,1){};
        \node at (3.5,2.5){(III)};
        \node at (3.5,1.75){(II)};
        \node at (3.5,.5){(I)};
        \node at (5.5,1){(IV)};
    \end{tikzpicture}
    \label{eq:spin6_1s_2flop}
\end{equation}

From the diagram, we read the following lengths, and area/tensions:

\begin{equation}
T_{\mathrm{(I)}}=\dfrac{1}{2}b_3\left(4b_1+4b_2+b_3+2m_0-m-2x\right)
\end{equation}
\begin{equation}
T_\mathrm{(II)}=\dfrac{1}{2}\left(b_2-b_3\right)\left(4b_1+3b_2-b_3+2m_0-m-2x\right)
\end{equation}
\begin{equation}
T_\mathrm{(III)}=\left(b_1-b_2\right)\left(2b_1+m_0-\dfrac{m}{2}\right)-\frac{1}{2}\left(b_2-x\right)^2
\end{equation}
\begin{equation}
T_\mathrm{(IV)}=-\dfrac{b_2^2}{2}-\dfrac{b_3^2}{2}+\left(b_2+b_3\right)x.
\end{equation}

A consistency check with the gauge theory prepotential reads:

\begin{align}
\dfrac{\partial\mathcal{F}_{SU(4)_\frac{1}{2}+1\textbf{F}}}{\partial \phi_2}&=T_\mathrm{(III)}
\\
\dfrac{\partial\mathcal{F}_{SU(4)_\frac{1}{2}+1\textbf{F}}}{\partial \phi_3}&=T_\mathrm{(II)}
\\
\dfrac{\partial\mathcal{F}_{SU(4)_\frac{1}{2}+1\textbf{F}}}{\partial \phi_1}&=2T_{\mathrm{(I)}}+T_\mathrm{(IV)}+T_\mathrm{(II)}\\
&= (b_2+b_3)(m_0-\frac{1}{2}m+2b_1+b_2+b_3)\nonumber
\end{align}

We find 
\begin{align}
x = \frac12 (b_1+b_2+b_3-2m).
\end{align}

\paragraph{}
\paragraph{Spin(6)+1S after four flops.}
The expected gauge theory prepotential is
\begin{equation}
    \mathcal{F}_{SU(4)_{\frac{1}{2}}+1{\bf F}}=m_0\sum_{i=1}^4 a_i^2+\frac{1}{6}\sum_{i<j}^4(a_i-a_j)^3-\frac{1}{12}\left(( a_1+m)^3-\sum_{i=2}^4( a_i+m)^3\right)+\frac{1}{12}\sum_{i=1}^4a_i^3.
\end{equation}

The corresponding 5-brane web is given in \eqref{eq:spin6_1s_3flop}.

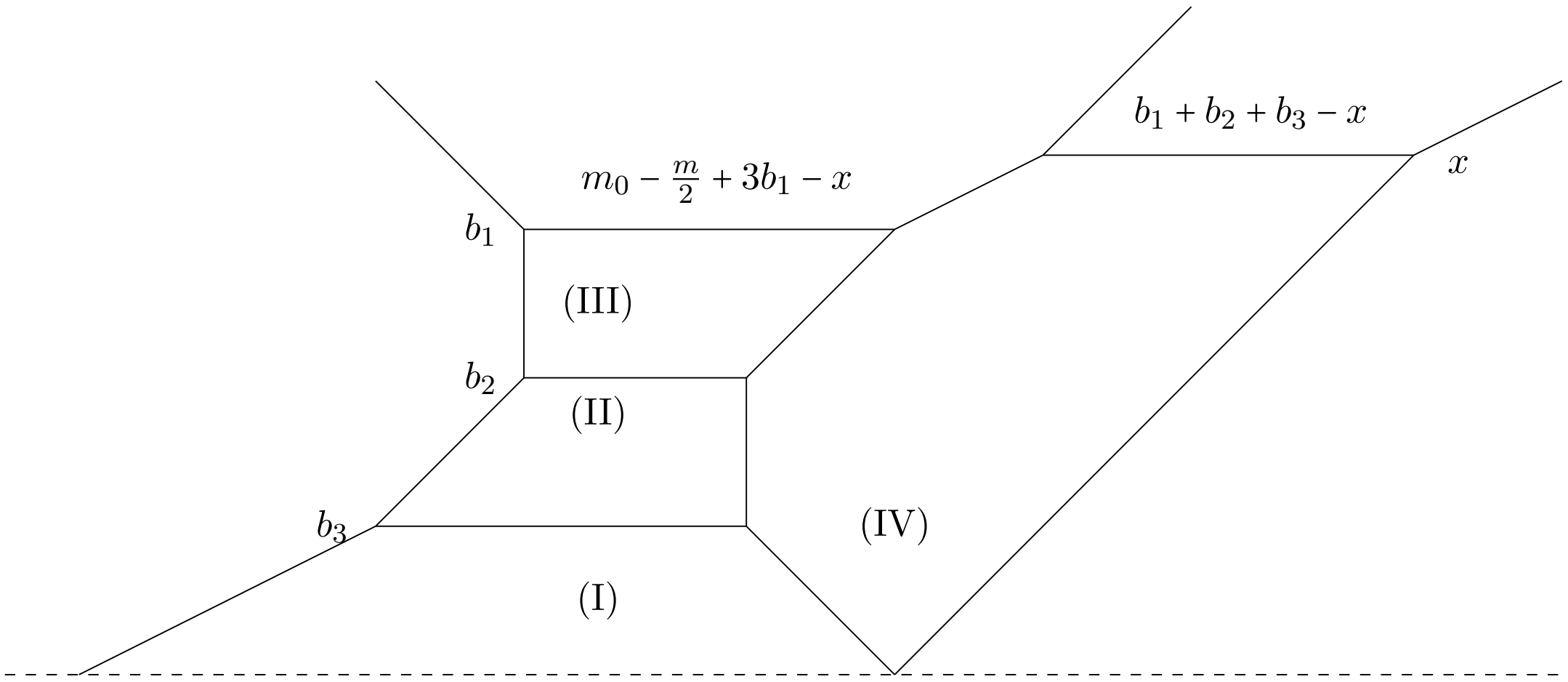
\begin{center}
\begin{equation}
\scalebox{0.9}{
    \begin{tikzpicture}[scale =1.5]
        \draw(-1,0)--(1,1);
        \draw(1,1)--(2,2);
        \draw(2,3)--(2,2);
        \draw(2,3)--(1,4);
        \draw(3.5,1)--(1,1);
        \draw(3.5,2)--(2,2);
        \draw(4.5,3)--(3.5,2);
        \draw(4.5,3)--(2,3);
        \draw(4.5,3)--(5.5,3.5);
        \draw(3.5,2)--(3.5,1);
        \draw(5.5,3.5)--(6.5,4.5);
        \draw(5.5,3.5)--(8,3.5);
        \draw(9,4)--(8,3.5);
        \draw(3.5,1)--(4.5,0);
        \draw(8,3.5)--(4.5,0);
        \draw[dashed](-1.5,0)--(9,0);
        \node[label=left:{$b_3$}] at (1,1){};
        \node[label=left:{$b_2$}] at (2,2){};
        \node[label=left:{$b_1$}] at (2,3){};
        \node[label=above:{$m_0-\frac{m}{2}+3b_1-x$}] at (3.3,3){};
        \node[label=above:{$b_1+b_2+b_3-x$}] at (6.9,3.5){};
        \node[label=above:{$x$}] at (8.3,3.2){};
        \node[label=below:{
        }] at (2.5,1){};
        \node at (2.5,2.5){(III)};
        \node at (2.5,1.75){(II)};
        \node at (2.5,.5){(I)};
        \node at (4.5,1){(IV)};
    \end{tikzpicture}}
    \label{eq:spin6_1s_3flop}
\end{equation}
\end{center}

\begin{align}
T_{\mathrm{(I)}}&=\dfrac{1}{2}b_3\left(4b_1+4b_2+b_3+2m_0-m-2x\right)\\
T_\mathrm{(II)}&=\dfrac{1}{2}\left(b_2-b_3\right)\left(4b_1+3b_2-b_3+2m_0-m-2x\right)\\
T_\mathrm{(III)}&=\frac12\left(b_1-b_2\right)\left(5b_1+b_2+2m_0-{m}-2x\right)\\
 T_\mathrm{(IV)}&=\left(b_1+b_2\right)x -\frac{1}{2}(x-b_3)^2-\dfrac{b_1^2+b_2^2}{2}\ .
\end{align}
A consistency check with the gauge theory prepotential reads:

\begin{align}
\dfrac{\partial\mathcal{F}_{SU(4)_\frac{1}{2}+1\textbf{F}}}{\partial \phi_2}&=T_\mathrm{(III)} \ ,\\ \dfrac{\partial\mathcal{F}_{SU(4)_\frac{1}{2}+1\textbf{F}}}{\partial \phi_3}&=T_\mathrm{(II)} \ , \\
\dfrac{\partial\mathcal{F}_{SU(4)_\frac{1}{2}+1\textbf{F}}}{\partial \phi_1}&=2T_{\mathrm{(I)}}+T_\mathrm{(IV)}+T_\mathrm{(II)}\\
&= -\frac{b_1^2}{2}
+ b_2^2+ b_3^2+ b_3 m_0
+ b_2 \left(2 b_3 + m_0 - \frac{m}{2}\right)
- \frac{b_3 m}{2}
- \frac{x^2}{2}
+ b_1 (2 b_2 + 2 b_3 + x).    \nonumber
\end{align}

We find 
\begin{align}
x = \frac12 (b_1+b_2+b_3-2m)\ .
\end{align}

\paragraph{}
\paragraph{Spin(6)+1S after five flops.}
The expected gauge theory prepotential is
\begin{equation}
    \mathcal{F}_{SU(4)_{\frac{1}{2}}+1\mathbf{F}}=m_0\sum_{i=1}^4 a_i^2+\frac{1}{6}\sum_{i<j}^4(a_i-a_j)^3+\frac{1}{12}\sum_{i=1}^4( a_i+m)^3+\frac{1}{12}\sum_{i=1}^4a_i^3\ .
\end{equation}

The corresponding 5-brane web is given in \eqref{eq:spin6_1s_4flop}.

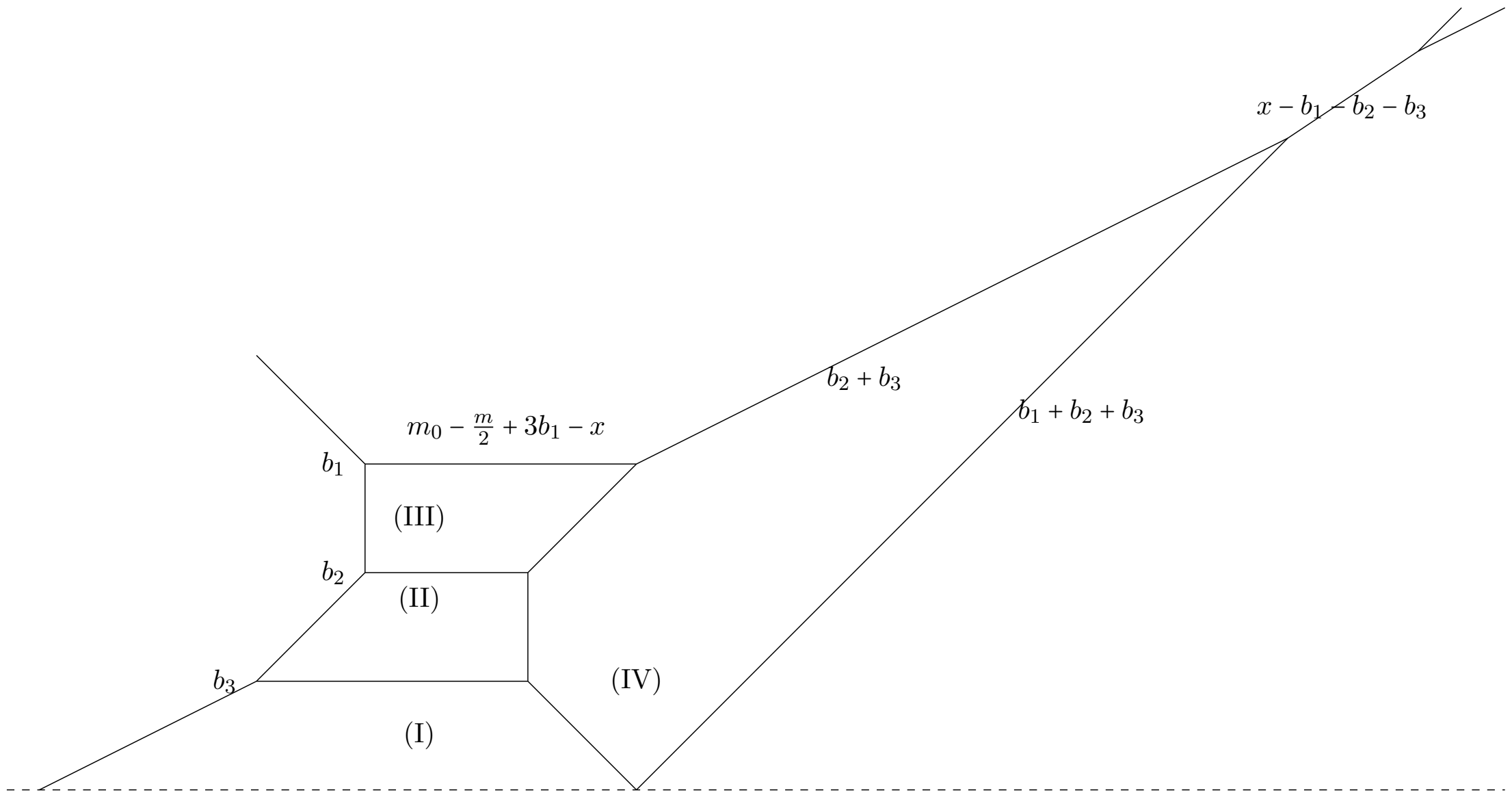
\begin{center}
\begin{equation}
\scalebox{0.7}{
    \begin{tikzpicture}[scale =1.5]
        \draw(-1,0)--(1,1);
        \draw(1,1)--(2,2);
        \draw(2,3)--(2,2);
        \draw(2,3)--(1,4);
        \draw(3.5,1)--(1,1);
        \draw(3.5,2)--(2,2);
        \draw(4.5,3)--(3.5,2); 
         \draw(4.5,3)--(2,3);
        \draw(4.5,3)--(5.5+5,3.5+2.5);
        \draw(3.5,2)--(3.5,1);
        \draw(11.7,6.8)--(10.5,6);
        \draw(11.7,6.8)--(12.5,7.2);
        \draw(11.7,6.8)--(12.1,7.2);
        \draw(3.5,1)--(4.5,0);
        \draw(10.5,6)--(4.5,0);
        \draw[dashed](-1.3,0)--(12.5,0);
        \node[label=left:{$b_3$}] at (1,1){};
        \node[label=left:{$b_2$}] at (2,2){};
        \node[label=left:{$b_1$}] at (2,3){};
        \node[label=above:{$m_0-\frac{m}{2}+3b_1-x$}] at (3.3,3){};
        \node[label=above:{$b_2+b_3$}] at (6.6,3.5){};
        \node[label=above:{$b_1+b_2+b_3$}] at (8.6,3.2){};
        \node[label=above:{$x-b_1-b_2-b_3$}] at (11,6){};
        \node[label=below:{
        }] at (2.5,1){};
        \node at (2.5,2.5){(III)};
        \node at (2.5,1.75){(II)};
        \node at (2.5,.5){(I)};
        \node at (4.5,1){(IV)};
    \end{tikzpicture}}
    \label{eq:spin6_1s_4flop}
\end{equation}
\end{center}

In \eqref{eq:spin6_1s_4flop}, 
$T_{\mathrm{(I)}}, T_\mathrm{(II)}$, and $T_\mathrm{(III)}$ are the same as in the {\bf Spin(6)+1S after four flops} section. 
The tension $T_\mathrm{(IV)}$ is given by
\begin{align}
 T_\mathrm{(IV)}&=b_1b_2+b_1b_3+b_2b_3\ ,
\end{align}
which is $x$-independent. A consistency check with the gauge theory prepotential reads:

\begin{align}
\dfrac{\partial\mathcal{F}_{SU(4)_\frac{1}{2}+1\textbf{F}}}{\partial \phi_2}&=T_\mathrm{(III)} \ ,\qquad\qquad \\
\dfrac{\partial\mathcal{F}_{SU(4)_\frac{1}{2}+1\textbf{F}}}{\partial \phi_3}&=T_\mathrm{(II)}\ , \\
\dfrac{\partial\mathcal{F}_{SU(4)_\frac{1}{2}+1\textbf{F}}}{\partial \phi_1}&=2T_{\mathrm{(I)}}+T_\mathrm{(IV)}+T_\mathrm{(II)}\\
&= \frac12 (b_2+b_3)(6b_1+3b_2+3b_3+2m_0-m-2x)  \ . \nonumber
\end{align}

We find 
\begin{align}
x = \frac12 (b_1+b_2+b_3-2m).
\end{align}

\bibliography{ref_line}
\end{document}